\documentclass[fleqn,usenatbib]{mnras}

\usepackage{graphicx}	
\usepackage{amsmath}	
\usepackage{amssymb}	
\usepackage{multicol}        
\usepackage{bm}		

\newcommand{\note}[1]{\textcolor{black}{#1}}
\newcommand{\newnote}[1]{\textcolor{black}{#1}}
\newcommand{\newnewnote}[1]{\textcolor{black}{#1}}

\usepackage[T1]{fontenc}
\usepackage{ae,aecompl}

\usepackage{newtxtext,newtxmath}

\title[Galactic chimney sweeping]{Galactic chimney sweeping: the effect of `gradual' stellar feedback mechanisms on the evolution of dwarf galaxies}
\author[Lilian Garratt-Smithson, Graham A. Wynn, Chris Power, C. J. Nixon]{Lilian Garratt-Smithson$^{1,2}$\thanks{lilian.garratt-smithson@uwa.edu.au}, Graham A. Wynn$^{1}$, Chris Power$^{2}$ and C. J. Nixon$^{1}$\\
  $^{1}$Department of Physics \& Astronomy,University of Leicester,Leicester,LE1 7RH,UK \\
  $^{2}$International Centre for Radio Astronomy Research,University of Western Australia,35 Stirling Highway,Crawley,Western Australia 6009 \\
  Australia \\
	 }

\begin{document}
\label{firstpage}
\pagerange{\pageref{firstpage}--\pageref{lastpage}}
\maketitle

\begin{abstract}
 We investigate the \newnote{impact} of time-resolved `gradual' stellar feedback processes \newnote{in high redshift dwarf spheroidal galaxies.} Here `gradual' feedback refers to individual stellar feedback events which deposit energy over a period of time. We conduct high resolution hydrodynamical simulations of dwarf spheroidal galaxies \newnote{with halo masses of 10$^7$ M$_{\odot}$ - 10$^8$ M$_{\odot}$, based on z = 6 progenitors of the Milky Way's dwarf spheroidal galaxies.} We also include a novel feedback prescription for individual massive stars, which includes stellar winds and a HMXB (High Mass X-ray Binary) phase, on top of supernovae. We find the mass of gas unbound across a 1 Gyr starburst is uniformly lowered if gradual feedback mechanisms are included \newnote{across the range of metallicities, halo concentration parameters and galaxy masses studied here.} Furthermore, we find including gradual feedback in the smallest galaxies delays the unbinding of the majority of the gas and facilitates the production of `chimneys' in the dense shell surrounding the feedback generated hot, pressurised `superbubble'. These `chimneys' vent hot gas from the galaxy interior, lowering the temperature of the central 10 kpc of the gaseous halo. Additionally, we find radiative cooling has little effect on the energetics of simulations that include a short, violent starburst compared with those that have a longer, less concentrated starburst. Finally, we investigate the relative impact of HMXB feedback and stellar winds on our results, finding the ubiquity of stellar winds throughout each starburst makes them a defining factor in the final state of the interstellar medium. 
\end{abstract}




\section{Introduction}
Stellar feedback is an important aspect of galaxy evolution. Locally, it drives supersonic turbulence in the inter-stellar medium (ISM) which inhibits the gravitational collapse of the cold molecular gas in molecular clouds (MCs) \citep[for a review of the processes driving interstellar turbulence see][]{Klessen2016}, as well as lowers the star formation efficiency of GMCs (giant molecular clouds) to a few percent \citep{Myers1986, Evans2009, Murray2010, Dale2011, Walch2012,  Krumholz2014, Skinner2015, Rahner2017}. On large scales supernova (SN) feedback has been found to drive galactic-scale outflows, affecting the global star formation history of a galaxy \citep[e.g.][]{Oppenheimer2010, Creasey2013, Hopkins2014, Peters2015, Girichidis2016}. This makes it important to investigate stellar feedback on resolved scales. However, this large dynamic range makes it difficult to numerically model stellar feedback effects. 

Stars can impact their environment through a number of processes; photo-ionisation \citep[e.g.][]{Dale2014, Geen2015, Walch2015}, radiation pressure \citep[e.g.][]{Murray2010, Krumholz2012, Kim2016}, line-driven stellar winds \citep[e.g.][]{Rogers2013, Mackey2015, Fierlinger2016}, SNe explosions \citep[e.g.][]{Martizzi2015, Walch2015a, Haid2016}. For a review on the numerical implementation of stellar feedback processes see \citep{Dale2015}. In this paper we focus on feedback from massive stars, motivated by the fact it is these stars that dominate the energy input from a stellar population \citep[e.g.][]{Agertz2013}. The efficiency of each feedback process at coupling to the ISM and ultimately suppressing star formation (on galactic and local scales) is highly dependent on environment, in particular the ISM density/ morphology \citep[e.g.][]{Dwarkadas2012, Martizzi2015, Li2015, Kim2015, Iffrig2015, Gatto2015}, which is altered by the action of prior feedback events \citep[e.g.][]{Rogers2013, Walch2015}. 

In this paper we are particularly interested in the interplay between `gradual' types of feedback and `instantaneous' feedback. We use the term `gradual' in order to refer to individual time-resolved stellar feedback events which are continuous over a set period; for example the energy input from stellar winds. By `instantaneous' feedback we mean an instant explosive event - i.e. a SN. This work follows on from work in \cite{Garratt-Smithson2018}, hereafter GS18, which investigated the gradual heating of the ISM via the jets associated with High Mass X-ray Binaries (HMXBs) and how this interplayed with SNe in GMCs (Giant Molecular Clouds). It was found the gradual heating of HMXBs acted to enhance the removal of gas via hot, low density chimneys in the cloud, which also acted to prolong star formation and increase star formation efficiency. 

During this paper we focus on stellar feedback in dwarf spheroidal galaxies. The dwarf galaxies of the Milky Way have a wide range of metallicities, luminosities, stellar populations and hence star formation histories \citep[e.g.][]{Tolstoy2009, McConnachie2012}. The shallow potential wells of dwarf galaxies means they can be significantly influenced by a number of external factors, for example; ram pressure stripping, cosmic re-ionisation and tidal effects \citep[e.g][]{Gatto2013, Emerick2016, Sawala2016, Zhu2016, Simpson2017}. Moreover, stellar feedback events such as SNe are thought to drive hot, metal-enriched gas from dwarf galaxies, as evidenced by a high abundance of metals in the IGM \citep{Schaye2003}, along with observations of `superbubbles' of hot, diffuse gas in star forming dwarf galaxies \citep[e.g.][]{Ott2005}. Stellar feedback is also thought to drive the cusp-core transformation of the underlying dark matter halo of dwarf galaxies \citep[e.g.][]{Trujillo2015}. For a review on this subject see \citet{Pontzen2014}. Furthermore, feedback from SNe is considered the primary cause of the observed flattening of the M$_{halo}$ - M$_{\star}$ relation seen with galaxies of mass $<$ 10$^9$ M$_{\odot}$ \citep[e.g.][]{Sawala2015}. Additionally, recent work has focussed on the stochasticity of SNe events and how this can drive the range of star formation histories seen in the dwarf spheroidal satellite galaxies of the Milky Way today \citep[e.g.][]{Ricotti2005, Weisz2014}. 

The dwarf galaxies we are investigating here are of relatively low density (the peak density in our canonical simulations is $\sim$ 10$^6$ M$_\odot$ kpc$^{-3}$), which results in longer cooling timescales and increases the efficiency of heating from shocked stellar winds and SNe. These two processes are expected to dominate over other stellar feedback processes in the creation of galactic-scale gaseous outflows \citep{Hopkins2012b}. 

Our choice to simulate individual isolated galaxies instead of a cosmological box simulation was due to our requirement to resolve individual massive stellar feedback events. Cosmological simulations have the benefit of being able to track galaxy formation from the initial formation of dark matter haloes, while also taking into account merger events and the influence of a galaxy's environment on its evolution.  However, the resolution in these large-scale simulations is such that stellar feedback is modelled as the integrated output of whole stellar populations, encompassing multiple SNe events. This means details of the smaller-scale physics are lost. For example, in their recent paper, \citep{Su2017} argue modelling SNe as discrete events in time and space is pivotal to capturing the multi-phase ISM, galactic winds and reasonable stellar masses in dwarf galaxies. Furthermore, \cite{Emerick2019} showed the importance of modelling feedback in isolated dwarf galaxies on a star-by-star basis, finding this is necessary to produce the observed trends in metallicity and star formation rate density. 

Our work also compliments \citet{Cashmore2017}, where the authors conducted a similar set of simulations of isolated dwarf galaxies and concluded dwarf galaxies would require an unusual set of properties in order to sustain star formation beyond a 1 Gyr starburst, if feedback from SNe was included. We aim to investigate whether the inclusion of `gradual' feedback mechanisms affects this result. Previous work by \citet{Artale2015} concluded the addition of HMXB feedback on top of SNe feedback led to an decrease in the star formation rate of low mass galaxies ($<$ 10$^{10}$ M$_{\odot}$), however an increase in overall star formation efficiency. This result was also hypothesised by \citet{Justham2012}, who concluded X-ray Binaries have the capacity to warm the ISM without unbinding it, leading to further star formation at later times. 

The mechanical luminosity of X-ray binary systems have gained interest over the last 10 years due to observations such as \citet{Gallo2005}, which found the kinetic power of the jets associated with Cygnus X-1 is comparable to its X-ray luminosity. This was further evidenced by observations of the relativistic jets associated with SS433, which point to a mechanical luminosity of $> 10^{39}$ erg/s  \citep[e.g.][]{Blundell2001,Mirabel2011,Goodall2011}. 

Furthermore, there is also evidence HMXBs play an enhanced role in higher redshift, lower metallicity environments \citep[e.g.][]{Basu-Zych2013a, Basu-Zych2013b, Fragos2013}, both due to these systems being more numerous \citep[e.g.][]{Dray2006, Douna2015}, as well as more luminous, due to a higher fraction following the ULX/Roche-Lobe overflow evolutionary pathway \citep[e.g.][]{Linden2010}. Blue Compact Dwarf galaxies (BCDs) are often used to investigate this hypothesis, since they are low metallicity and are thought to be analagous to high redshift environments. Recent work into these systems has found a significant increase in the HMXB population of BCDs, compared with solar metallicity environments \citep[e.g.][]{Kaaret2011, Brorby2014}.

Projects such as BPASS \citep{Eldridge2017} have also highlighted the importance of considering binaries when implementing stellar feedback. BPASS is a stellar population synthesis code which includes the effects of binary evolution. For example, using BPASS it has been found binary interactions could play an important role during re-ionisation \citep[e.g.][]{Stanway2016}, increasing the population of massive stars beyond 3 Myr into star formation and hence providing a supply of ionising photons that can escape through channels/ chimneys carved in the surrounding ISM by previous stellar feedback events \citep{Ma2016}. Moreover, GS18 found the formation of these chimneys is also enhanced by the heating from HMXBs.

In this paper we present simulations that include stellar wind feedback phase and a SN phase for individual massive stars, and also a HMXB phase and a second SN (where a HMXB is also considered present) for the relevant binary stars. In order to investigate the combined effect of `gradual' and `instant' feedback, we have conducted very high resolution simulations of individual dwarf galaxies; this enables us to resolve the effects of single massive stellar feedback events. Previous work has found the action of stellar winds on the surrounding ISM prior to a SN can reduce the coupling between the SN energy and the ISM \citep[e.g.][]{Rogers2013}. Furthermore, in their recent paper, \cite{Emerick2018} also focussed on modelling individual stellar feedback events in isolated dwarf galaxies, finding pre-SNe feedback events are integral for SNe to produce galactic-scale outflows. We aim to investigate the impact the inclusion of gradual feedback has on the ability of a star-forming dwarf galaxy to retain gas to fuel later episodes of star formation. 

\section{Numerical Model}\label{sec:Num}
This work uses GADGET-3, a modified version of the hybrid N-body/ SPH (smoothed particle hydrodynamics) code GADGET-2 \citep{Springel2005}. We use the SPHS method \citep{Read2012} in order to help model mixing of feedback-generated multiphase gas. Furthermore, we use a Wendland-2 kernel with 100 neighbours for the gas \citep{Wendland1995, Dehnen2012}, coupled with an ideal gas equation of state. This is governed by the relationship: $P=(\gamma-1)\rho u$ (where $P$ is the gas pressure, $\gamma$ is the adiabatic constant, set to $5/3$ in our simulations, $u$ is gas internal energy and $\rho$ is particle density). The gas particles have both an adaptive smoothing and softening length, with a minimum softening length of 0.1 pc. We also model cooling down to 20 K using two different schemes. Firstly, we use look-up tables generated using MAPPINGS III \citep{Sutherland1993} for gas temperatures down to 10$^4$ K, based on the metallicity of each simulation. Below 10$^4$K we use the method outlined in \citet{Mashchenko2008} in order to model the fine-structure metal line cooling of the metals in the gas. For the runs at primordial metallicity we take an [Fe/H] value of -6 (corresponding to 10$^{-6}$ times solar). 

In general we use a gas particle mass resolution of 9 M$_\odot$. In order to model the dark matter halo of the galaxy we used 10$^{5}$ N-body particles. For our canonical runs, this means a mass resolution of 1100 M$_\odot$. These have a set softening length of 10 pc (based on R$_{200}$/N$_{200}^{0.5}$, where R$_{200}$ is the virial radius of the halo and N$_{200}$ is the number of dark matter particles within R$_{200}$ - which we take to be the full 10$^{5}$). We verify that our choice of softening length is not a defining factor in our results in Appendix \ref{appendix:soft}. Furthermore, we include stellar populations containing massive stars as star particles within our simulation. These have a fixed softening length of 0.1 pc, corresponding to the minimum softening length of the SPH particles. A sink particle is included at the centre of the galaxy. This repositions on the point of minimum potential within its 100 neighbours and is included to remove gas particles with prohibitively small times.pdf by accreting any particles within 0.5 pc, that are also gravitationally bound to the sink particle.

We also include a sink particle formation criterion which sets the Jeans mass of the gas particle as the minimum resolvable mass, which in SPH is 2 N$_{neigh}$m$_{p}$, \newnote{where m$_{p}$ is the gas particle mass and N$_{neigh}$ is the number of SPH neighbours} \citep{Bate1997}. In order to form a sink particle, gas particles must have a density greater than
\begin{equation}\label{eqn:Jeans_rho}
\rho_{\rm J} = \left(\frac{\pi k_{\rm B} T}{G\mu m_{\rm H}}\right)^{3} \frac{1}{(2N_{\rm neigh}m_{\rm p})^{2}}
\end{equation}
where $k_{\rm B}$ is the Boltzmann constant, \newnote{$T$ is the temperature of the gas particle}, $\mu$ is the mean molecular weight and $m_{\rm H}$ is the mass of Hydrogen. We also require that the gas must be converging (i.e. $\nabla \cdot \textbf{v} < 0$) and that the temperature of the gas particle must be $< 500$\,K (in order to ensure high temperature and high density gas particles are not considered star forming). We set the mean molecular weight in equation \ref{eqn:Jeans_rho} to 1.24 in all simulations (note this is not the case when calculating the cooling rates, where $\mu$ is calculated self-consistently based on the electron fraction). We do not have the mass resolution to follow the formation of individual stars in our simulations, however this sink criterion exists to remove high density SPH particles with prohibitively small times.pdf. 

\section{Initial Conditions}\label{ICs}
\begin{table*}
 \caption{A table summarising the initial conditions of the simulations run in this paper. The columns listed are: name of run, \newnote{total dark matter halo mass M$_{dm}$, total gas mass M$_{g}$, mass of a single gas particle m$_p$,} dark matter halo concentration parameter $c$, the virial radius of the dark matter halo (R$_{vir}$), gas metallicity, given as [Fe/H] (the log$_{10}$ of the ratio between the metal content of the galaxy compared with that of our sun), the standard deviation in the `wake-up' times of the star particles ($\sigma_{star}$) and finally the types of feedback included.}
  \begin{tabular}{|l|l|l|l|l|l|l|l|l|}\label{tab:Runs}
    Run&M$_{dm}$ (10$^7$ M$_{\odot}$)&M$_g$ (10$^7$ M$_{\odot}$)&m$_p$ (M$_{\odot}$)&$c$&R$_{vir}$ (kpc)&[Fe/H]&$\sigma_{star}$ (Gyr)&Feedback Included\\
    \hline
    1&11.0&1.8&8.8&3.54&7.78&-6&0.13&SNe, HMXBs, stellar winds\\
    \hline
    2&11.0&1.8&8.8&3.54&7.78&-6&0.13&SNe\\
    \hline
    3&11.0&1.8&8.8&7.08&7.78&-6&0.13&SNe, HMXBs, stellar winds\\
    \hline
    4&11.0&1.8&8.8&7.08&7.78&-6&0.13&SNe\\
    \hline
    5&1.5&0.24&1.2&3.68&4.04&-6&0.13&SNe, HMXBs, stellar winds\\
    \hline
    6&1.5&0.24&1.2&3.68&4.04&-6&0.13&SNe\\
    \hline
    7&11.0&1.8&8.8&3.54&7.78&-1.2&0.13&SNe, HMXBs, stellar winds\\
    \hline
    8&11.0&1.8&8.8&3.54&7.78&-1.2&0.13&SNe\\
    \hline
    9&11.0&1.8&8.8&3.54&7.78&-6&0.06&SNe, HMXBs, stellar winds\\
    \hline
    10&11.0&1.8&8.8&3.54&7.78&-6&0.06&SNe\\
    \hline
    NoHMXB&11.0&1.8&8.8&3.54&7.78&-6&0.13&SNe, stellar winds\\
    \hline
    NoWinds&11.0&1.8&8.8&3.54&7.78&-6&0.13&SNe, HMXBs\\
    \hline
   
  \end{tabular}
\end{table*}
Our initial conditions are summarised in Table \ref{tab:Runs}. We chose the initial conditions of this paper to represent z$\sim$ 6 progenitors of the classical dwarf spheroidal satellite galaxies of the Milky Way. At this point the galaxies are massive enough to support cooling predominately via atomic and molecular Hydrogen (given the virial temperatures of 6000 K and 1600 K for the largest and smallest haloes respectively) \citep[e.g.][]{Glover2005, Moore2006, Power2014} and the majority of the gas has had time to cool and virialise. Current day halo masses range from 10$^{8-9}$ M$_{\odot}$ \citep[e.g.][]{Walker2007}. Depending on their merger tree, from \citet{Power2014} we expect the 10$^{8}$ M$_{\odot}$  redshift z=0 halos to have a mass of $\sim$ 1.5$\times$ 10$^{7}$ M$_{\odot}$ at z$=$6 and the 10$^{9}$ M$_{\odot}$ (z$=$0) haloes to have a mass of 1.1$\times$ 10$^{8}$ M$_{\odot}$ at z$=$6. We set up each halo according to a Hernquist density profile \citep{Hernquist1990}, with the virial radius (r$_{200}$) set according to 
\begin{equation}\label{eqn:rvir}
\centering r_{200} = \left(\frac{M_{200} G}{100 H^{2}}\right)^{\frac{1}{3}}
\end{equation}
where M$_{200}$ is the virial halo mass (which we set accordingly) and H is the Hubble constant. \newnote{The number of gas particles used throughout our simulations is 2 $\times$ 10$^6$}. We based the concentration parameters on equation 20 of \citet{Correa2015}, which is a fitting function dependent on just redshift and halo mass. The function was fit using WMAP5 cosmology and is valid for z$>$4, at all halo masses. Using this fitting function, we obtained concentrations of 3.54 and 3.68 for halo masses of $1.1\times 10^{8}$ M$_{\odot}$ and 1.5$\times$10$^{7}$ M$_{\odot}$ respectively. We also perform a simulation with double the halo concentration for the halo of mass 1.1$\times$10$^{8}$ M$_{\odot}$, in order to investigate the effect this has on our results. 

As discussed in the introduction, the SFH (star formation history), primarily obtained using synthetic colour magnitude diagrams (CMDs) \citep[see review by][]{Tolstoy2009}, varies widely between dwarf spheroidal galaxies. This means there is likely to be a variation in the level of metal enrichment in the galaxies, dependent largely on the number of massive stars that have left the main sequence up until this point. We investigate the effect of altering the metallicity, Z, of the gas in the galaxy by varying it between two values: [Fe/H] = -6 and [Fe/H] = -1.2. This alteration in metallicity will manifest as a difference in the cooling rate of the gas, along with a difference in the lifetimes of the massive stars; which in turn will alter the duration of stellar wind feedback and HMXB feedback. We do not follow metal enrichment in our simulations, however this is something we would like to implement in future simulations.

Furthermore, we assume the baryons in the galaxy correspond to the 0.16 baryon fraction of the universe \citep{Planck2014}. We further assume the gas follows the same underlying density profile as the dark matter. The temperature of the gas in the centre of the halo is set as virial, with a gradual drop at larger radii in accordance with \citet{Komatsu2001}. We insert 100 star particles into each simulation, each representing a stellar population containing one massive star in a binary system \citep[this assumption is based on the high multiplicity seen in papers such as][]{Sana2013}. \newnote{This number is conservative and chosen to be in line with work by \citet{Cashmore2017}, to enable a direct comparison.}

\newnote{We introduce these star particles, as opposed to forming them in situ, due to the fact we are primarily concerned with the ability of the dwarf galaxies to retain gas given alterations in starburst parameters (for example the violence of the burst), which we are able to control with a pre-existing population. Moreover, although we do include a star formation prescription (see Section \ref{sec:Num}), given the typical mass resolution of our simulations is 8.8 M$_{\odot}$ (see Table \ref{tab:Runs}), following equation \ref{eqn:Jeans_rho} we see, at a minimum temperature of 20 K, the required Jeans density for star formation to occur is 2 $\times$ 10$^{-23}$ gcm$^{-3}$ (assuming an average neighbour number of 100). This is 3-4 orders of magnitude higher than the maximum gas density seen in the simulations (e.g. see figures \ref{fig:TempVdens_R1v2}). At the highest resolution runs (Runs 5 and 6, with gas particle mass 1.2 M$_{\odot}$), the corresponding Jeans density is 1.1$\times$10$^{-21}$ gcm$^{-3}$, which is 4-5 orders of magnitude higher than the highest gas density seen during the starburst (see figure \ref{fig:R56_TvD}). We therefore see no star formation during these simulations.}

The mass of each star particle is set to 30 M$_{\odot}$ and they are placed at random positions consistent with a stellar bulge that follows a Hernquist density profile with a scale radius of 0.1 times that of the halo. In this way the stellar population of the galaxy contributes to $<$ 0.02$\%$ of the total galaxy mass and is included purely to \newnote{act as tracer particles that} represent the locations of massive star feedback events within the galaxy.

We allowed our initial conditions to relax for 1 Gyr and we also ran a set of simulations with no feedback included in order to ascertain the changes in the baryons and dark matter we see in this paper are due to stellar feedback. We show the results of these in Appendix \ref{appendix:Nofb}.

\subsection{Massive Star Population}\label{sec:Massive_stars}
The properties of the massive stars are set at the beginning of each simulation, using the Monte-Carlo approach described in GS18. We will briefly summarise this method here. Each star particle is assumed to host at least one massive star. The mass of this star is sampled from a Kroupa IMF between 8 M$_\odot$ -- 100 M$_\odot$ (based on the progenitors of Type-II SNe). \newnote{This means some star particles will be assigned virtual primary star masses which are greater than that of the particle itself. This effects 18 out of the 100 star particles and the slight alteration in gravitational interactions is expected to have a minimal result on the results of this paper. As stated above, the purpose of these particles is to act as tracers for the massive stellar population locations.} Beyond this, the star is given a probability of 0.14 of becoming a HMXB. This number is based on Table 2 of GS18, which in turn was found using flat mass distribution of binary mass ratios \citep{Sana2013}, along with a survival criterion dependent on whether or not the binary system maintains more than half its mass during the SN of the primary star. If the primary star is considered to be in a HMXB system, the secondary mass is sampled using the same method as the primary star.

Both primary and secondary stars were assigned lifetimes according to look-up tables based on their mass and metallicity. As in GS18, the [Fe/H] = -1.2 runs used Table 46 from \citet{Schaller1992}. For the runs at primordial metallicity, we used Table 2 from \citet{Ekstrom2008}. In both cases we added the lifetimes of the Hydrogen and Helium burning phases in order to estimate the total stellar lifetime. 

\section{Feedback Prescriptions}\label{sec:feedback_prescription}
We assigned `wake-up' times for each star particle, based on a Gaussian distribution of set standard deviation, or $\sigma_{star}$, with a mean set to 0.5 Gyr. The majority of the simulations were run with $\sigma_{star}$ equal to 0.13 Gyr (see Table \ref{tab:Runs}). However a subset were also run with a $\sigma_{star}$ equal to 0.06 Gyr. The smaller the value of $\sigma_{star}$, the shorter and more violent the starburst (given the energy injected across the starburst is the same). Once the simulation has progressed to a star particle's `wake-up' time, the star particle will initialise \note{stellar feedback}. \note{All star particles undergo stellar wind feedback prior to SNe feedback and then a subset will also go on to a further HMXB feedback phase and a final SN event. The lifetime of the stellar wind feedback and HMXB feedback is determined by the lifetime of the primary and companion star respectively.}

We implement the shock-heating from stellar winds as a thermal energy injection into the 100 SPH neighbours of each star particle, at constant power and across multiple times.pdf until the end of the lifetime of the primary star. The energy injected into an individual SPH particle is kernel-weighted, while the total internal energy injected by a star particle is determined by a set power 10$^{35}$ erg/s and is proportional to its timestep. We chose this power input based on a wind velocity of 1000 km/s \citep{Leitherer1992} and a mass outflow rate of 10$^{-6}$ M$_\odot$/yr \citep{Repolust2004}.

Once the lifetime of the primary star has been reached, the star particle then undergoes a single SN feedback event and injects 10$^{51}$ erg of thermal energy into its surrounding 100 neighbours in one time-step. Beyond this, if the massive star has been determined to be part of a HMXB \note{(see section \ref{sec:Massive_stars})}, the particle then undergoes HMXB feedback, which lasts the lifetime of the companion star. If not, the star particle ceases all feedback. 

HMXB feedback is also implemented as an internal energy injection of set power; 10$^{36}$ erg/s. We chose this power based on estimations of the power output of the wind-fed jet in Cygnus X-1, which is between 10$^{35}$ to 10$^{37}$ erg/s. This value is conservative when considering the estimated power output of SS433, which is considered to be a ULX on its side \citep{Begelman2006} and outputs $\sim$ 10$^{39}$ erg/s into the ISM. \newnote{Using this method we have ensured HMXB feedback is occuring in a pre-processed environment, which is physical since HMXBs are typically linked with star-forming regions \citep[e.g.][]{Mineo2012, Gilfanov2004}.}

\newnote{In this way the key difference between stellar winds and HMXB feedback is their number (N$_{HMXB} = $ 0.14 N$_{winds}$), relative time delay (stellar winds act earlier, prior to SNe feedback, however HMXBs are always preceded by both SNe and stellar winds) and power (HMXBs are ten times more powerful). Furthermore, looking at Figure \ref{fig:Lifetimes} we can see HMXBs all have lifetimes (set by the companion lifetimes, lower plot) close to the upper tail of the stellar wind lifetime distribution (in other words the primary lifetime, upper plot), hence act over longer periods on average.}

This work neglects the stellar luminosities of the massive main sequence stars, which for a star with a mass of 8 M$_\odot$ is $\sim$ 10$^{37}$ erg/s \citep[following calculations by][at solar metallicity]{Bressan1993}. However the impact of this radiation and its ability to produce momentum-driven outflows depends on the coupling of this energy to the ISM. For example, \citet{Krumholz2009} find if the expansion of massive star wind-fed bubbles are energy driven, the kinetic energy of the momentum-driven shell produced by the stellar radiation is $\sim$ 100 times smaller. Moreover, \citet{Rogers2013} argue, based on \citet{Krumholz2009}, provided the leakage of wind energy is comparable to the leakage of stellar photons, radiation pressure will be of the same order as the wind pressure.

At the end of the companion lifetime \note{all HMXB-containing} star particles will then undergo one more SN feedback event and feedback will cease for these particles. A limitation to our method is that we do not investigate the anisotropy of the jet feedback associated with HMXBs - which is dependent on the jet precession, power and the density of the surrounding ISM \citep[e.g.][]{Goodall2011}. In future it would be of interest to compare and contrast wind feedback and HMXB feedback by including this anisotropy as an additional variable.  

\begin{figure} 
  \includegraphics[width=\columnwidth]{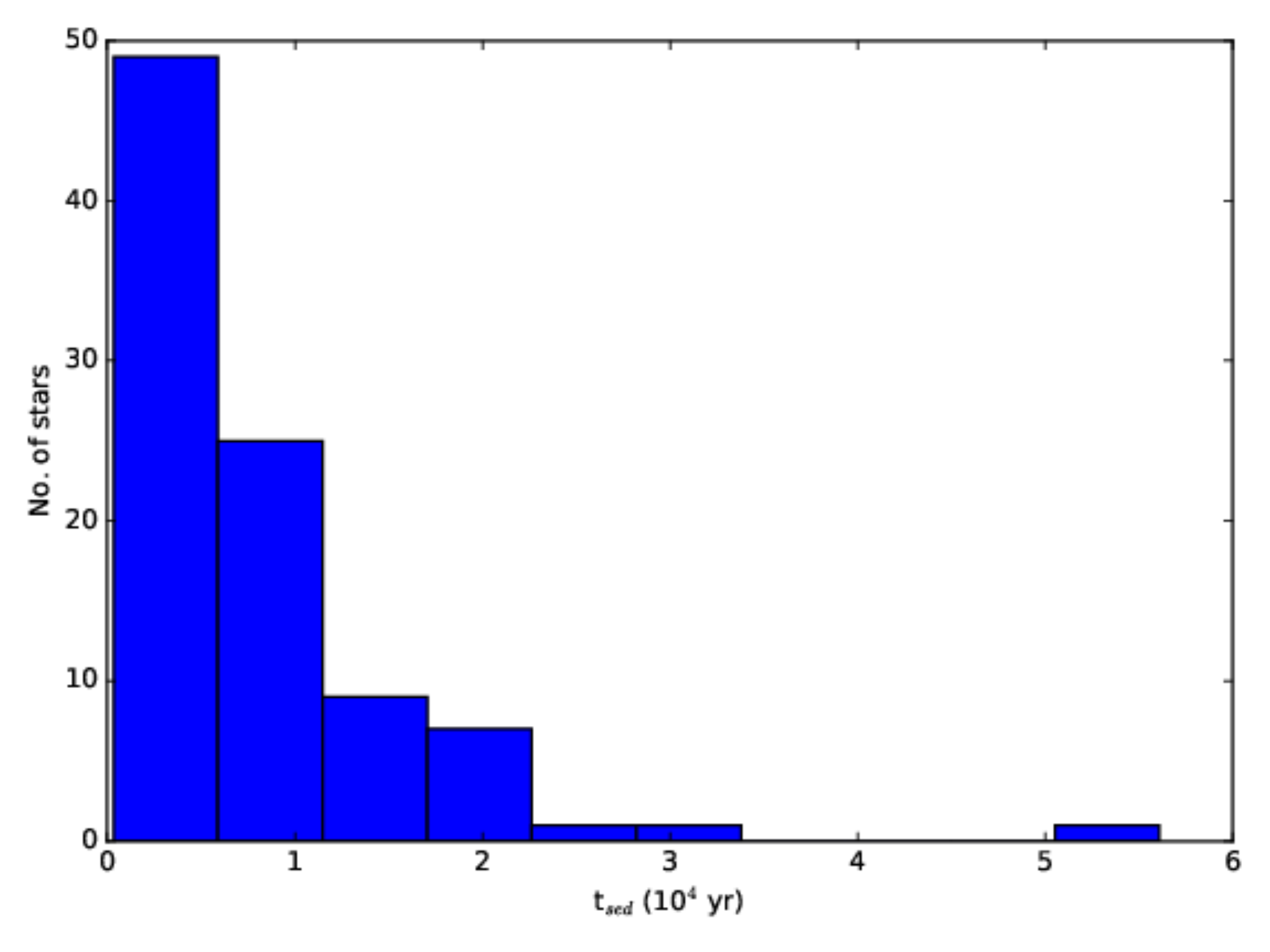}
  \caption{A histogram showing the time it would take a SNR (SN remnant), following the Sedov-Taylor analytic solution for shockwave expansion, to reach the radius of the 100th neighbour of each star particle (labelled t$_{sed}$). The values were calculated using a snapshot of Run 1 taken 508 Myr into the simulation, therefore during the middle of the starburst. Only stars within the virial radius of the galaxy were plotted, since any beyond this radius do not undergo feedback.}
  \label{fig:tsed_check} 
\end{figure}

\note{In order to avoid spurious effects associated with stars beyond the outer radii of the gaseous halo attempting to heat SPH particles that are well beyond their radius of influence, we choose to ignore stellar feedback for particles beyond the virial radius of the cloud. This only affects 6 star particles across each simulation. We checked the average distance between a star particle and its 100 neighbours 509 Myr into Run 1, finding the mean value is 200 pc. As a further check, we calculated the time it would take a shock wave, produced by a SN of the canonical energy 10$^{51}$ erg, to reach the radius of influence of each star particle (set by the relative distance of its 100th neighbour), assuming it followed the Sedov-Taylor analytic solution \citep{Taylor1950, Sedov1959}. We plotted these times in a histogram (see Fig. \ref{fig:tsed_check}), finding a mean value of $7.8 \times 10^{3}$ yrs, which is of the same order as a typical simulation timestep ($\sim 10^4$ yrs). We therefore conclude each star particle is not influencing an un-physically large radius in our simulations.}

\section{Results}\label{Results}
\subsection{The massive star population of each galaxy}
\begin{figure} 
  \includegraphics[width=\columnwidth]{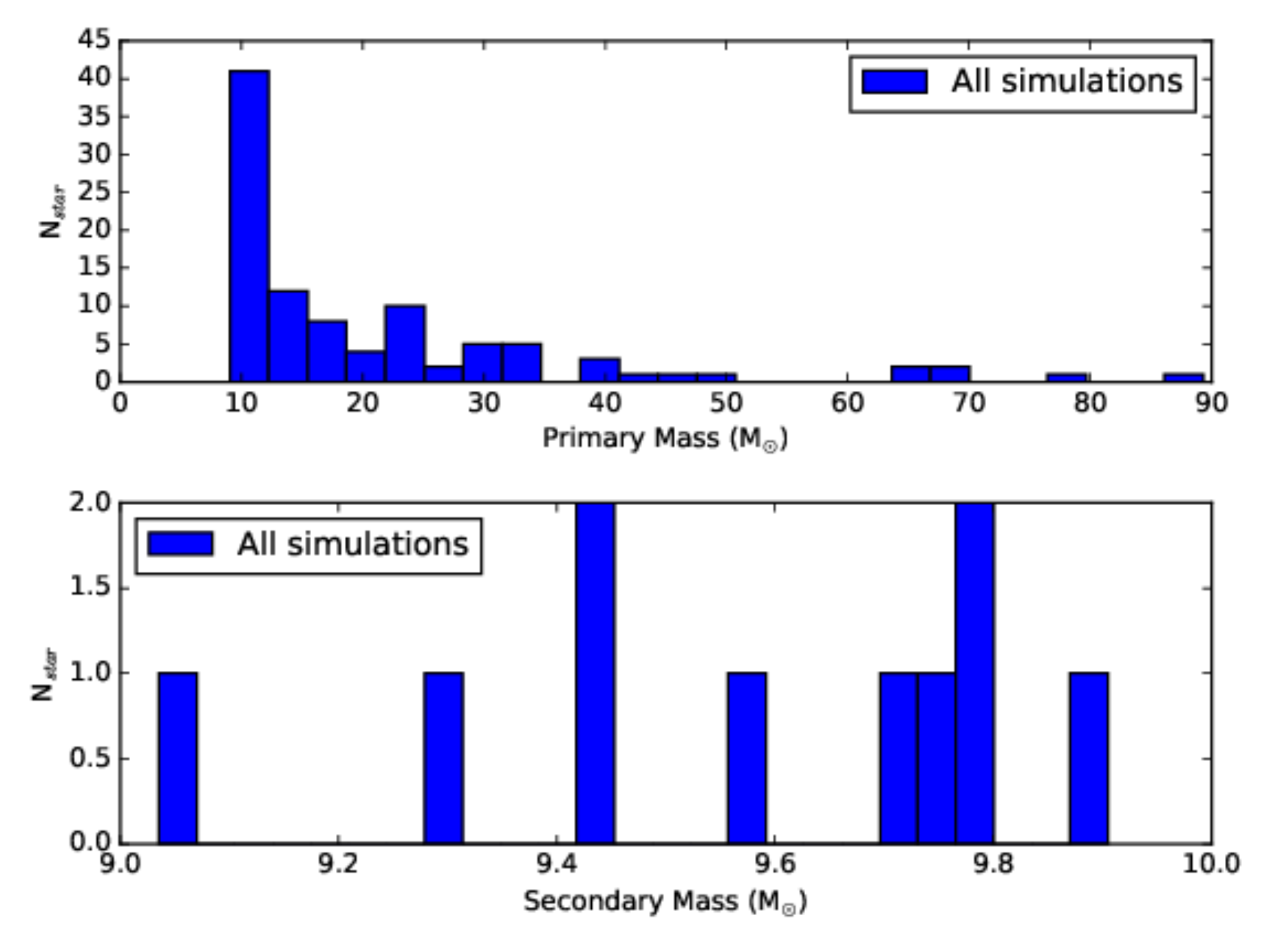}
  \caption{Top plot - the masses of the primary stars in the 100 binary systems contained in each simulation. Bottom plot - the corresponding masses of the companion stars in HMXB systems.}
  \label{fig:Pmass_smass} 
\end{figure}

In Fig. \ref{fig:Pmass_smass} we plot the primary and companion stellar masses (top and bottom plot respectively) assigned to the star particles for all simulations. As expected from the IMF, the vast majority of the stars have a primary mass of $\sim$ 10 M$_{\odot}$, while the secondary stellar masses are clustered between 9-10 M$_{\odot}$. These stars were then assigned metallicity-dependent lifetimes, along with wake-up times (which in turn were dependent on the standard deviation used for the underlying Gaussian profile).

\begin{figure}
	\includegraphics[width=\columnwidth]{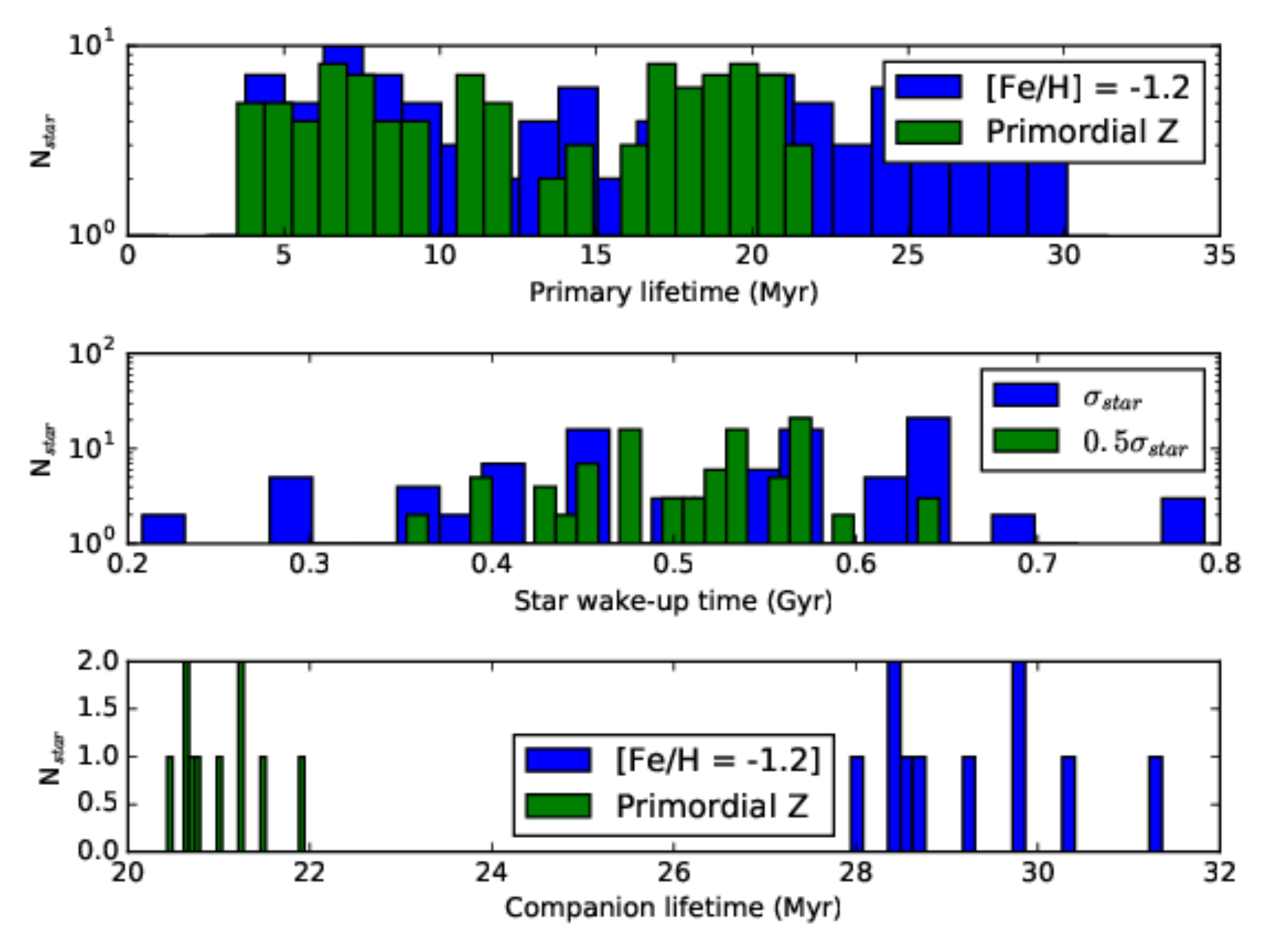}
    \caption{Upper plot - histograms to show the distribution of assigned primary star lifetimes for the stellar population, either at primordial metallicity (green) or [Fe/H] = -1.2 (blue). Middle plot - the wake-up times assigned to star particles, set using a Gaussian with a standard deviation of either $\sigma_{star}$ or 0.5 $\sigma_{star}$. Bottom plot - a histogram to show the distribution of assigned companion star lifetimes.}
    \label{fig:Lifetimes}
\end{figure}

In figure \ref{fig:Lifetimes} we plot the lifetimes of both the primary and companion stars in all simulations, along with the wake-up times of the star particles in simulations with either $\sigma_{star}$ (0.13 Gyr) and 0.5$\sigma_{star}$ (0.06 Gyr). By increasing the metallicity of the gas in the galaxy, the assigned primary lifetimes have increased in range from $\sim$ 22 Myr to $\sim$ 30 Myr. Additionally the companion lifetimes have also increased from $\sim$ 21 Myr to $\sim$ 30 Myr. This will mean more energy will be deposited into the ISM of the \newnote{higher} metallicity galaxies across the lifetime of the HMXB phase and the stellar winds phase. Moreover, the delay time until the onset of SNe feedback will be greater. 

\subsection{Gradual versus instantaneous feedback - Runs 1 and 2}
\begin{figure}
	\includegraphics[width=\columnwidth]{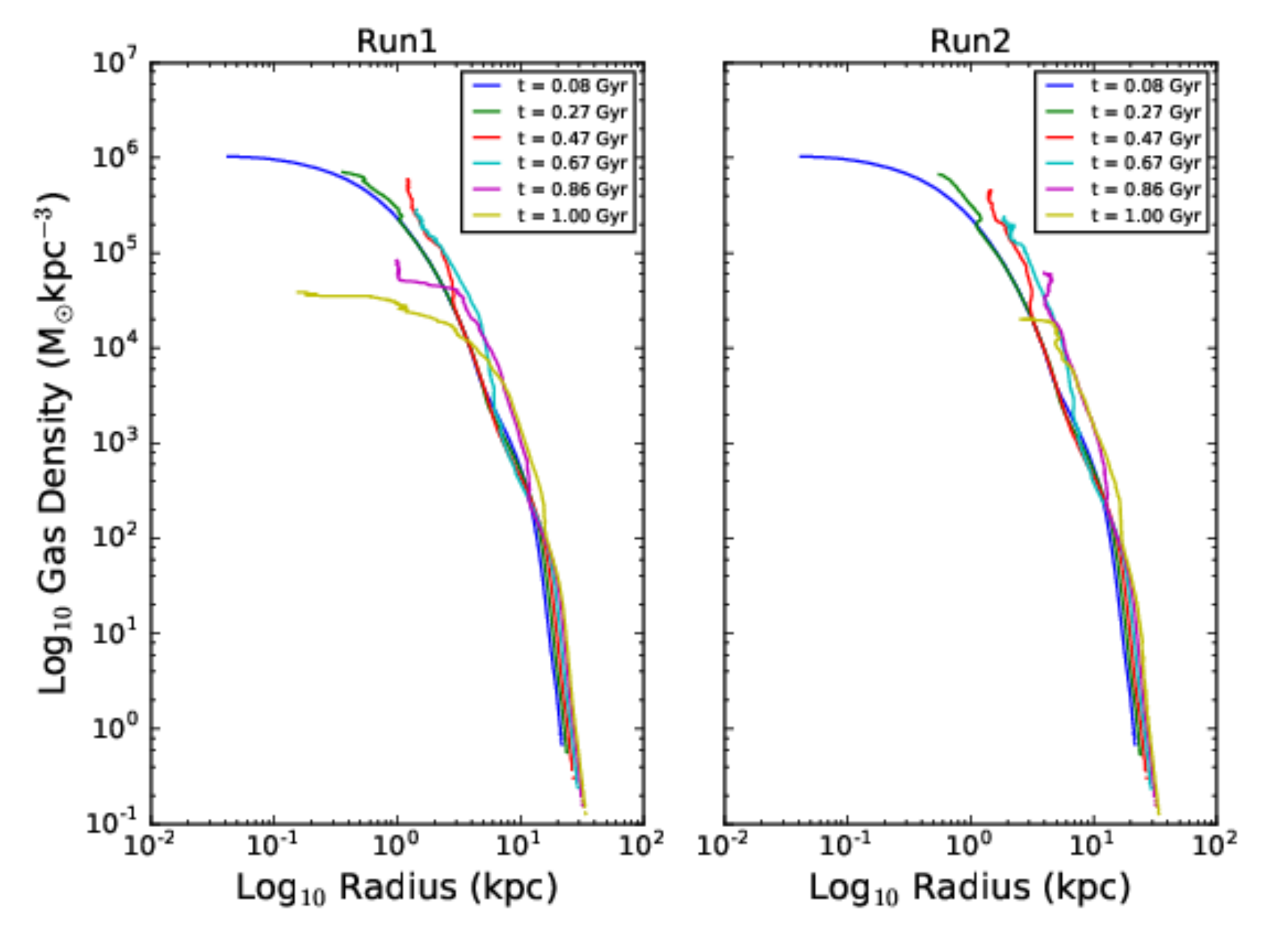}
    \caption{The density profile for the gas at varying times across the the simulation for Runs 1 (left) and 2 (right). Run 1 includes stellar winds and HMXB feedback on top of SNe feedback, while Run 2 just includes SNe feedback. }
    \label{fig:GasDensity_R12}
\end{figure}

In this section we compare the effects of including just instantaneous feedback (SNe, Run 2) or a combination of gradual (HMXB and stellar winds) and instantaneous stellar feedback (Run 1) during a Gyr starburst in a $1.1\times 10^{8}$ M$_{\odot}$ primordial metallicity dwarf galaxy. In figure \ref{fig:GasDensity_R12} we compare the gas density profiles of Runs 1 and 2. It is clear the central kpc of the galaxy in Run 2 has been efficiently cleared of gas by 1 Gyr, however gas has survived down to a radius of a few tens of parsec in Run 1, although with a $\sim$ 2 orders of magnitude drop from the initial central gas density. Furthermore, there is evidence for inflowing gas in both runs, given the density profile for both runs at 1 Gyr extends down to a smaller radius than the corresponding profiles taken at 0.86 Gyr.

\begin{figure*}
	\includegraphics[width=\textwidth]{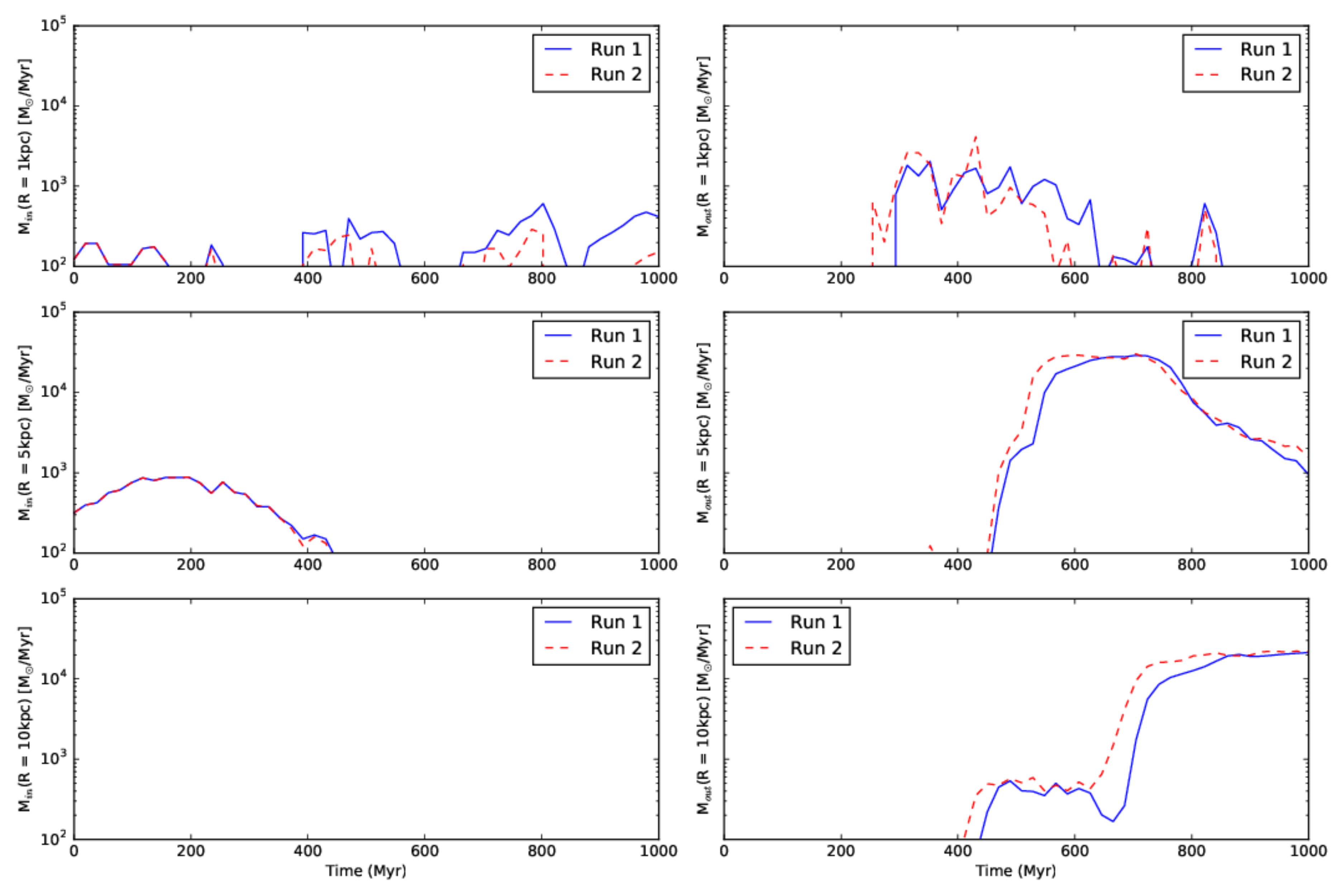}
    \caption{Left column - the mass inflow rate (M$_{in}$) of bound gas at radii of 1, 5 and 10 kpc, evaluated at snapshot times across Run 1 (blue, solid lines) and Run 2 (red, dashed lines). Right column - the mass outflow rate (M$_{out}$) of unbound gas, evaluated at 1, 5 and 10 kpc at snapshot times for Runs 1 (again, blue solid lines) and 2 (red,dashed lines).  The mass inflow rate only includes gas that is bound to the galaxy, while the mass outflow rate only includes unbound gas.}
    \label{fig:Mass_in_out_R12}
\end{figure*}

We explore this idea in Fig. \ref{fig:Mass_in_out_R12} which plots the mass inflow and outflow at various radii, evaluated at snapshot times across Runs 1 and 2. The mass inflow into the central kpc of the galaxy is consistently higher in Run 1 and mass continues to inflow until the end of the starburst at 1 Gyr. This  indicates cooling (and the subsequent re-accretion of gas) is less efficient when gradual feedback is ignored.  On the other hand, the mass outflow at 1 kpc is confined to the period of stellar feedback (between $\sim$ 200 Myr to 900 Myr) for both Runs 1 and 2, however it initiates sooner in Run 2. The majority of the mass outflow occurs at 5 kpc in both runs, while the mass outflow rate for Run 2 is once again higher at earlier times. The mass outflow rate at 10 kpc is consistently higher in Run 2 than Run 1, until $\sim$ 900 Myr. Neither simulation shows inflowing gas at this radius, indicating the prescence of galaxy-wide galactic outflows.

\begin{figure}
	\includegraphics[width=\columnwidth]{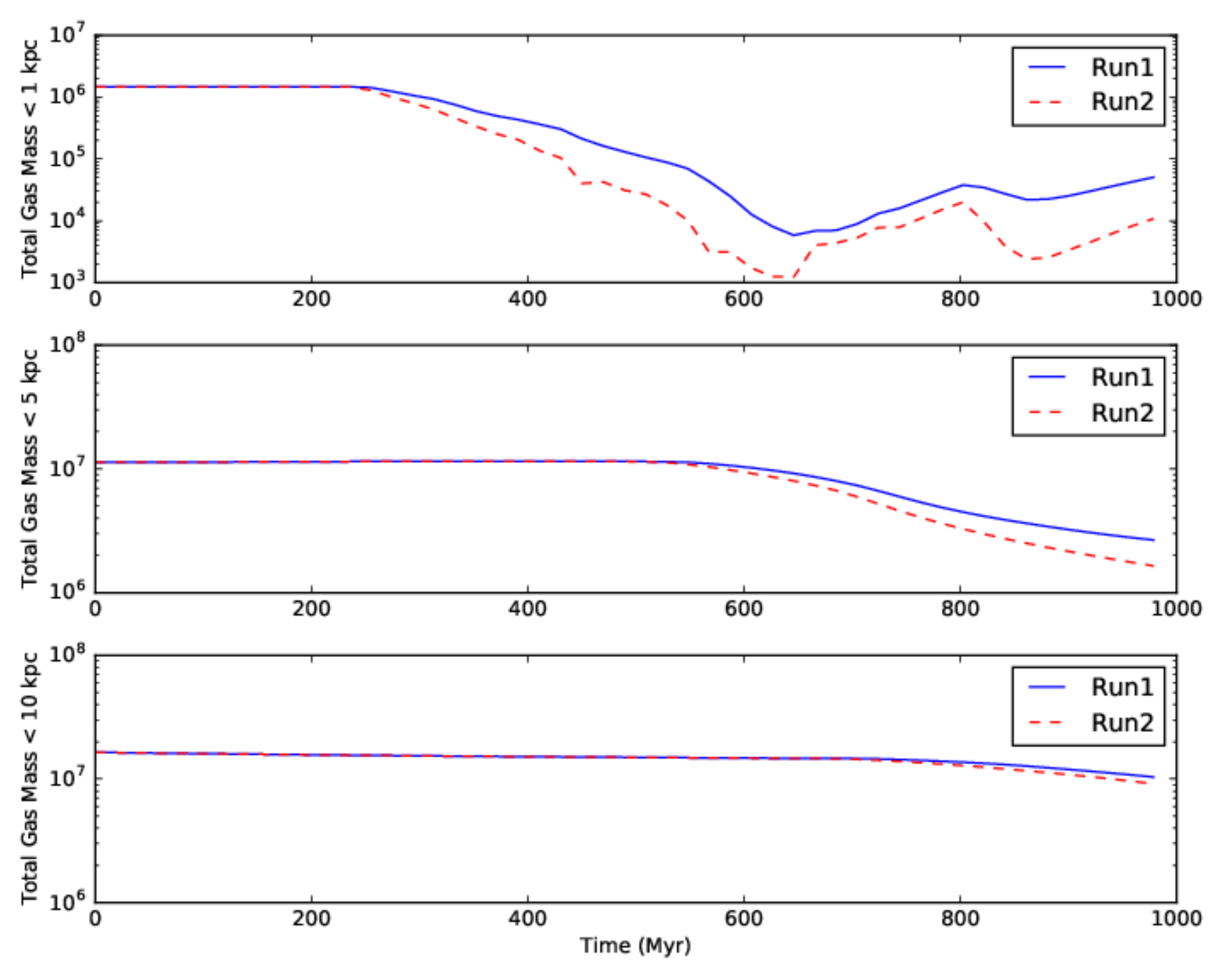}
    \caption{The total \newnote{gas} mass contained within a radius of 1 kpc (top plot), 5 kpc (middle) and 10 kpc (bottom) in Runs 1 (blue, solid line) and 2 (red, dashed line), at various times across the simulation. }
    \label{fig:Mass_in_radii_R1v2}
\end{figure}

Moreover, in Fig.  \ref{fig:Mass_in_radii_R1v2} we plot the total gas mass below each radii (10 kpc, 5 kpc and 1 kpc) for Runs 1 and 2. As expected from the higher mass inflow rates and lower mass outflow rates seen for Run 1 in Fig. \ref{fig:Mass_in_out_R12} ,  the gas mass below each radius is higher in Run 1.  As well as this, the total gas mass inside the inner 1 kpc of Run 2 drops to 10$^3$ M$_\odot$ at a time of 650 Myr, compared to 10$^4$ M$_\odot$ for Run 1. Therefore, by including gradual feedback, a higher gas mass has been retained at the centre of the halo, which could fuel more star formation.

However, \newnote{Fig. \ref{fig:Mass_in_radii_R1v2}} shows the total gas mass within 5 and 10 kpc is decreasing beyond $\sim$ 550 Myr and 650 Myr respectively, indicating a global outflow of gas. This is further evidenced by the high mass outflow rates (comparative to the inflow rates) seen at these radii in both Runs 1 and 2. This is expected since the energy injection of 100 SNe - 10$^{53}$ erg - is greater than the total binding energy of the dwarf galaxy ($\sim$ 10$^{52}$ erg). The mass outflow rate inside Run 2 is larger than inside Run 1 between $\sim$ 450 - 600 Myr at a radius of 5 kpc and between 400 - 900 Myr at a radius of 10 kpc, indicating a persistent large-scale outflow of gas. However, the mass outflow rates of Runs 1 and 2 have converged at 10 kpc by the end of each simulation. 

\begin{figure}
	\includegraphics[width=\columnwidth]{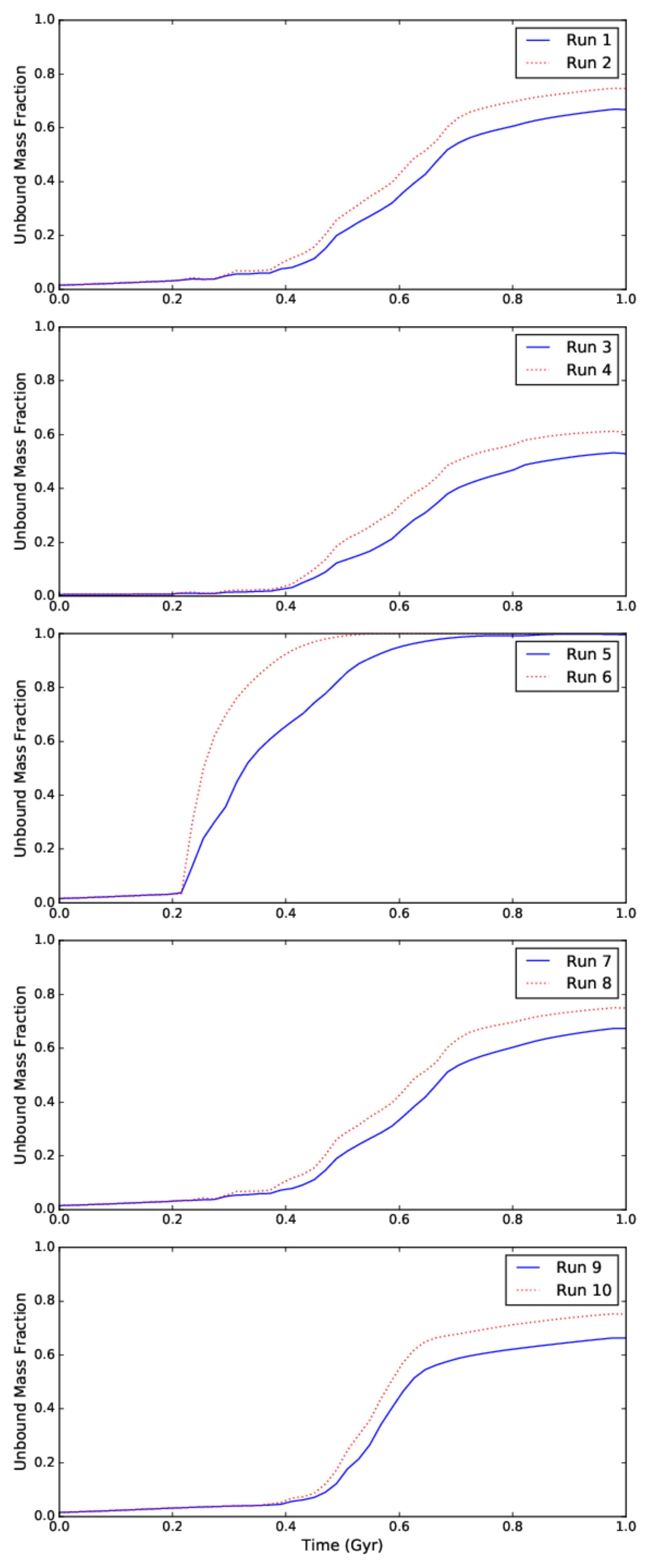}
    \caption{Plot to show the fraction of the gas which is unbound in runs containing gradual types of feedback alongside SNe feedback (blue, solid lines), as well as those containing just SNe feedback (red, dotted lines). The simulations on each individual plot have identical initial conditions. Gas is defined as unbound if it satisfies the condition $E_{therm} + E_{kin} + E_{pot} > 1$. }
    \label{fig:Ubd_R12}
\end{figure}

\begin{figure}
	\includegraphics[width=\columnwidth]{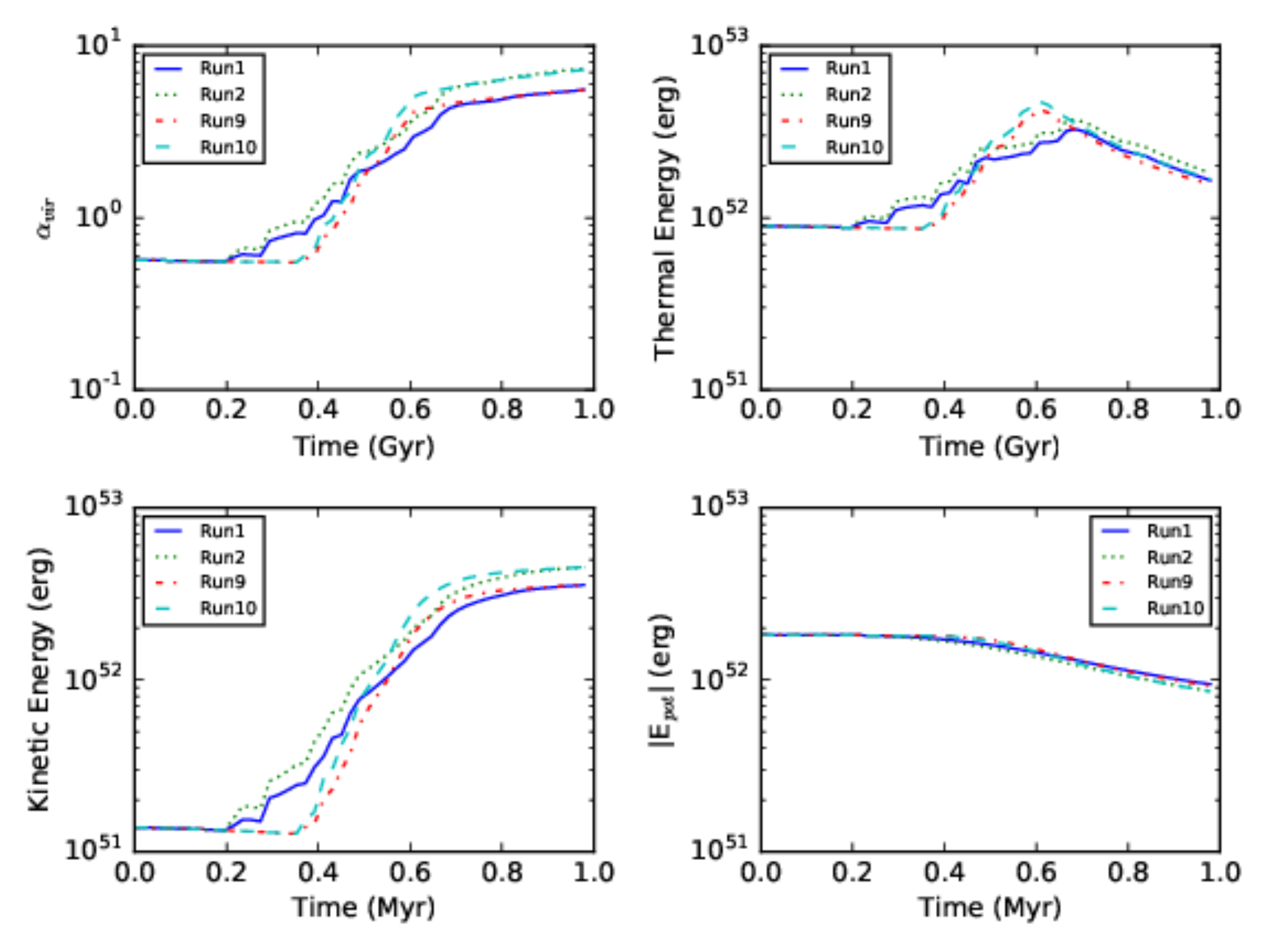}
    \caption{Plots to compare the global energetics of Runs 1 (navy blue, solid line), 2 (green, dotted line), 9 (red, dot-dash line) and 10 (light blue, dashed line). Top left plot - the time evolution of the virial parameter. Top right plot - the time evolution of the total thermal energy of the gas. Bottom left - the time evolution of the total kinetic energy of the gas. Bottom right - the time evolution of the total potential energy of the gas.}
    \label{fig:Eplot_sigma}
\end{figure}

As well as this, the top plot of Fig. \ref{fig:Ubd_R12} plots the fraction of unbound mass in both Runs 1 and 2. We can see, despite less energy being injected across the simulation, Run 2 has a consistently higher mass fraction of unbound gas throughout the Gyr starburst. By the end of the simulation, including gradual feedback has lowered the unbound mass fraction by 8 $\%$. Moreover, in Fig. \ref{fig:Eplot_sigma} we plot the virial parameters ($\alpha_{vir} = E_{therm} + E_{kin} / |E_{pot}|$) of both simulations, along with the total kinetic, thermal and potential energies of the gas. We can see Run 2 has a higher total thermal and kinetic energy throughout the starburst compared with Run 1, which results in a higher global virial parameter. These plots indicate the gas in Run 1 is able to cool more efficiently than the gas in Run 2. 

\begin{figure}
	\includegraphics[width=\columnwidth]{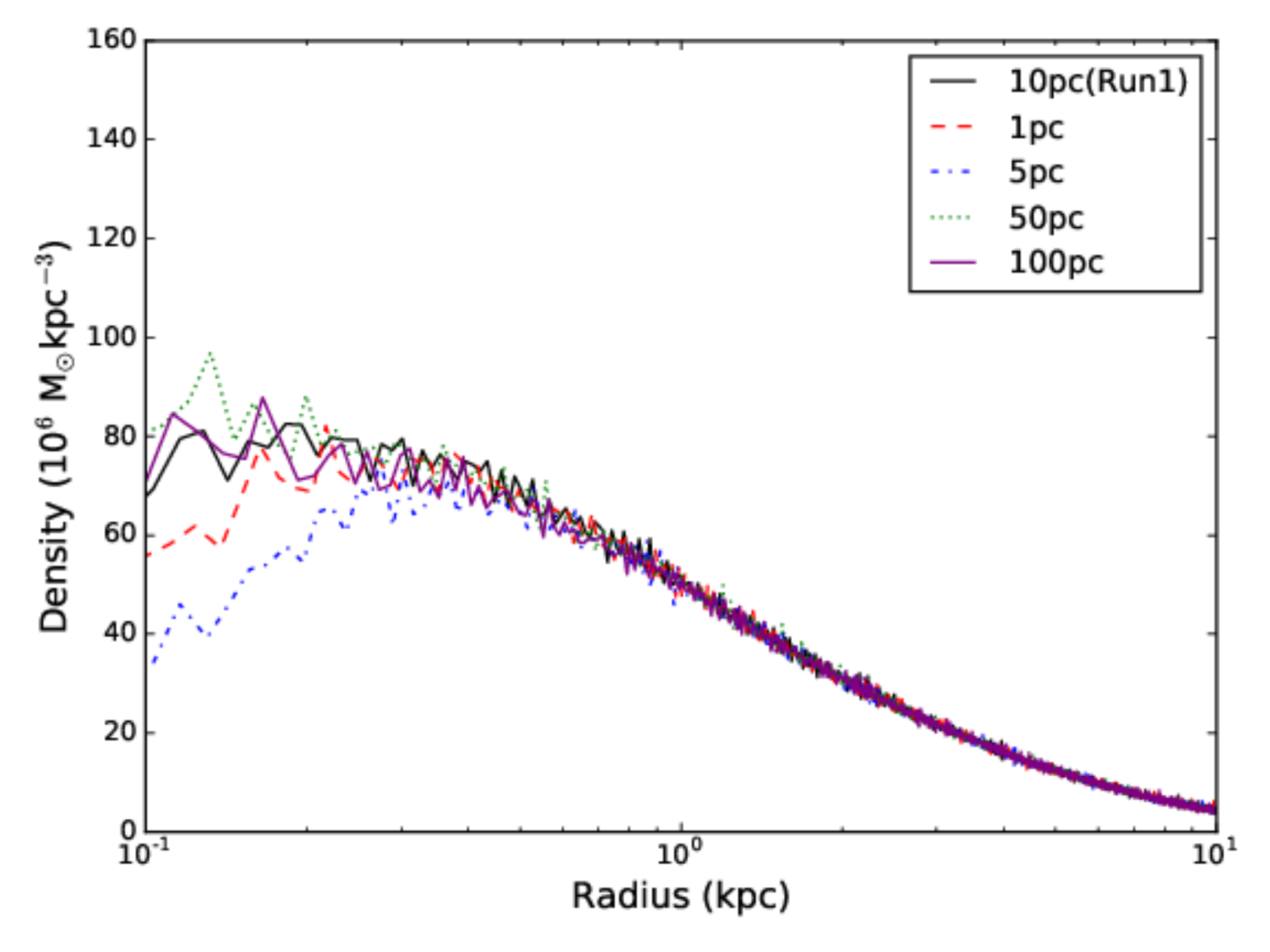}
    \caption{A plot of the mass contained in 20 temperature bins taken across the total gas particle temperature range of Runs 1 (hatched) and 2 (solid filled).}
    \label{fig:TvMass_R12}
\end{figure}

\begin{figure}
	\includegraphics[width=\columnwidth]{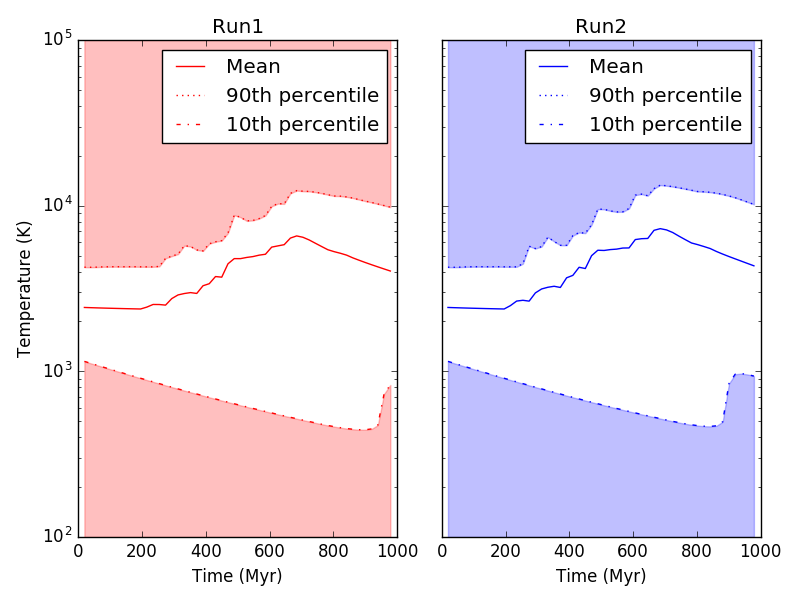}
    \caption{A plot to show the evolution of the mean (solid lines), 10th percentile (lower shaded areas) and 90th percentile (upper shaded areas) temperatures across the 1 Gyr starburst in Runs 1 (left plot, red) and 2 (right plot, blue).}
    \label{fig:R12_Temp}
\end{figure}

In figure \ref{fig:TvMass_R12} we plot the mass contained in 20 temperature bins between $10-10^{5.5}$ K for Runs 1 and 2. From this plot we can see a general trend of higher gas mass at temperatures $\lesssim 10^{3.5}$ K for Run 1 when compared with Run 2, along with a lower gas mass at `warm' temperatures above 10$^{4}$ K. Moreover, beyond 10$^{4}$ K the mass in each temperature bin is larger in Run 2 than Run 1. However, in both runs the majority of the gas mass is at temperatures between 10$^{3-4}$ K, corresponding to the point when collisional excitations become rare. Run 1 also contains an order of magnitude lower gas mass at temperatures above $\sim 10^5$ K compared with Run 2 (however this represents a very small fraction of the overall gas mass; $\sim$ 0.01$\%$). Below $10^{5.5} K$ cooling is dominated by collisional excitations of electrons in atoms, followed by their subsequent de-excitation and emittance of radiation. The lower densities present in Run 2 (figure \ref{fig:GasDensity_R12}) and the resulting decrease in the number of collisions at a specific temperature compared with Run 1, is therefore a likely cause of the inefficient cooling seen in figures \ref{fig:Eplot_sigma} and \ref{fig:TvMass_R12}. However, figure \ref{fig:R12_Temp} plots the mean temperature across both runs, alongside the the 90th and 10th percentile temperatures. This figure shows there is marginal difference between the mean temperatures of the gas in both runs, only in the mass of gas occupying each temperature bin. 

\begin{figure}
	\includegraphics[width=\columnwidth]{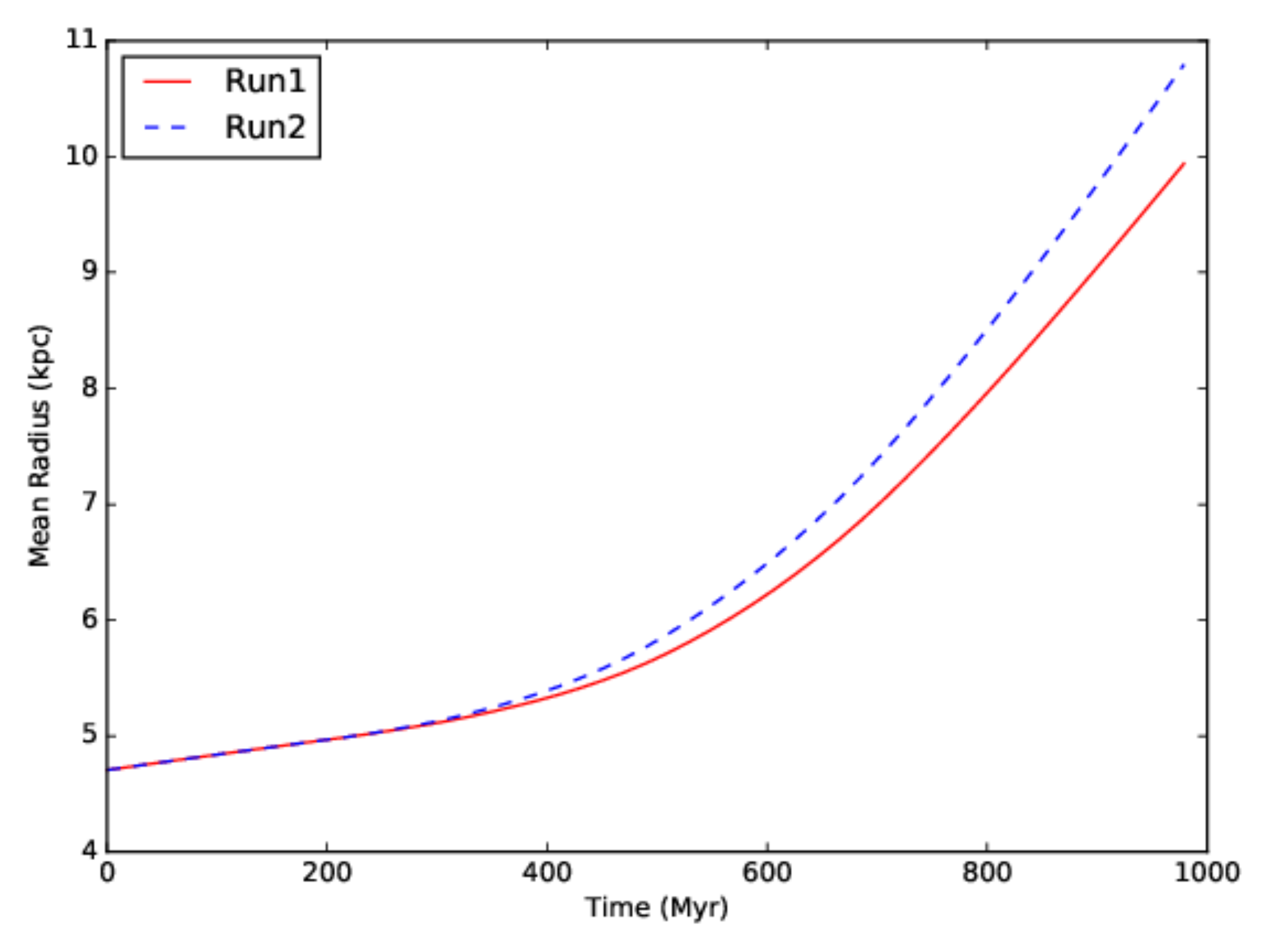}
    \caption{The mean radius of gas particles across 1 Gyr in Runs 1 (solid, red line) and 2 (dashed, blue line).}
    \label{fig:MeanR_R12}
\end{figure}

\begin{figure}
	\includegraphics[width=\columnwidth]{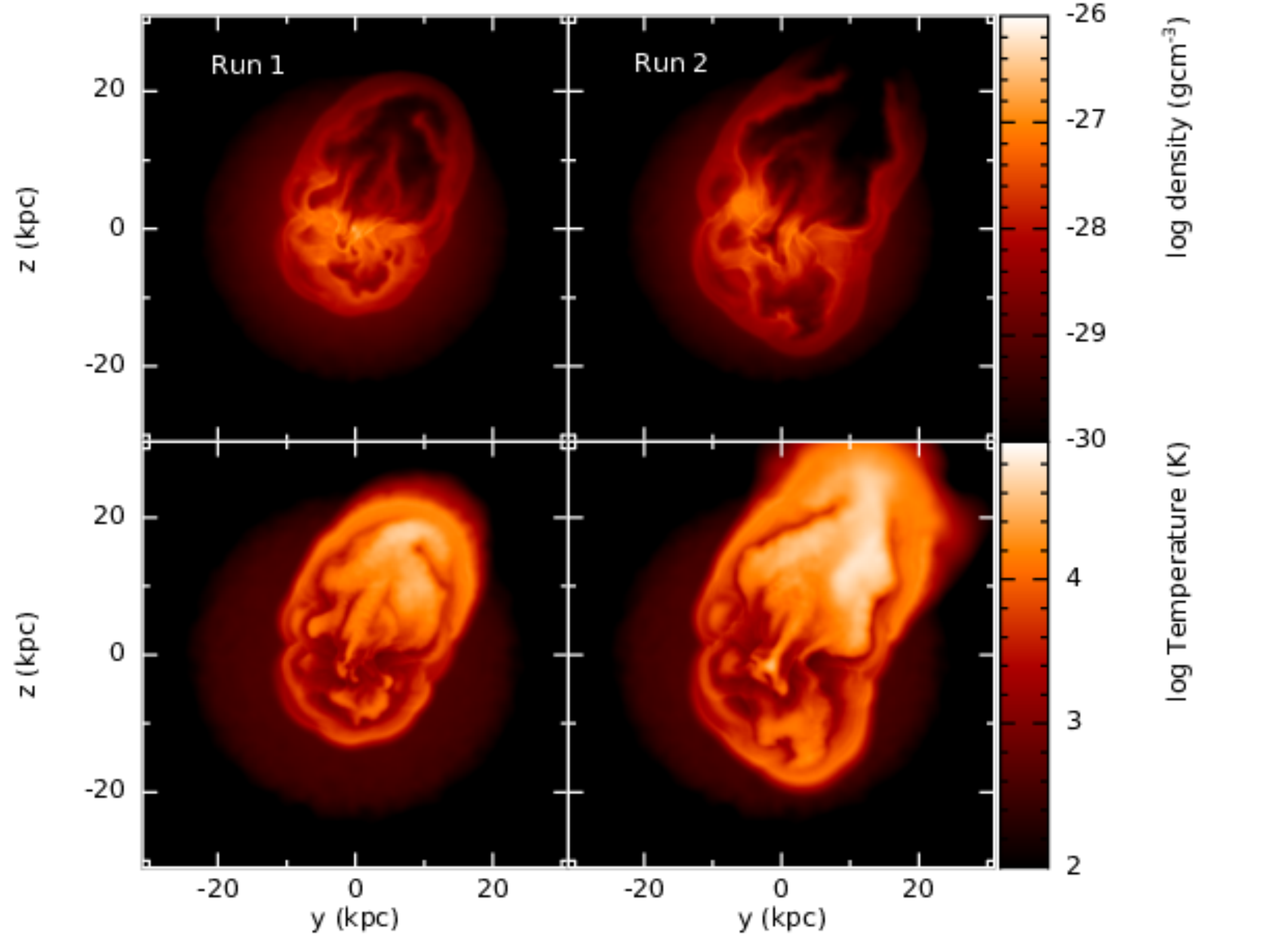}
    \caption{Top row - density slices taken at x = 0 in the y-z plane 1 Gyr into Runs 1 (left) and 2 (right plot). Bottom row - temperature slices taken at x = 0 in the y-z plane 1 Gyr into Runs 1 (left) and 2 (right).}
    \label{fig:TD_xz_plane_R12}
\end{figure}

We also plotted the mean radius of the gas particles across the simulation for both Runs 1 and 2 (see figure \ref{fig:MeanR_R12}), finding beyond $\sim$ 0.5 Gyr the mean radius is larger in Run 2 than Run 1 and the gap between the two increases with time. By 1 Gyr the mean radius in Run 2 is $\sim$ 1 kpc larger than in Run 1. The fact the mean radius is increasing in both runs, is further evidence of gaseous outflows. We investigated this by plotting density/temperature slices of Runs 1 and 2 in all three planes at the end of each simulation, taken with the origin at the centre of each galaxy. The most pronounced difference occurred in the y-z plane (see Fig. \ref{fig:TD_xz_plane_R12}). Here it can be seen both the SNe in Run 2 and the combination of SNe/HMXBs/stellar winds in Run 1 have inflated a 20 kpc low density bubble filled with gas between $10^{4-5}$ K, bordered by a ring of higher density ($\sim 10^{-28}$ gcm$^{-3}$) gas. However the ring is broken in Run 2, while in Run 1 it is intact. In Run 2 this has allowed hot gas to escape into regions of lower density and this process can be seen occurring in the bottom right plot of figure \ref{fig:TD_xz_plane_R12}. 

\begin{figure}
	\includegraphics[width=\columnwidth]{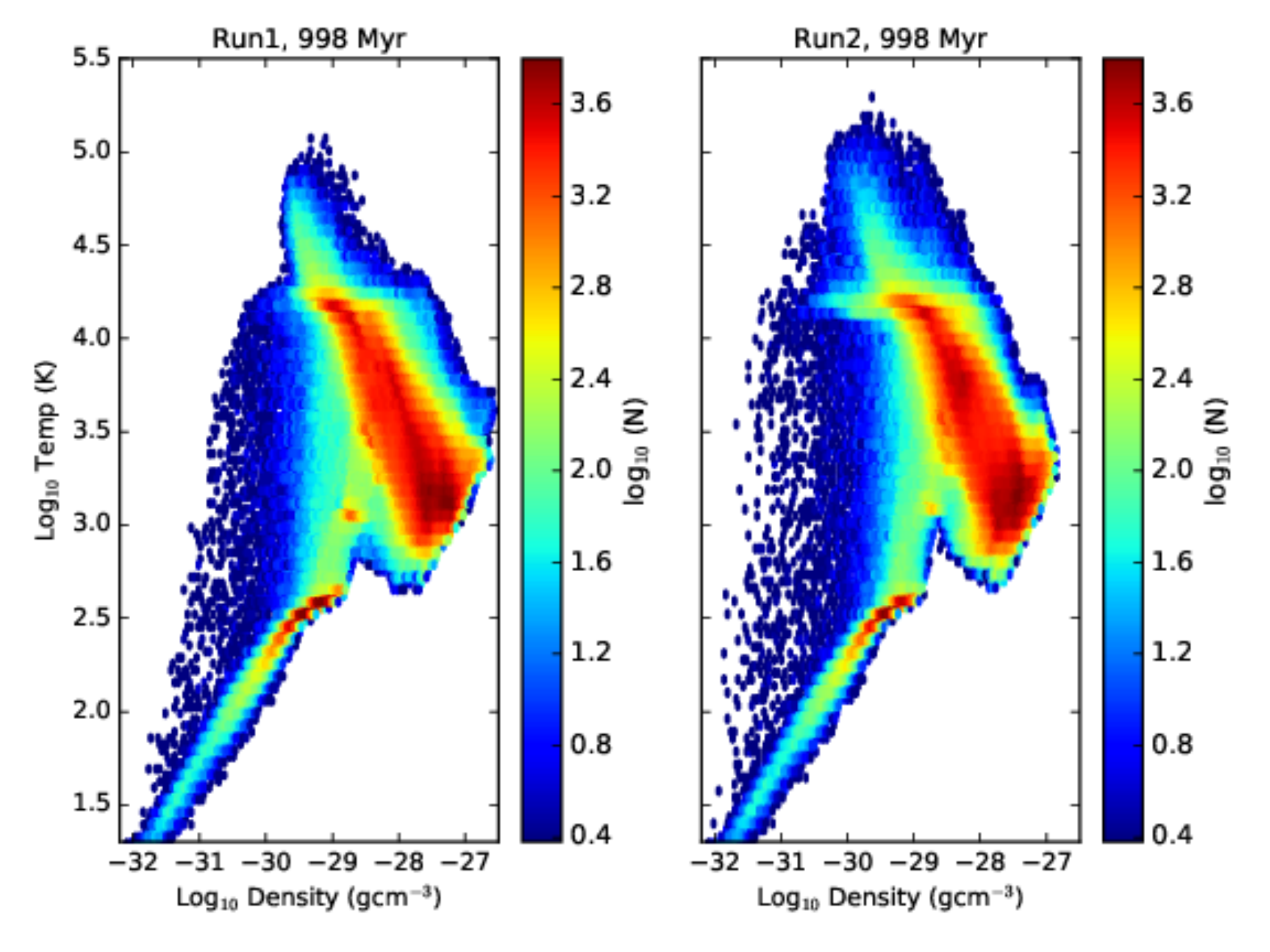}
    \caption{Plots to show the temperature and density of the gas in Runs 1 (left) and 2 (right) at the end of each simulation, rendered according to particle number. }
    \label{fig:TempVdens_R1v2}
\end{figure}

This excess of hot/warm, low density gas seen in Run 2 compared with Run 1, is highlighted in Fig. \ref{fig:TempVdens_R1v2}, which plots the temperature versus density of the gas in Run 1 and 2 at 998 Myr, which has been rendered according to particle number. Comparing the populations of gas particles with densities between 10$^{-32}$ $-$ 10$^{-30}$ gcm$^{-3}$, we see Run 2 has a higher number of gas particles within this density range, with a wider range of temperatures. In particular, Run 2 has an excess of gas particles with temperatures of between $\sim$ 15 000 K to 17 000 K, which correspond to a peak in the primordial metallicity cooling curve, due to the collisional excitation of H$^{0}$ and He$^{+}$ \citep{Mo2010}. However, Run 1 contains no gas in this temperature range that has a density less than 10$^{-31}$ gcm$^{-3}$. This suggests the cooling of the gas is being suppressed in Run 2 compared with Run 1 due to the comparative low density of the gas (which is also shown in Fig. \ref{fig:TD_xz_plane_R12}). This cooling suppression means the hot gas will likely retain enough energy to escape the galaxy entirely and hence constitutes a galactic wind.

\begin{figure}
	\includegraphics[trim={0 5cm 30cm 0}, clip,width=\columnwidth]{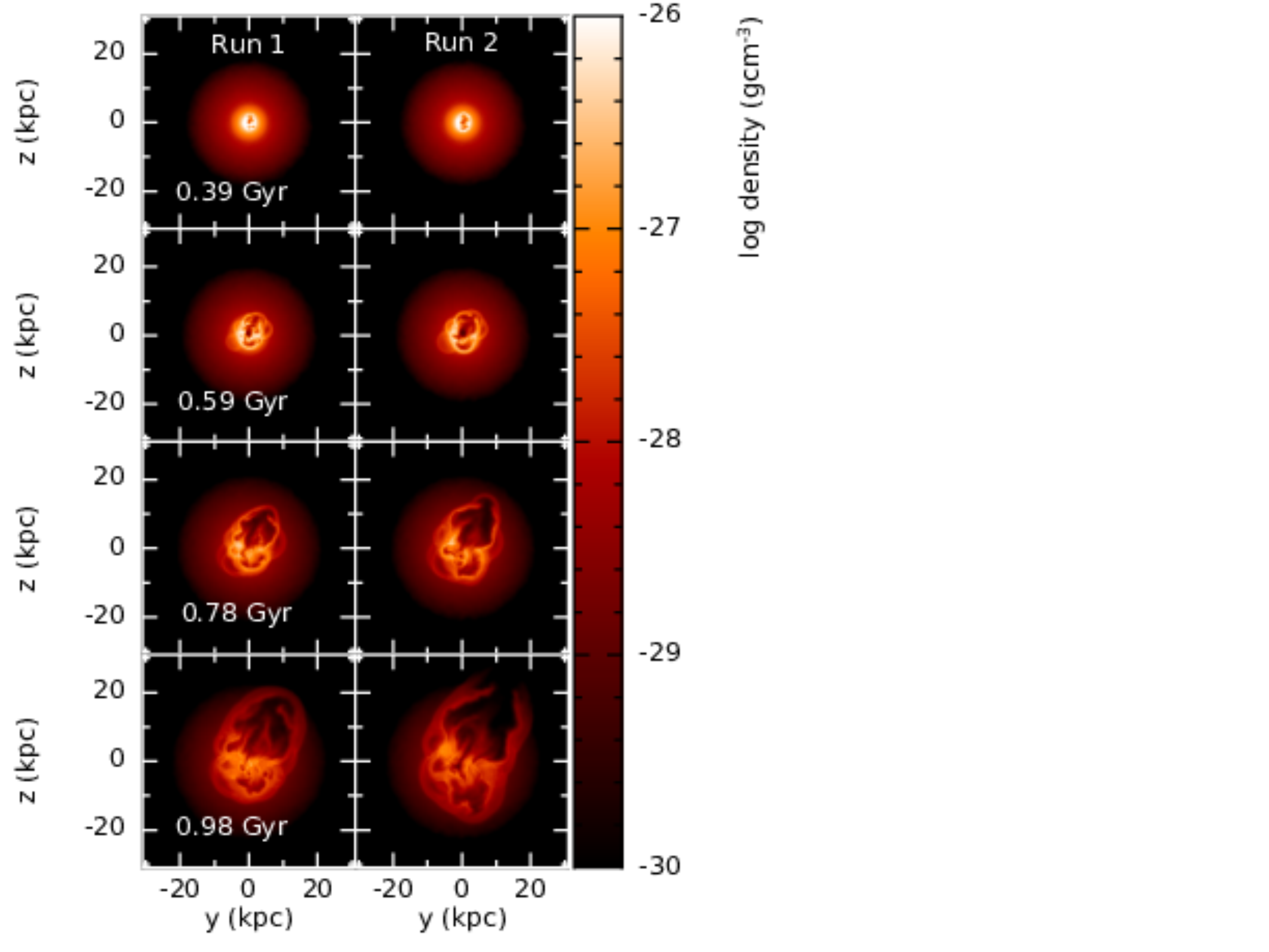}
    \caption{Plots to show the evolution of the density of the gas contained in a x = 0 slice of the y-z plane in Runs 1 (left column) and 2 (right column), at varying times. }
    \label{fig:Fountain_evoln_R1v2}
\end{figure}
In Fig. \ref{fig:Fountain_evoln_R1v2} we show the evolution in density of the outflow seen in Fig. \ref{fig:TD_xz_plane_R12} for Runs 1 (left column) and 2 (right column). By 0.59 Gyr, we can see the interplay of  multiple low density bubbles, surrounded by rings of high density gas and driven by the shock-heating from stellar feedback events. The location of SNe events in Runs 1 and 2 largely correspond due to the underlying stellar populations being identical. However, current and previous gradual feedback events in Run 1 have acted to limit the radii of these low density bubbles (for example, comparing the lower portion of Runs 1 and 2 at 0.59 Gyr). This links back to work by \cite{Rogers2013}, which showed the action of prior feedback events can reduce the coupling of SN energy to the surrounding ISM. Furthermore, more high density filaments have been retained inside the virial radius (7.78 kpc) of Run 1 than Run 2, suggesting the mixing of the hot feedback-generated bubbles is more efficient in Run 2 than Run 1. Beyond this, by 0.78 Gyr spatially and temporally coincident feedback events have acted to inflate two bubbles in the y-z plane of Run 2; one on the bottom right of Fig. \ref{fig:Fountain_evoln_R1v2} and another in the top right. On the other hand, Run 1 also contains low density bubbles at the corresponding locations, however they are less spatially extended and the interior gas is higher in density. 

\begin{figure}
	\includegraphics[width=\columnwidth]{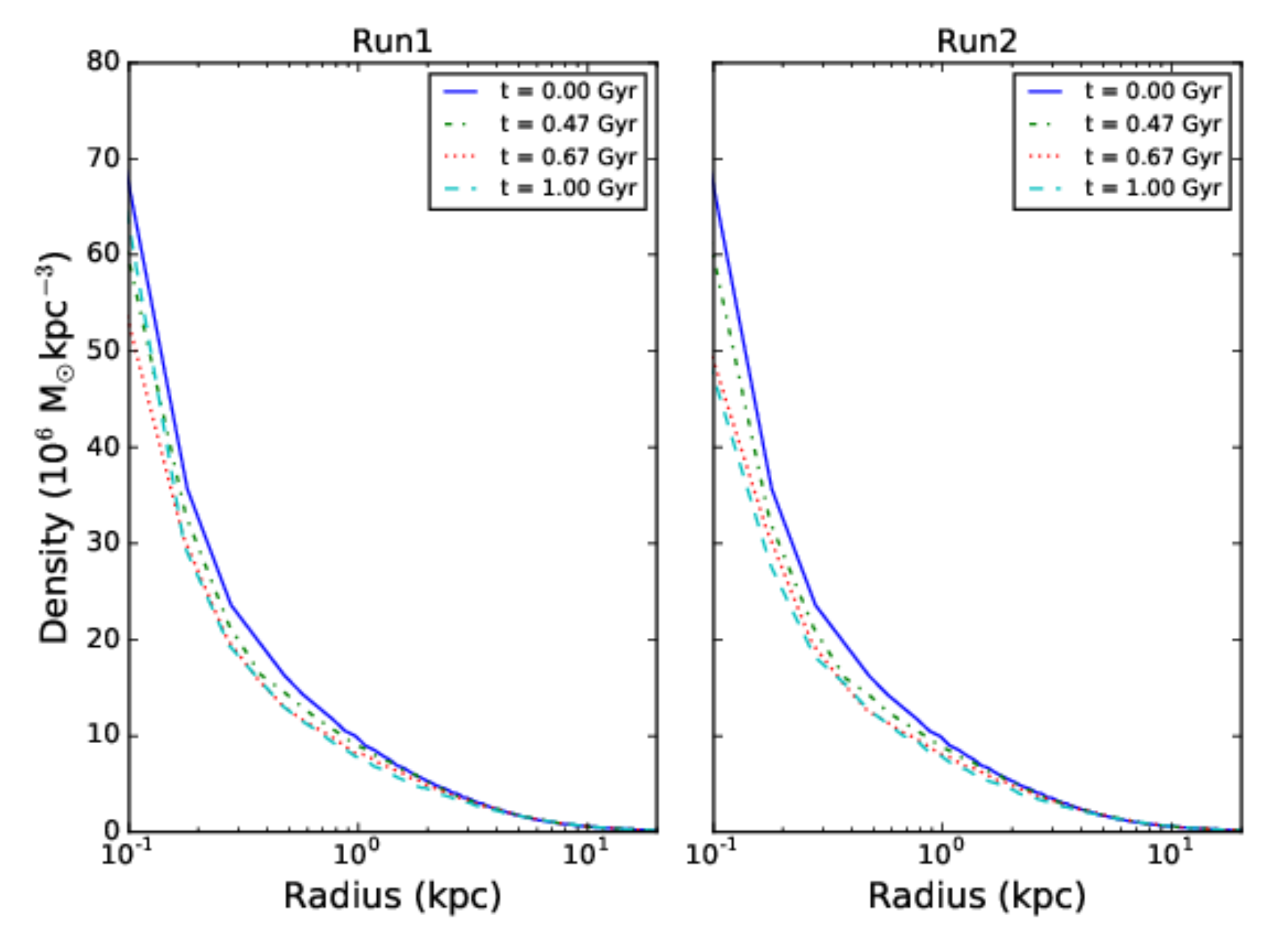}
    \caption{Dark matter halo density profile taken for Runs 1 (left) and 2 (right) at varying times into each simulation. }
    \label{fig:DMDens_R12}
\end{figure}

\note{In order to ascertain any effect of the feedback on the underlying dark matter halo, in figure \ref{fig:DMDens_R12} we plot the density profile for the dark matter in Runs 1 and 2 at varying times. Run 1 shows a 22$\%$ drop in density at 100 pc, between 0 and 0.67 Gyr, however it also shows a $\sim$ 20 $\%$ rise between 0.67 Gyr and 1 Gyr. This means by the end of the simulation the inner dark matter profile has only dropped in density by $\sim$ 7 $\%$. On the other hand, the density of the dark matter above a radius of $\sim$ 200 pc is lower at 1 Gyr than at all other times. Looking instead at Run 2, the main difference arising from neglecting gradual feedback is the inner $\sim$ 200 pc of the density profile, which is lower than Run 1 beyond 0.67 Gyr and also does not show a rise in density to correspond with the density increase seen in Run 1 between 0.67 to 1 Gyr. This increase in the inner density profile of Run 1 could indicate that, post-starburst, recovery of the original dark matter halo profile is possible in the runs which include gradual feedback. However, this increase in density inside the inner 200 pc is comparative to the noise level seen in Appendix \ref{appendix:soft}, hence it is difficult to ascertain if this result is a result of the gradual feedback or noise in the dark matter profile. Overall, the density drop in both runs is marginal when compared with the dark matter density cores seen in work such as \citet{Teyssier2013}, along with the density drop seen in the smaller haloes investigated in \citet{Cashmore2017}. }

To conclude, the addition of gradual types of feedback on top of instantaneous feedback in Run 1 has resulted in a lowering of the total internal energy and kinetic energy of the gas in the galaxy, along with a decrease in the mass fraction of gas unbound during the 1 Gyr starburst. Moreover, the lack of any type of gradual feedback has also led to a higher mass of gas at temperatures above $10^5$ K, along with a decrease in gas mass below $\sim 300$ K. Additionally, the mass inflow rate was increased when HMXB and SNe feedback was included. \note{Regarding the dark matter profile, our simulations lack the dark matter mass resolution to rigorously investigate the impact of gradual versus instantaneous feedback events, as indicated by Appendix \ref{appendix:soft}, hence in this paper we will focus on the fate of the ISM.}

\subsection{Decreasing $\sigma_{\star}$ - Runs 9 and 10}
\begin{figure}
	\includegraphics[width=\columnwidth]{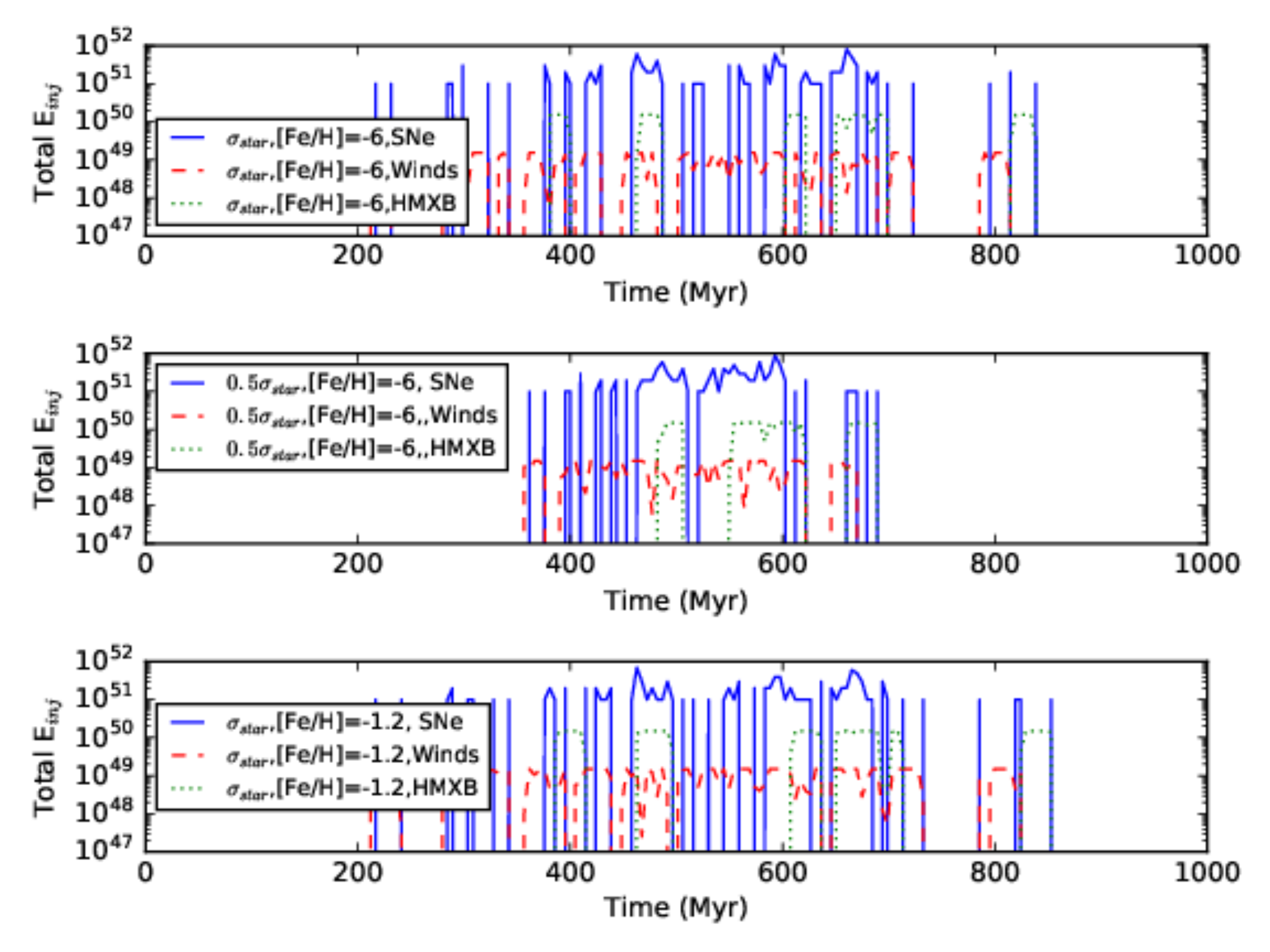}
    \caption{The total energy injected by SNe (solid, blue lines), stellar winds (red, dashed lines) HMXBs (dotted, green lines) per 5 Myr across Run 1 (top plot), Run 9 (smaller $\sigma_{star}$, middle plot) and Run 7 (higher metallicity, bottom plot).}
    \label{fig:Rel_Energy}
\end{figure}

In the top two plots of Fig. \ref{fig:Rel_Energy} we plot the energy injection by the 3 different mechanisms; stellar winds, HMXBs and SNe, temporally binned in units of 5 Myr for both the $\sigma$ and $0.5 \sigma$ primordial metallicity runs where all three feedback types are included (Runs 1 and 9). As expected from Fig. \ref{fig:Lifetimes}, by decreasing $\sigma_{star}$ we have created a more concentrated starburst and by maintaining the same stellar population as the $\sigma_{star}$ runs, the net result is a more concentrated injection of energy, focused between 350-650 Myr. In both cases the stellar winds create a low-level, near-constant energy injection between SNe, with HMXBs contributing an order of magnitude more energy, however over shorter timescales towards the second half of the starburst. Given stellar winds are present prior to the first SN, along with their ubiquity throughout the starburst, this likely means they have a larger overall impact on the simulation than the higher powered, less frequent, HMXB events. We go on to investigate their comparative impact in section \ref{sec:NoWinds}. The main difference between the $\sigma_{star}$ and $0.5 \sigma_{star}$ runs is the unsurprisingly larger gaps between feedback events (particularly stellar winds) in the $\sigma_{star}$ run (Run 1). 

Despite the longer times between massive star feedback events in Run 9, Fig. \ref{fig:Ubd_R12} indicates there is only marginal difference between the mass fraction of unbound gas between runs with corresponding feedback mechanisms at different $\sigma_{star}$. On further analysis, Run 9 had a higher unbound gas mass fraction than Run 1 of 1$\%$, while the fraction in Run 2 was 0.5$\%$ higher than that of Run 10. This indicates the feedback energy injected into Runs 1 and 2 has not been lost to radiative cooling, causing the energetics of the gas in Runs 1 and 2 to converge with Runs 9 and 10 respectively, since the total feedback energy injected is the same in both cases.

Furthermore, we can use Fig. \ref{fig:Eplot_sigma} to compare the global energetics of the Runs 1, 2, 9 and 10. As expected the total kinetic and internal energies of the gas in Runs 1 and 2 initially dominate that of the $0.5 \sigma$ runs (Run 9 and 10), due to the fact feedback kicks in earlier (see figure \ref{fig:Rel_Energy}). However, at $\sim$ 0.5 Gyr the total internal energies and kinetic energies (along with $\alpha_{vir}$) of Runs 9 and 10 overtake that of the corresponding larger $\sigma_{star}$ runs. As was seen at $\sigma_{star} = 0.13$ Gyr, the run at $0.5 \sigma_{star}$ (or $\sigma_{star} = 0.06$ Gyr) that lacks any type of gradual feedback (Run 10) has a higher virial parameter, along with global kinetic and thermal energy, than the corresponding run containing stellar winds and HMXBs (Run 9). However, the total kinetic energy of both Runs 1 and 9, along with 2 and 10 converge around 700 Myr. As well as this, the total thermal energy of the gas in both Runs 9 and 10 is marginally lower than in their higher $\sigma_{star}$ counterparts, indicating the gas in the runs that include a shorter, more violent starburst is able to cool more efficiently. Both Fig. \ref{fig:Eplot_sigma} and Fig. \ref{fig:Ubd_R12} indicate changing the timescale of the starburst has had a marginal impact on the multiphase ISM of the galaxy; the main factor being the presence of stellar winds and HMXB feedback. 
\begin{figure}
	\includegraphics[trim={0 0 0 0}, clip,width=\columnwidth]{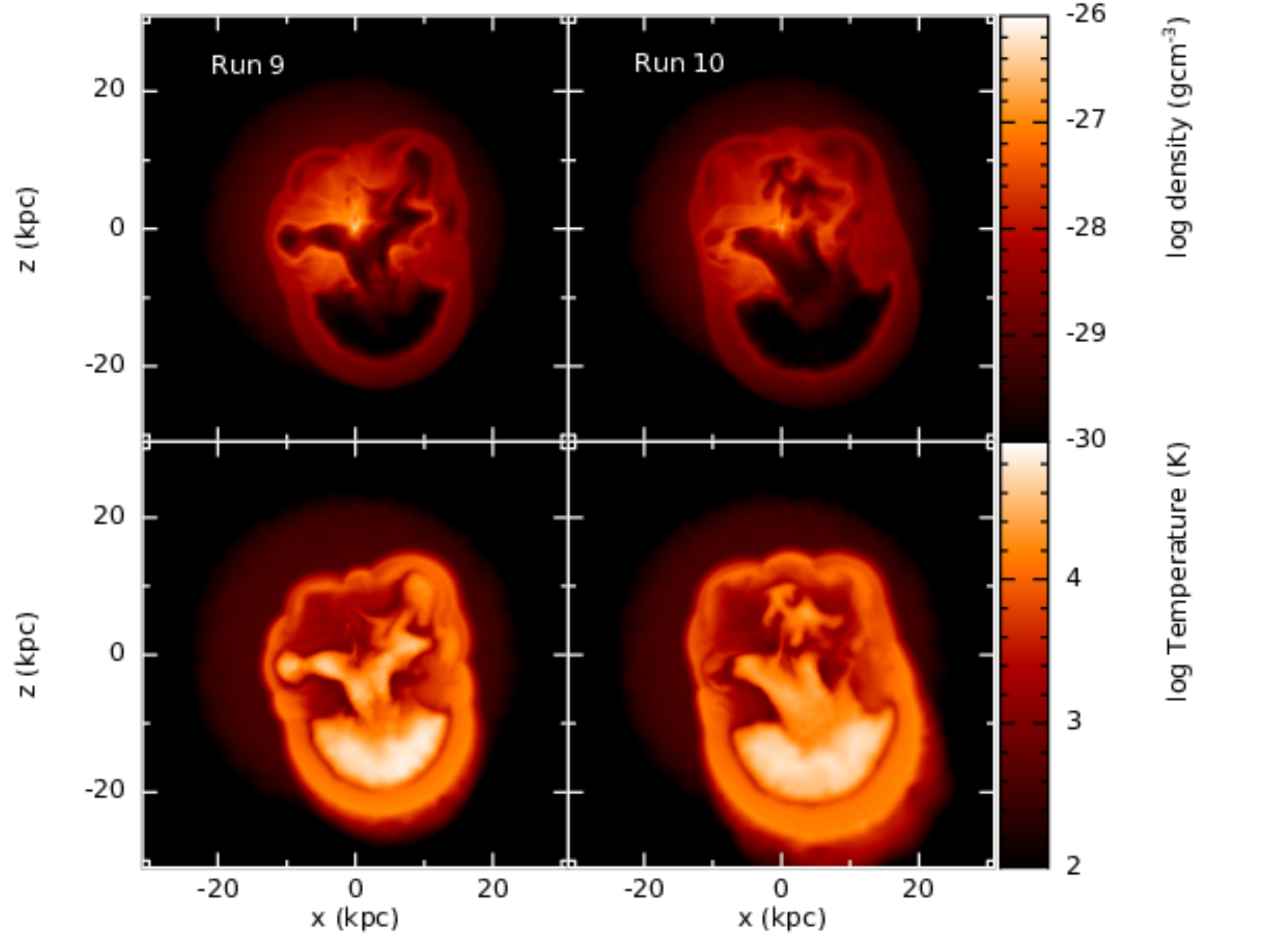}
    \caption{Density (top row) and temperature (bottom row) slices taken in the x-z plane at y = 0, for Runs 9 (left column) and 10 (right column) at 1 Gyr.}
    \label{fig:R9v10xz}
\end{figure}
\begin{figure}
	\includegraphics[width=\columnwidth]{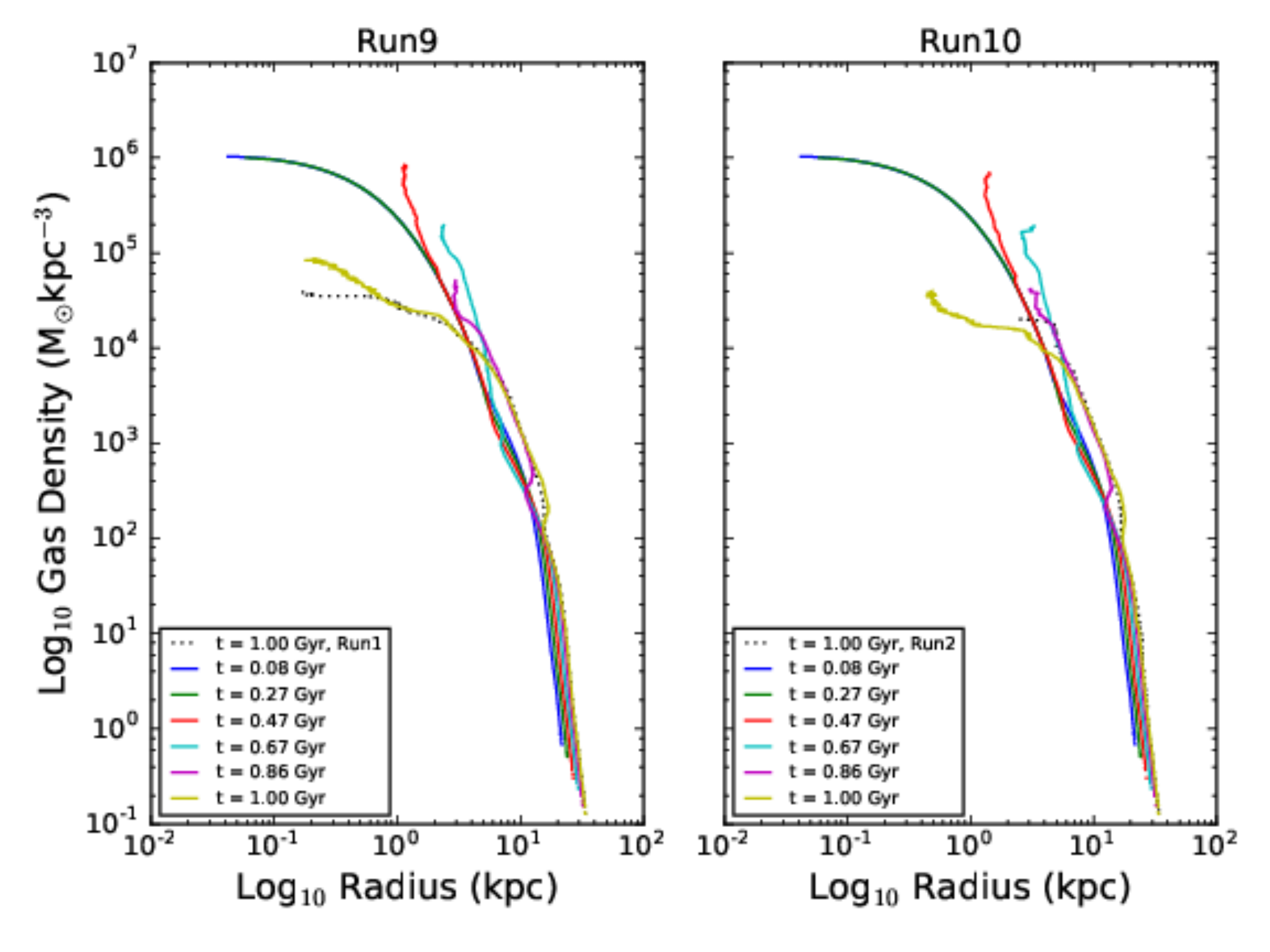}
    \caption{The gas density profiles taken at various times into Run 9 (left) and 10 (right), with the final density profiles of Runs 1 and 2 overlaid onto Run 9 and 10 respectively (dotted lines).}
    \label{fig:Gas_dens_R89}
\end{figure}

In order to investigate the formation of galactic winds in Runs 9 and 10, we plotted slices rendered in density and temperature in all three planes, finding the y = 0 slice in the x-z plane (Fig. \ref{fig:R9v10xz}) contained low density `superbubbles' akin to those seen in Fig. \ref{fig:TD_xz_plane_R12}. Comparing Run 9 with Run 10, we see in both cases there appears to be a mushroom-shaped low density bubble filled with gas at $\sim$ 10$^5$ K, towards the bottom half of each plot. In both cases the bubble is surrounded by a shell of higher density ($\sim$ 10$^{-28}$ - 10$^{-29}$ gcm$^{-3}$), slightly cooler ($\sim$ 10$^{4}$ K) gas. The bubble in Run 10 is larger than Run 9, indicating excluding gradual feedback aids the development of these large-scale ($\sim$ 20 kpc), wide-angle outflows. On the other hand, Run 9 shows two smaller-angle `chimneys' of low density, hot gas on the left and right of Fig. \ref{fig:R9v10xz} . These are not present in Run 9, in this case the hot, low density gas associated with these regions has been effectively funneled into the growing larger-scale `superbubble'. In this way the addition of gradual feedback on top of feedback from SNe appears to have facilitated the production of smaller `chimneys' of hot/warm, low density gas. The energy being funneled through these chimneys would otherwise have been added to the work done on the main `superbubble' and further driven the large-scale galactic wind. 

Fig. \ref{fig:Gas_dens_R89} plots the gas density profiles of Runs 9 and 10 at various times into each simulation, alongside the gas density profiles taken at the end point of Runs 1 and 2 (corresponding to 1 Gyr). We see there is marginal difference between the density profiles of Runs 1 and 9 above 1 kpc, however below this the gas density is higher in Run 9 than Run 1. Therefore, a shorter, more violent starburst has resulted in the retention of a higher gas mass in the centre of the galaxy. As well as this, Run 10 also shows a significantly higher gas mass at radii of $>$ 1 kpc than Run 2. By decreasing the starburst duration, we have also minimised the variation between runs which include gradual types of the feedback and runs which just include SNe feedback; Runs 9 and 10 differ less than Runs 1 and 2. Furthermore, Runs 1 and 9 differ less than Runs 2 and 10, indicating gradual feedback acts to reduce the differences in gas radial distributions, which arise by altering the period of a starburst. However, despite the mean radius of the inner density profile being smaller in Run 10, the unbound gas mass fraction is not significantly altered from Run 2. These results indicate the bound gas in Run 10 is located closer to the centre of the galaxy than the same gas in Run 2, likely through efficient cooling or higher resistance to heating via stellar feedback processes.  

To conclude, altering the duration of the starburst in the galaxy has had only a marginal impact on the final state of the gas at 1 Gyr. The main factor is still the presence of gradual feedback, which has acted to reduce the unbound gas mass fraction (see Fig. \ref{fig:Ubd_R12}) and decrease the total kinetic and thermal energy of the gas in the system (Fig. \ref{fig:Eplot_sigma}). Furthermore, the gradual feedback in the shorter starburst has facilitated the production of $\sim$ 10 kpc, comparatively narrow `chimneys' which have funneled hot, volume-filling gas away from the growing wide-angle super-bubble that exists in both Run 9 and 10.

\subsection{Changing metallicity (Runs 7 and 8)}
Changing the metallicity of the gas in the galaxy has had very little impact on the unbound mass fraction (see figure \ref{fig:Ubd_R12}). As was seen at primordial metallicity, the [Fe/H] = -1.2 run containing gradual feedback on top of SNe feedback (Run 7) has a lower unbound mass fraction across the simulation than equivalent Z run containing just SNe feedback (Run 8). 

\begin{figure}
	\includegraphics[width=\columnwidth]{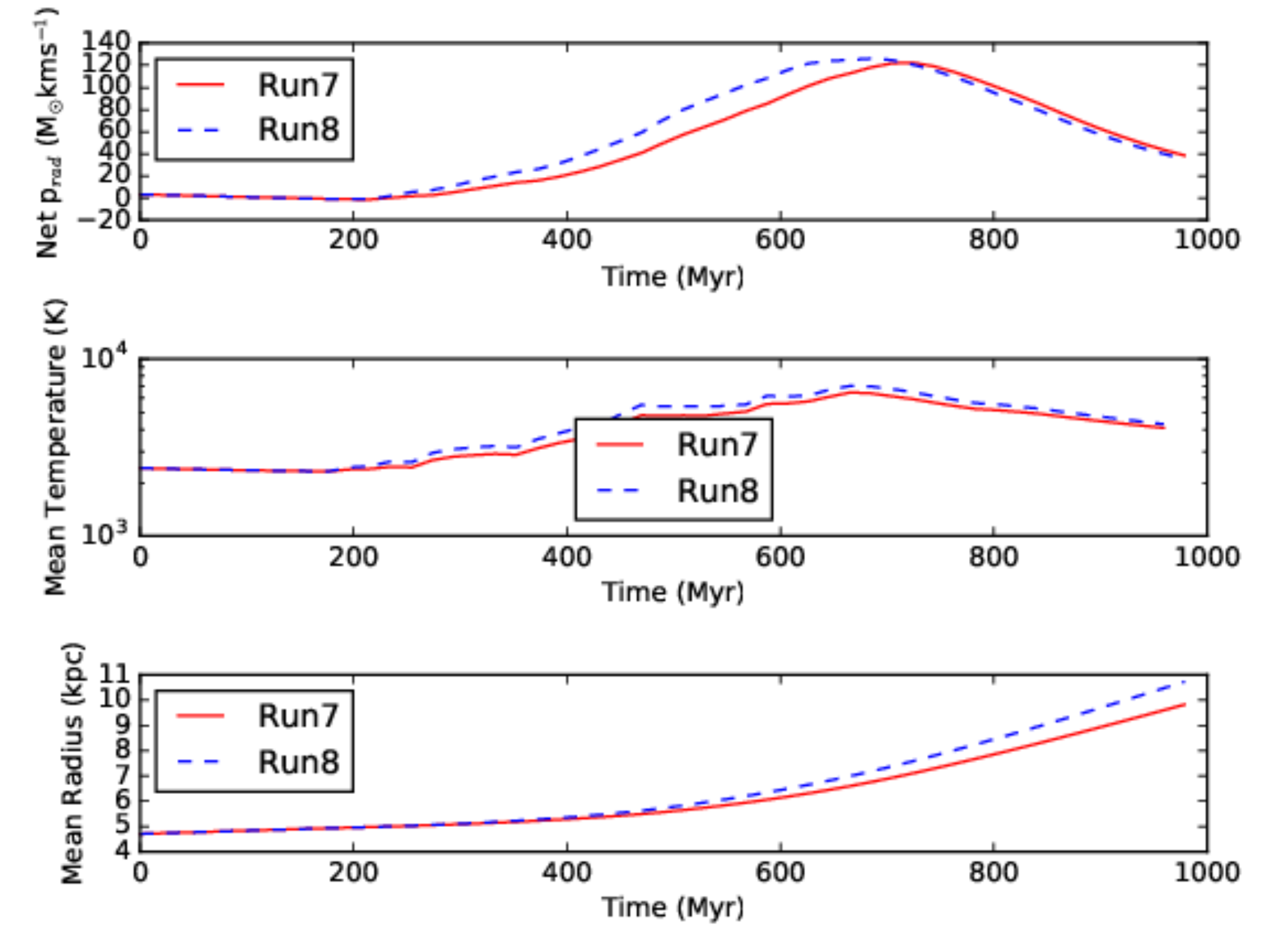}
    \caption{Top plot - the time evolution of the net radial momentum of the gas in Run 7 (solid, red line) and 8 (blue, dashed line). Middle plot - the time evolution of the mean temperature of the gas in Runs 7 and 8. Bottom plot - the time evolution of the mean radius of the gas in Runs 7 and 8.}
    \label{fig:Run7v8_MaxT}
\end{figure}

In Fig. \ref{fig:Run7v8_MaxT} we can see the net radial momentum, mean temperature and the mean radius of the gas in Run 8 is greater than Run 7, once again showing including gradual feedback can lessen the efficiency of galactic winds. Furthermore, the mean temperature is the same for runs at primordial metallicity as those at [Fe/H] = - 1.2. In both runs the majority of the gas is at $\sim$ 6000 K, which corresponds to the virial temperature of the halo. The fact the mean temperature of the gas is similar between different metallicity runs indicates the bulk of the gas is at a high enough density to have a similar cooling timescale. 

\begin{figure}
	\includegraphics[width=\columnwidth]{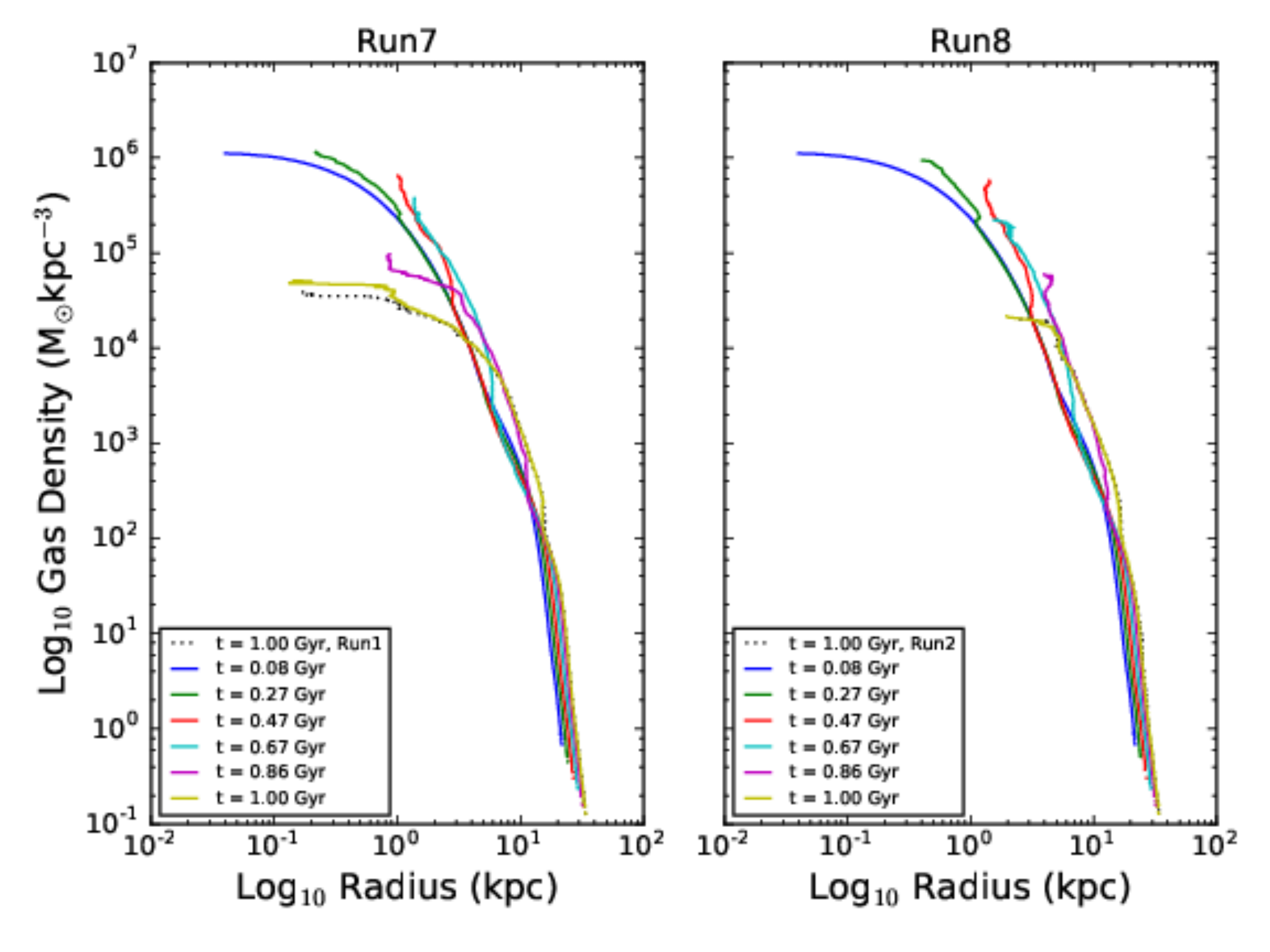}
    \caption{The gas density profiles of Run 7 (left) and 8 (right) taken at varying intervals into each simulation. The final gas density profiles of Run 1 and 2 (taken at 1 Gyr) were over-plotted on Run 7 and 8 respectively.  }
    \label{fig:Gasdens_R7v8}
\end{figure}

Moreover, Fig. \ref{fig:Gasdens_R7v8} shows the gas density of Runs 7 and 8 do not vary significantly for their low metallicity counterparts (Runs 1 and 2). The same trends that are present in Runs 1 and 2 are also present at higher metallicity, namely the inclusion of gradual feedback has prevented the efficient clearing of gas below 1 kpc.  

Overall, altering the metallicity of the dwarf galaxy has had very little impact on both the gas phase and the dark matter/ gaseous halo. This is most likely due to the fact the density of the ISM at the location of the SNe is high, resulting in similar cooling timescales between different metallicity runs and hence the gas in both runs converging on the same temperature (see Fig. \ref{fig:Run7v8_MaxT} and Fig. \ref{fig:R12_Temp}). Furthermore, since by altering the metallicity we have effectively altered the lifetimes of the stellar wind and HMXB phase (see Fig. \ref{fig:Lifetimes}), these results indicate the additional energy input from winds and HMXB has had little impact on the state of the gas in the galaxy. 

\subsection{Increasing the concentration parameter (Runs 3 and 4)}
Firstly, from Fig. \ref{fig:Ubd_R12} it is clear by increasing the concentration parameter, the amount of gas that has been unbound in the galaxy has dropped by 14$\%$. This is expected since the potential in the centre of the dark matter halo will be higher for a more concentrated halo, hence feedback will have to do more work on the ISM to unbind the gas. However, Runs 3 and 4 also show the addition of gradual feedback has reduced the amount of gas being unbound, despite more energy being injected into the ISM.

\begin{figure}
	\includegraphics[trim={0 0 0 0}, clip, width=0.5\textwidth]{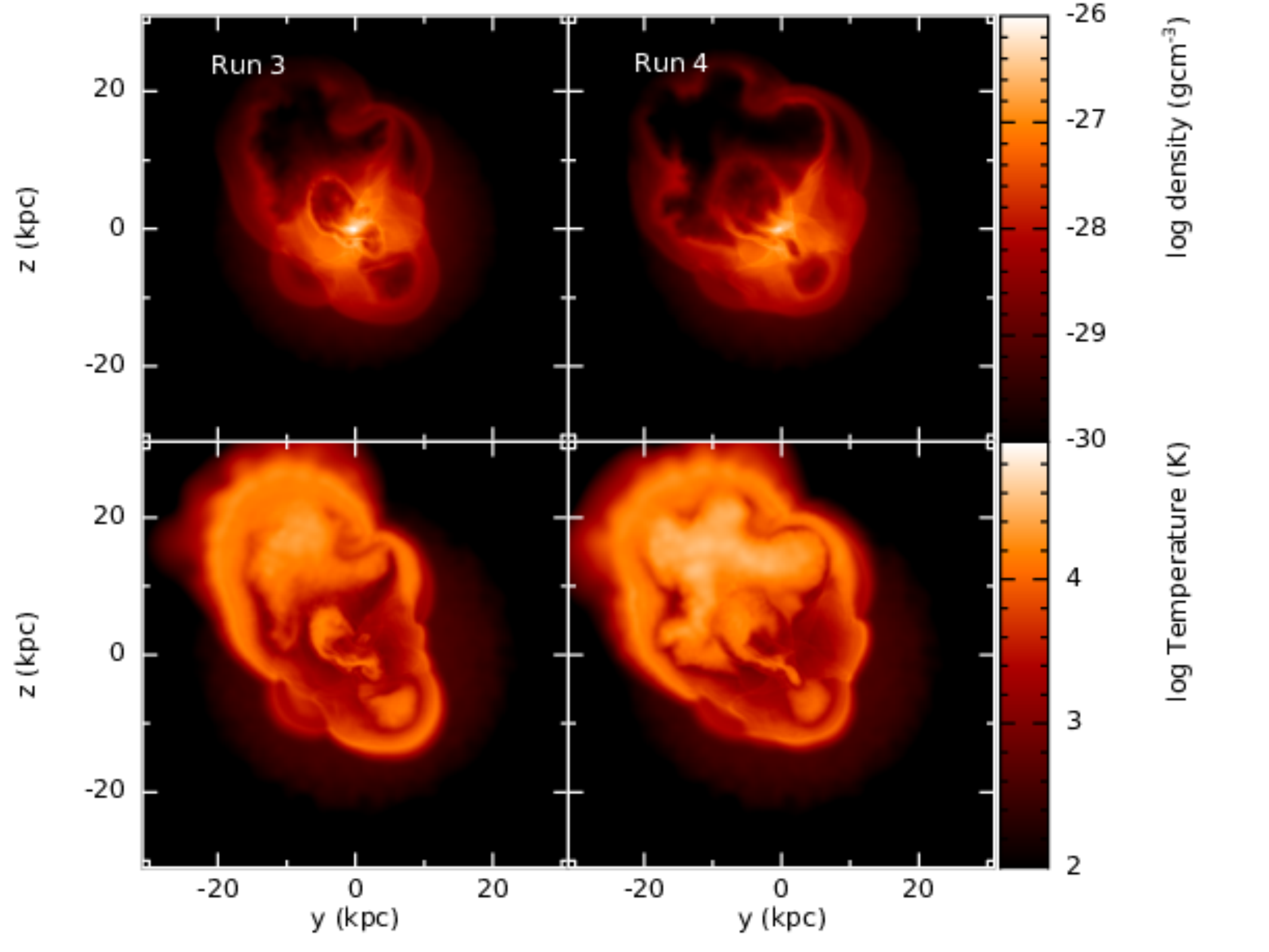}
    \caption{Density (top row) and temperature (bottom row) slices taken in the y-z plane at x = 0, for Runs 3 (left column) and 4 (right column) at 1 Gyr.}
    \label{fig:R3v4_yz}
\end{figure}

In order to investigate whether or not the `superbubbles' seen in Fig. \ref{fig:R9v10xz} and Fig. \ref{fig:TD_xz_plane_R12} also occur in Runs 3 and 4, we plotted density and temperature slices in the x, y and z planes. In Fig. \ref{fig:R3v4_yz} we show an example in the y-z plane at x = 0. Here we see both the SNe and HMXB/stellar winds have successfully produced warm/hot (10$^{4}$ - 10$^5$ K) lobes of low density gas in the upper left hand corner of the plot. Unlike in previous instances, these lobes still contain smaller ring-like structures of dense gas less than 10 kpc in scale. Furthermore, two smaller ($\sim$ 5 kpc in diameter) bubbles of warm gas at densities of $\sim$ 10$^{29}$ gcm$^{-3}$ can be seen in the bottom right of both Runs 3 and 4. As seen previously, the run which excludes gradual types of feedback (4), contains a larger `superbubble'. 

\begin{figure}
	\includegraphics[width=\columnwidth]{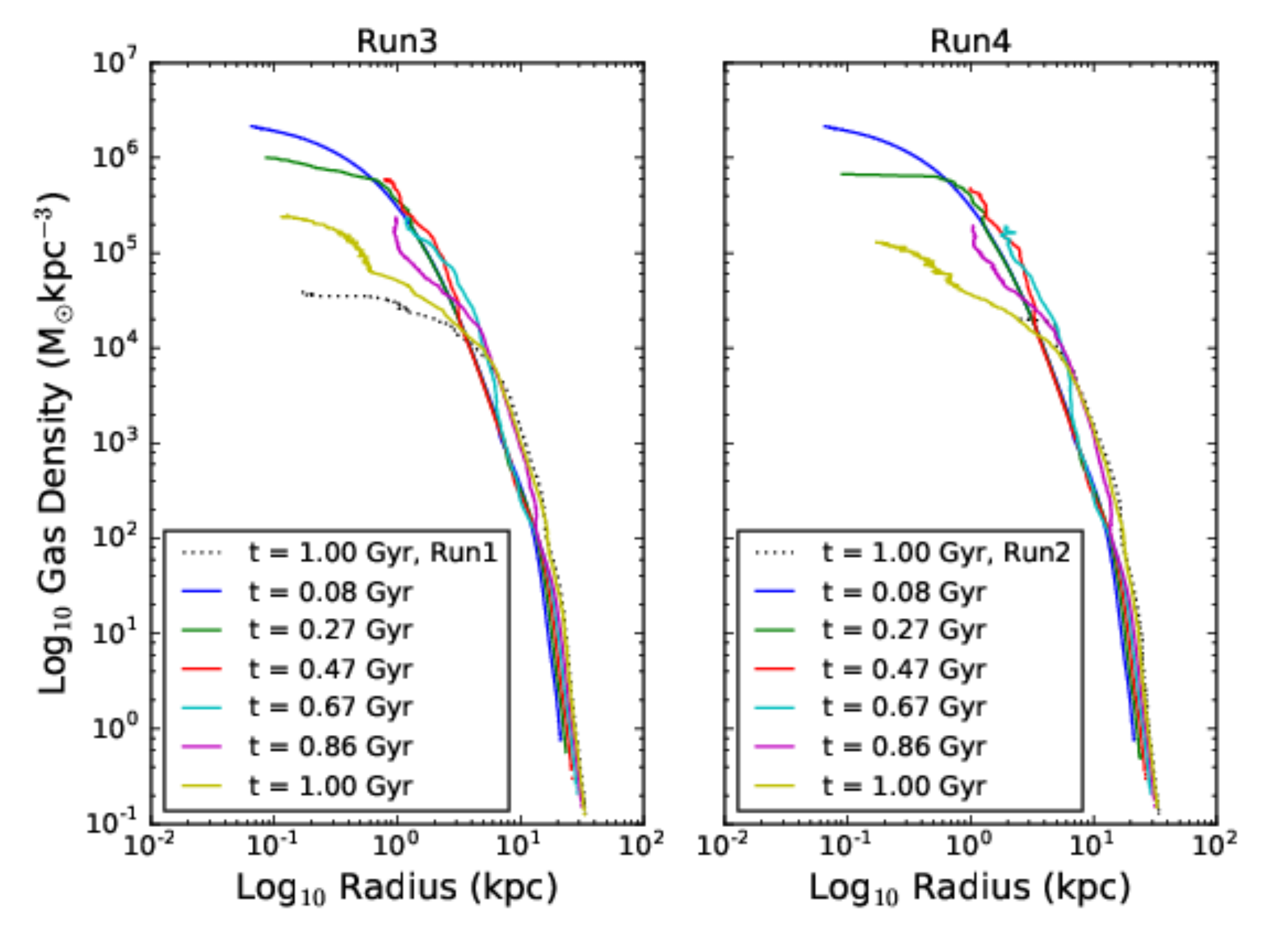}
    \caption{The density profile of the gas in Runs 3 (left) and 4 (right) taken at various times into each simulation. The final density profiles (taken at 1 Gyr) for Run 1 and 2 (dotted lines) have been over-plotted on Run 3 and 4 respectively.}
    \label{fig:Gdens_34}
\end{figure}

From Fig. \ref{fig:Gdens_34} we can see the gas density profile is much less affected by feedback than the runs that have a lower dark matter halo concentration. As seen previously, once feedback has switched off (prior to 850 Myr) the gas has re-accreted onto the inner 1 kpc of the halo, increasing the density of the gas in this region between 0.86 Gyr to 1 Gyr. Furthermore, the gas density inside the inner 1 kpc of Run 3 is higher than Run 4. In this way, once again including gradual feedback has increased the re-accretion of gas onto the centre of the halo once the starburst has ended. The density profiles of Run 2 and Run 4 differ significantly at the end of the simulation; with the gas in Run 4 extending to an order of magnitude smaller radius than Run 2. The difference is less drastic when comparing Runs 1 and 3, however, the density is approximately an order of magnitude greater inside the inner kpc of Run 3. 

\begin{figure}
	\includegraphics[width=\columnwidth]{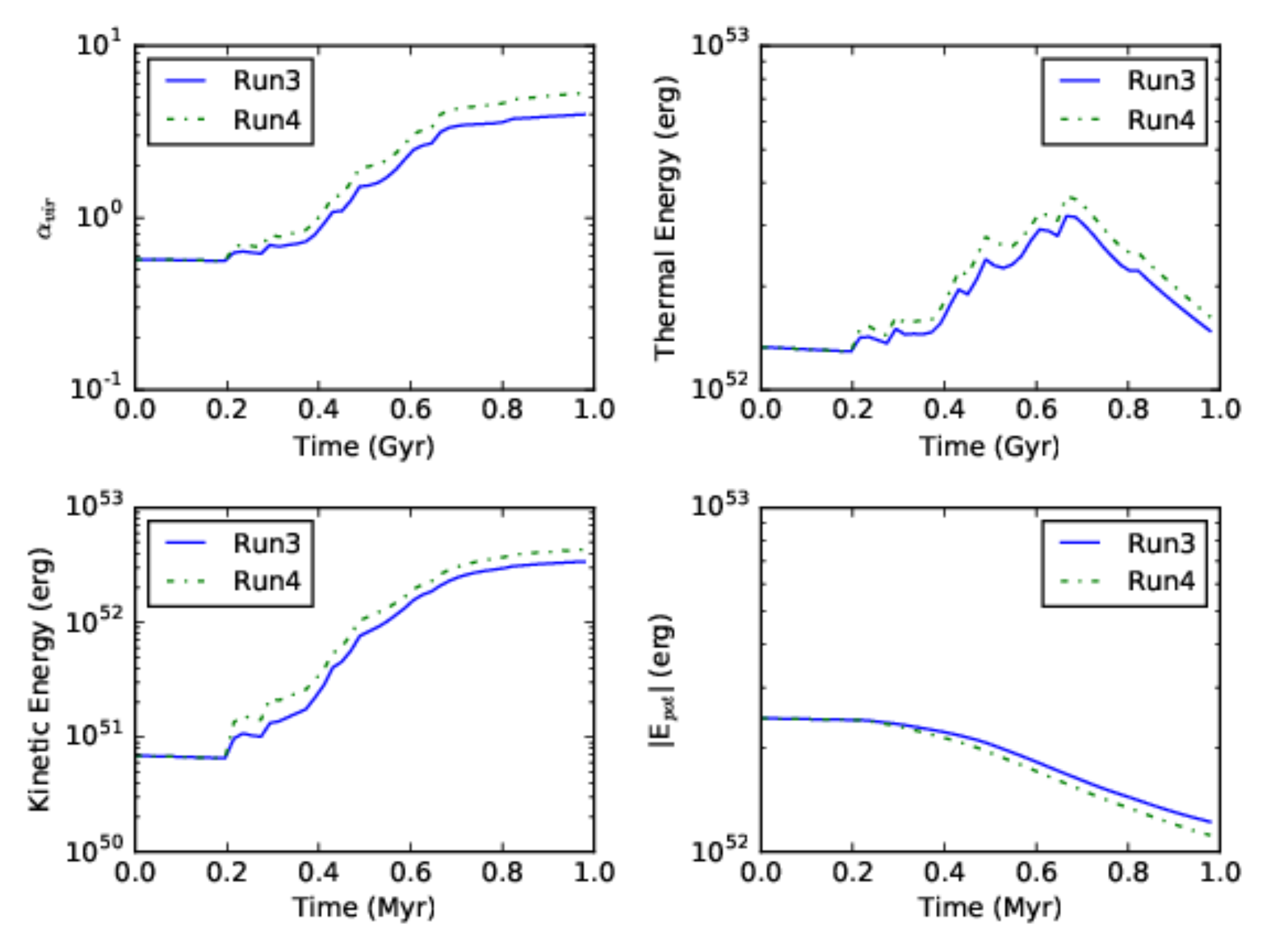}
    \caption{The time evolution of the virial parameter (top left plot, where $\alpha_{vir}$ = E$_{kin}$ + E$_{therm}$ / |E$_{pot}|$) and the total thermal (top right), kinetic (bottom left) and potential (bottom right) energy of the gas in Run 3 (solid, blue lines) and 4 (dot-dashed, green lines).}
    \label{fig:Eplots_34}
\end{figure}

We next plot the time evolution of the global energetics of the gas in Run 3 and 4 in Fig. \ref{fig:Eplots_34}. As expected from Fig. \ref{fig:Ubd_R12}, the total kinetic and thermal energy of the gas in Run 4 is larger than Run 3, along with the virial parameter, while the potential energy is correspondingly lower. This is due to more efficient cooling in Run 3, which is most likely caused by the increased density of the majority of the gas (as seen in Fig. \ref{fig:Gdens_34}).

Overall, as is expected, increasing the concentration of the dark matter halo (and thereby increasing the potential at the centre of the halo) has resulted in less gas being unbound by stellar feedback. These runs also show the same trend that has been seen throughout this paper; gradual feedback has acted to decrease the amount of gas unbound by the feedback and also increase the central gas density (below 1 kpc) of the halo at the end of the simulation. 

\subsection{A smaller galaxy (Run 5 and 6)}
From Fig. \ref{fig:Ubd_R12} we can see altering the size of the galaxy has had the largest impact on the amount of gas unbound from the halo. In both Run 5 and 6 the fraction of the initial gas mass that has been unbound approaches unity by the end of the simulation. However, including gradual feedback has had a large impact on the timescale over which the majority of the gas has been unbound; by 0.5 Gyr $\sim$ 100$\%$ of the gas has been unbound in Run 6, however, approximately 85 $\%$ has been unbound in Run 5.
\begin{figure}
	\includegraphics[trim={0 0 0 0}, clip, width=0.5\textwidth]{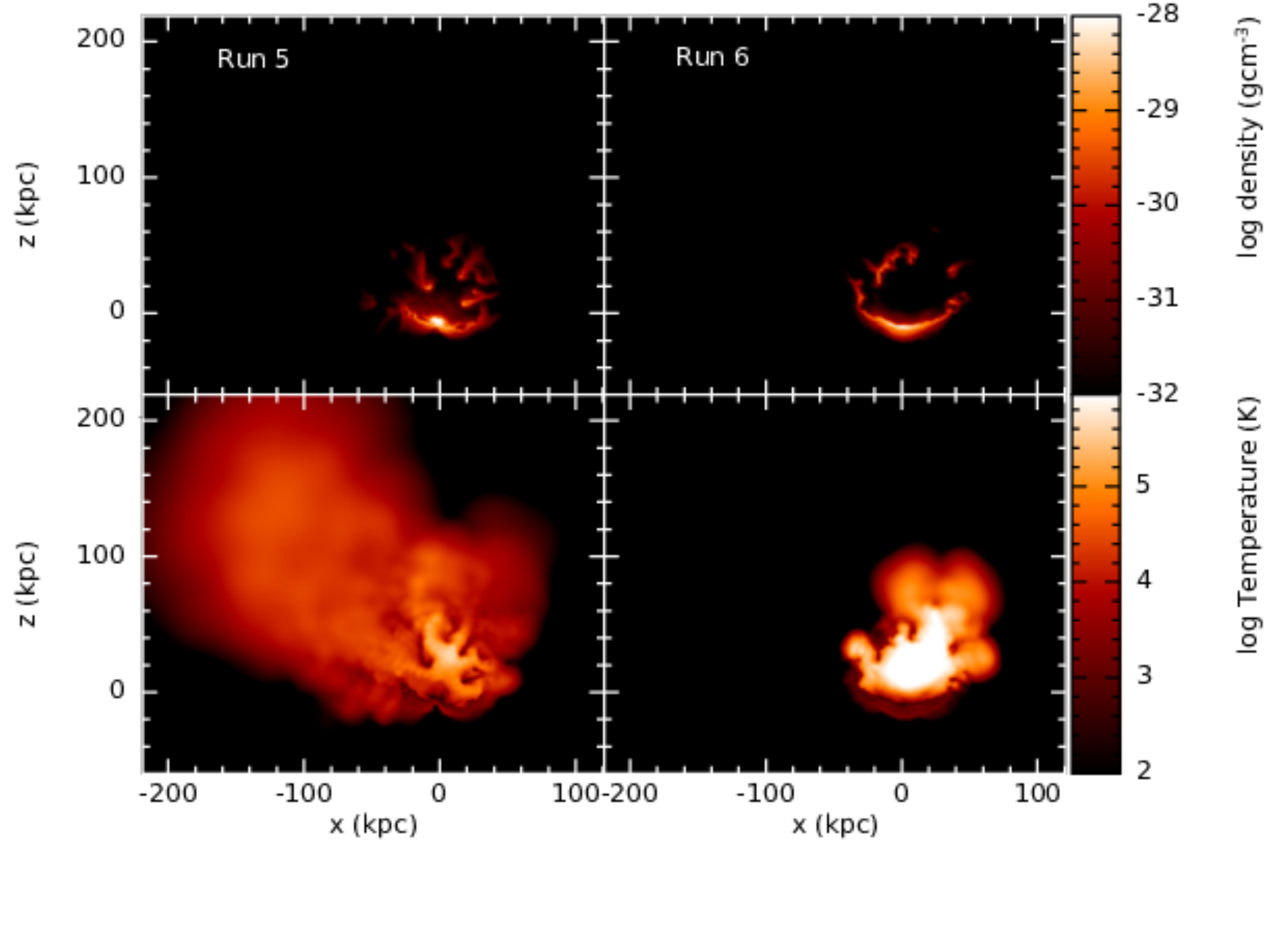}
    \caption{Density and temperature slices (upper row and bottom row respectively) for Runs 5 (left column) and 6 (right column), taken at 1 Gyr in the x-z plane at y = 0.}
    \label{fig:R56_yz}
\end{figure}

Looking at the corresponding temperature and density slices taken in the x-z plane at y=0 (Fig. \ref{fig:R56_yz}), we see Run 5 has gas extending to beyond 200 kpc, while the gas in Run 6 is confined to below 100 kpc. The central 40 kpc of Run 6 has been uniformly heated to 10$^6$ K, while the same radius in Run 5 contains regions of cooler ($<$ 10$^3$ K) gas. Run 5 appears to be in the process of funneling hot (10$^5$ - 10$^6$ K) gas away from the galactic centre, via multiple lower density ($<$ 10$^{-32}$ gcm$^{-3}$) chimneys or channels, similar to those seen in GS18. This chimneys can also been seen in the corresponding y-z and x-y planes. 
\begin{figure}
	\includegraphics[width=\columnwidth]{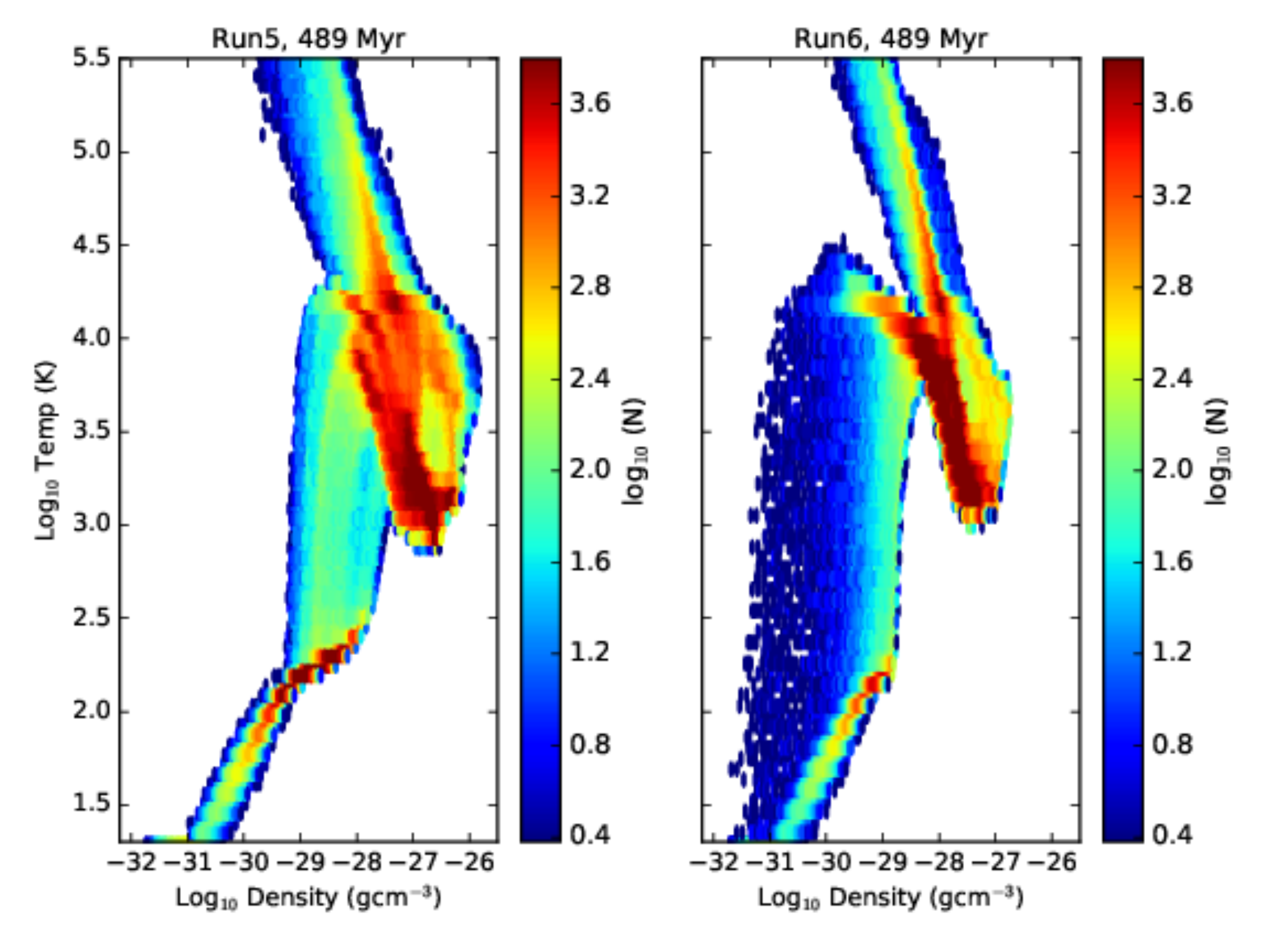}
    \caption{The density and temperature of gas particles 489 Myr into Runs 5 (left) and 6 (right), rendered according to particle number.}
    \label{fig:R56_TvD}
\end{figure}

To investigate the large difference between the mass fraction of unbound gas seen for Run 5 and 6 in Fig. \ref{fig:Ubd_R12} approximately half way through the simulation, we plot the temperature and density of the gas in both runs for a snapshot at 489 Myr (Fig. \ref{fig:R56_TvD}). \newnote{Comparing the two runs, we see there is more high density ($>$ 10$^{-27}$ gcm$^{-3}$), low temperature ($<$ 10$^4$ K) gas in the run that includes gradual types of feedback (Run 5). Given we do not have the resolution to follow the fragmentation of this gas into pc-scale cold clumps and the ensuing star formation, this result simply indicates gradual feedback has the potential to facilitate further star formation in dwarf galaxies beyond the onset of massive stellar feedback during a starburst.} 

\begin{figure}
	\includegraphics[width=\columnwidth]{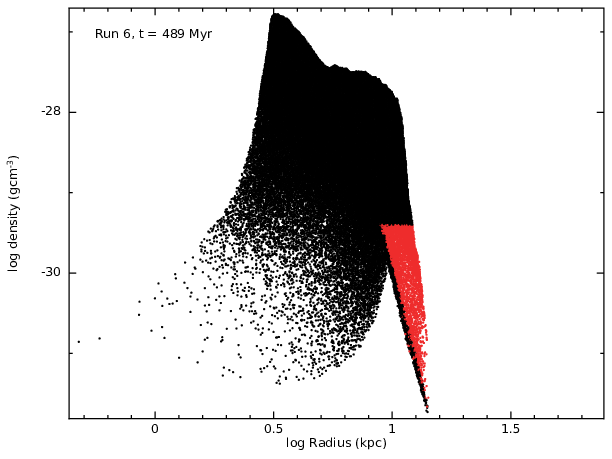}
    \caption{The density versus radius of all gas particles in Run 6 at 489 Myr, with a subset colored red according to their position in Fig. \ref{fig:R56_TvD}. These particles have a density lower than 10$^{-29}$ gcm$^{-3}$ temperatures spanning 20 K to 3 $\times$ 10$^{4}$K. Particles with this temperature and density range are missing from Run 5. }
    \label{fig:R6_DensVrad}
\end{figure}

\begin{figure}
	\includegraphics[width=\columnwidth]{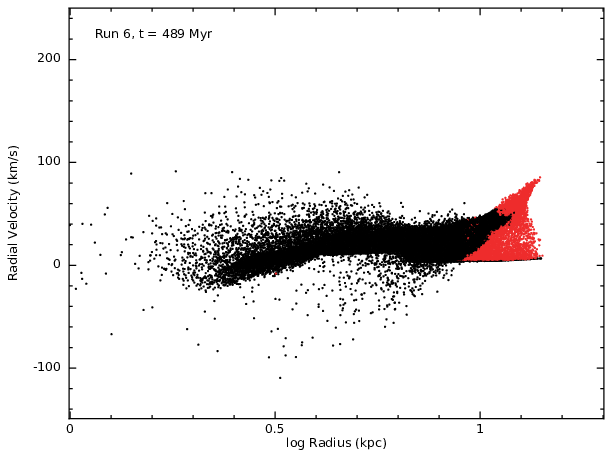}
    \caption{The radial velocity versus radius of all gas particles in Run 6 at 489 Myr, with a subset colored red according to their position in Fig. \ref{fig:R56_TvD}. These particles have a density lower than 10$^{-29}$ gcm$^{-3}$ temperatures spanning 20 K to 3 $\times$ 10$^{4}$K.}
    \label{fig:R6_vr}
\end{figure}

\begin{figure}
	\includegraphics[width=\columnwidth]{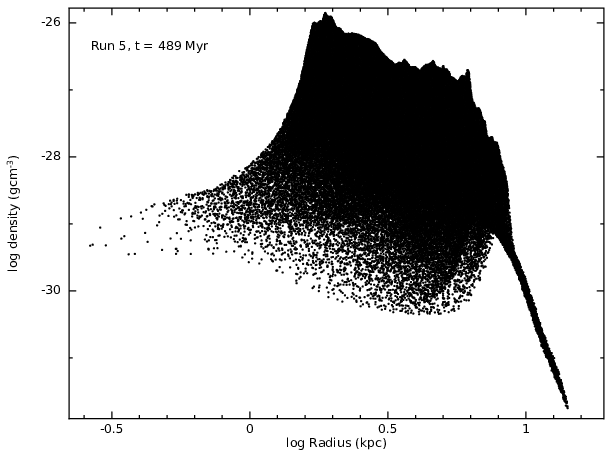}
    \caption{The density versus radius of all gas particles in Run 5 at 489 Myr.}
    \label{fig:R5_DvR}
\end{figure}

\newnote{Fig. \ref{fig:R56_TvD} also shows a} branch of low density ($<$ 10$^{-29}$ gcm$^{-3}$) gas spanning a temperature range from 20 K to 3 $\times$ 10$^{4}$ K in Run 6 that is not present in Run 5. We explore the origins of this gas in Run 6 by coloring a subset of these particles red and plotting both the radial velocity and the density of all particles in Run 5 and Run 6, as a function of radius (Fig. \ref{fig:R6_DensVrad} and Fig. \ref{fig:R6_vr}) respectively. Here we see multiple SNe in the centre of the galaxy have resulted in a roughly spherical shock front, indicated by the density peak in Fig. \ref{fig:R6_DensVrad}. The low density particles we are interested in constitute the swept-up mass of the shock. Here the gas has been accelerated radially (shown by the positive radial velocity) to the outskirts of the gas halo. If we also plot the density profile for the particles in Run 5 (Fig. \ref{fig:R5_DvR}), this swept-up mass is absent, while the global density peak is more fragmented, consisting of multiple shocks that are out of phase with one another (as opposed to the smoother density peak of Run 6).  

\begin{figure}
	\includegraphics[width=\columnwidth]{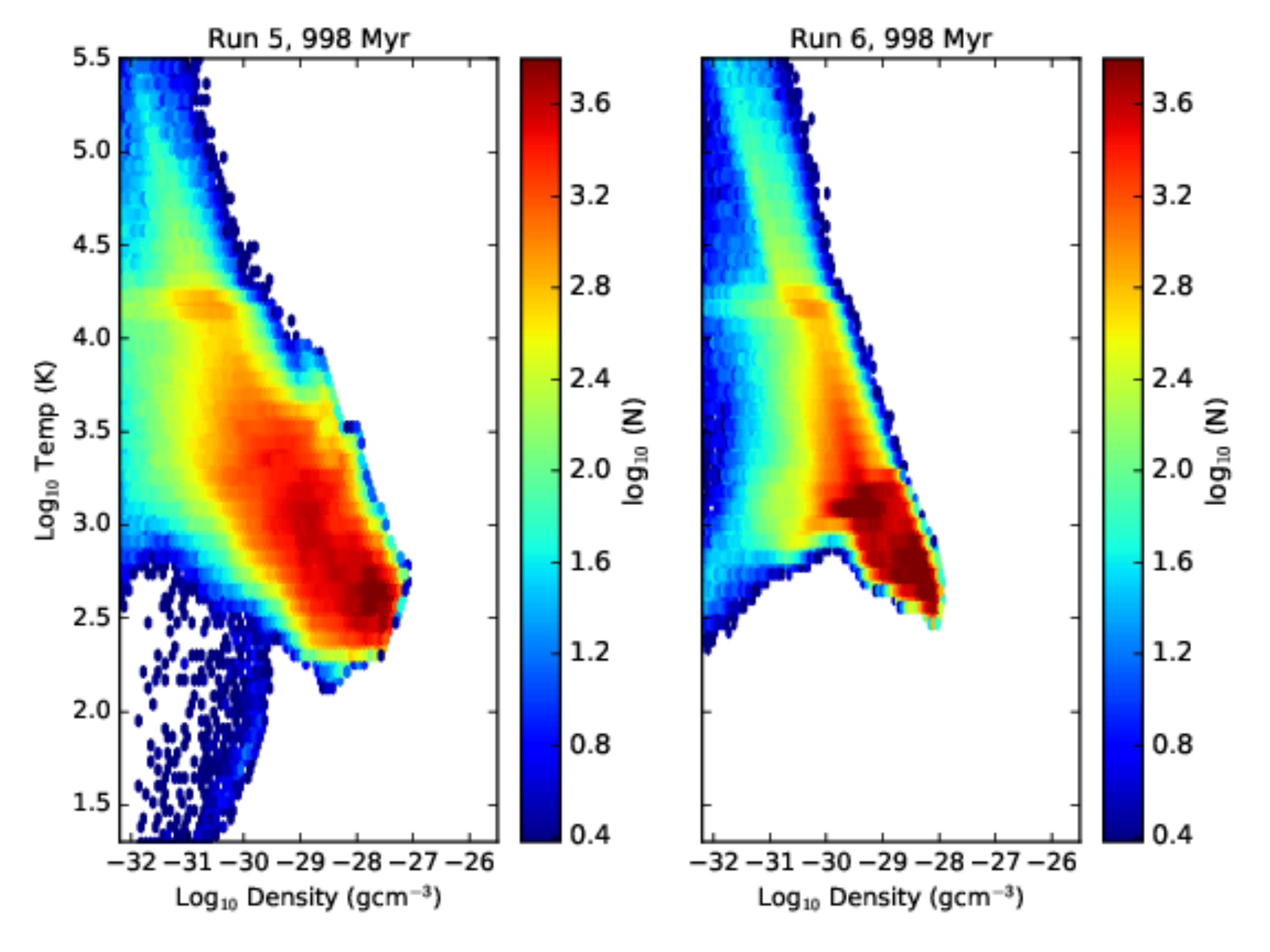}
    \caption{\newnote{The density and temperature of gas particles 998 Myr into Runs 5 (left) and 6 (right), rendered according to particle number.}}
    \label{fig:R56_TvD_1Gyr}
\end{figure}

\newnewnote{Moreover, we re-plotted Fig. \ref{fig:R56_TvD} at $\sim$ 1 Gyr instead, see Fig. \ref{fig:R56_TvD_1Gyr}, in order to further explore the changes in gas phase seen across the simulation. Here we can see there is no gas above 10$^{-27}$ gcm$^{-3}$ present in either simulation, while the bulk of the gas has cooled in both Runs 5 and 6 - the mean gas temperature has dropped by 1000 K and 3000 K respectively. Furthermore, Run 5 shows a broader range of density bins than Run 6; with gas occupying density bins of $>$ 10$^{-28}$ gcm$^{-3}$ and $<$ 10$^{-34}$ gcm$^{-3}$, which are not present in Run 6. This indicates the ability of gaseous chimneys to allow the co-existence of a wide range of gas phases in a galaxy.}

Run 5 also contains an arm of low density, cold gas with temperatures below 10$^2$ K \newnewnote{in Fig. \ref{fig:R56_TvD_1Gyr},} which is not present in Run 6. Upon further investigation, this arm consisted of $\sim$ 1000 gas particles at radii between 10-40 kpc. We tracked these particles across the simulation, finding the majority are part of the tails in the density distributions of multiple swept up shells. As we saw previously, these shells are less pronounced in Run 6 due to the fact individual SNe remnants have combined to form a more coherent shock front (see Fig. \ref{fig:R6_DensVrad}). The particles were also heated and cooled multiple times across the simulation. In this way, the low temperature, low density arm of this diagram is populated by different gas particles at any one snapshot during the starburst. This means the gas is being heated, cooled and heated again on a timescale shorter than tens of Myr (the snapshot frequency).

\begin{figure}
	\includegraphics[width=\columnwidth]{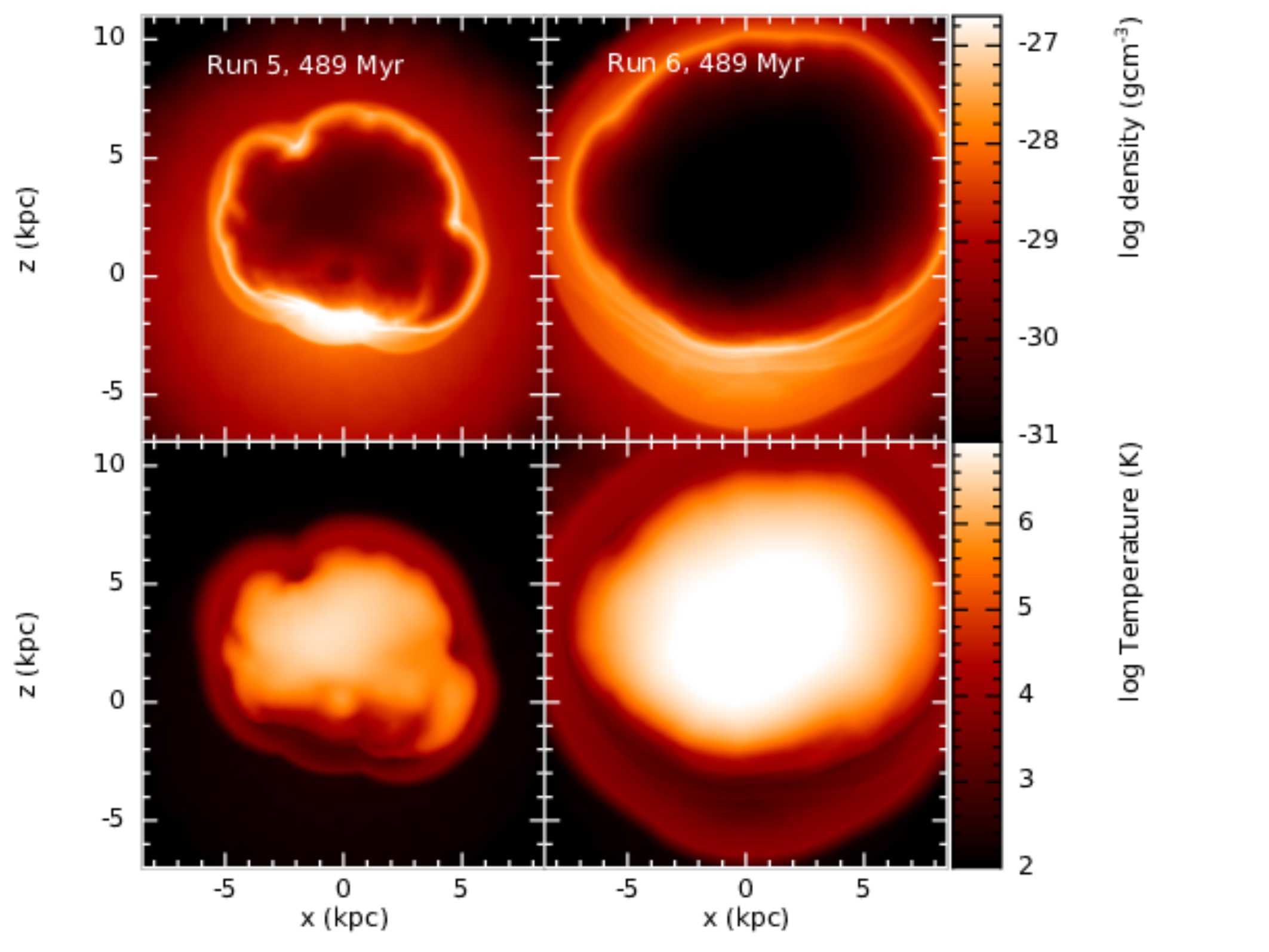}
    \caption{Density and temperature slices (upper row and bottom row respectively) for Runs 5 (left column) and 6 (right column), taken at 489 Myr in the x-z plane at y = 0.}
    \label{fig:R56_xz_05Gyr}
\end{figure}

We also plot the x-z density and temperature y = 0 slices for Run 5 and Run 6 at 489 Myr (Fig. \ref{fig:R56_xz_05Gyr}), to investigate the origin of the swept-up mass that is present in Run 6 (Fig. \ref{fig:R6_DensVrad}) and absent in Run 5 (Fig. \ref{fig:R5_DvR}). Here we see, as was indicated by Fig \ref{fig:R6_DensVrad}, multiple SNe have created a more or less isotropic, hot bubble of gas, which is expanding and sweeping up the surrounding ISM to produce a dense, cool/warm ($\sim$ 10$^4$-10$^2$ K, as was seen in Fig. \ref{fig:R56_TvD}) shell of gas, which is 4 kpc thick in places. On the contrary, adding HMXB and wind feedback to Run 5 has resulted in a bubble that is less spatially extended and `pinched' in places. These `pinches' represent regions of high density, cold gas that have survived the expansion of the feedback-heated ISM. In this way, the gradual feedback has acted to reduce the mixing of the feedback-heated gas and the colder, denser gas. In particular, dense regions have formed between feedback-heated bubbles, representing areas where shells of swept-up mass have collided. In the case of Run 6 these have been efficiently heated and hence destroyed by the more powerful SNe explosions. However, since the stellar winds are less powerful, the warm gas inside the bubbles has instead escaped through the low density channels either side, carving low density chimneys as they do so. These then provide a path of least resistance for the SNe energy to escape the galaxy and less efficiently couple to the cold, dense ISM.
\begin{figure*}
	\begin{center}
	\includegraphics[trim={0 1cm 0 0}, clip, width=\textwidth]{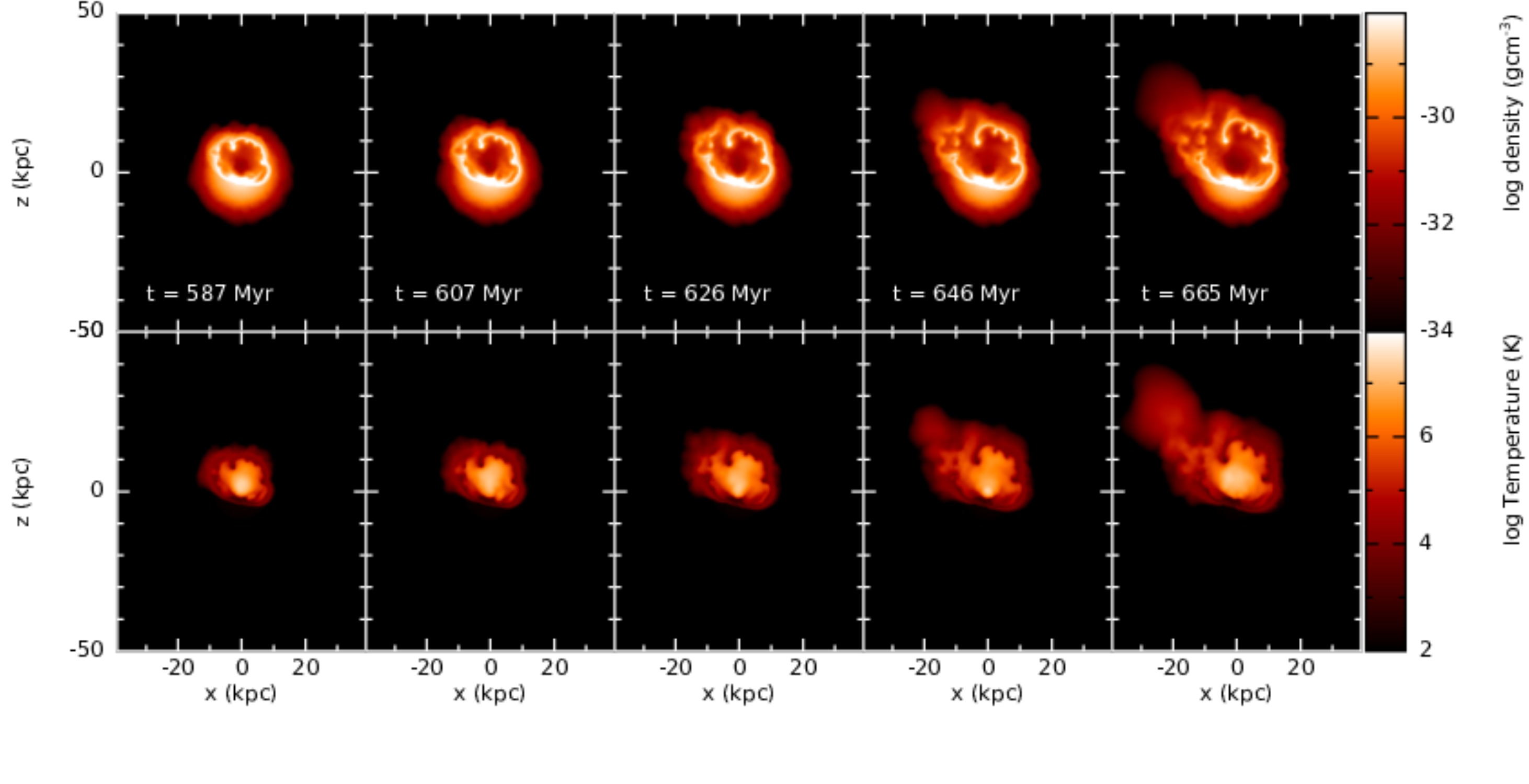}
    \caption{Density and temperature slices (upper row and bottom row respectively) for Run 5, taken at times ranging from 587 Myr to 665 Myr in the x-z plane at y = 0.}
    \label{fig:R5_xz_evoln}
    \end{center}
\end{figure*}

\begin{figure}
	\includegraphics[width=\columnwidth]{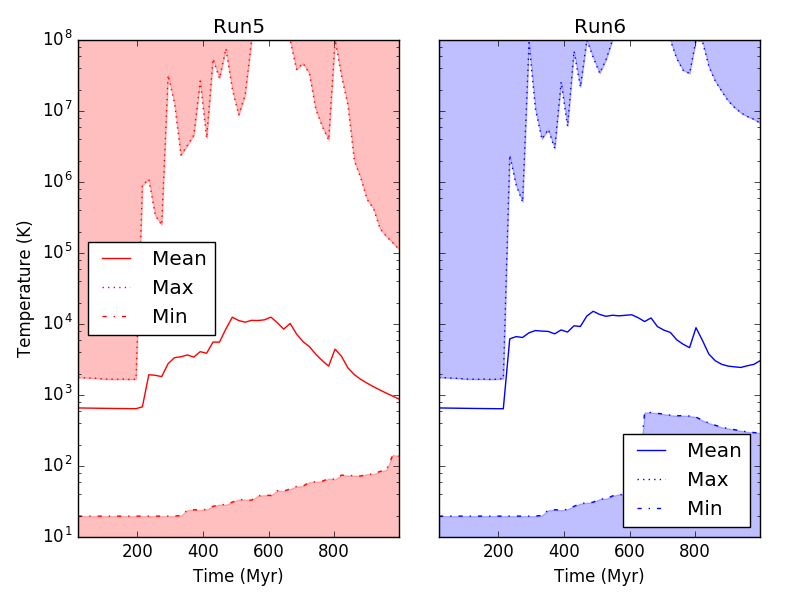}
    \caption{A plot to show the evolution of the mean (solid lines), minimum (dot-dashed line, bordering the lower shaded areas) and maximum (dotted line, bordering the upper shaded areas) temperature of the gas inside a radius of 10 kpc, across the 1 Gyr starburst in Runs 5 (left plot, red) and 6 (right plot, blue).}
    \label{fig:R56_Temp_Analysis}
\end{figure}

In Fig. \ref{fig:R5_xz_evoln} we follow the evolution of Run 5 across 78 Myr from 587 Myr to 665 Myr in order to track the development of the chimney seen in Fig. \ref{fig:R56_yz}. Comparing with Fig. \ref{fig:Rel_Energy}, we see winds are ubiquitous throughout this period, however there are also two episodes of HMXB feedback between 600 Myr and 700 Myr. It is at this point the hot, pressurised gas inside the bubble punches through the least dense areas of the surrounding shell (to the left of the plots) and beyond this hot gas is in the process of being vented through these gaps. This suggests the more powerful HMXB events facilitate the formation of `chimneys' in the swept-up high density shell, which act to vent hot gas. Looking at Fig. \ref{fig:R56_yz}, at 1 Gyr it appears these chimneys have acted to reduce the temperature of the gas in the centre of the halo, as well as maintain regions of high density, cold gas in the central 10 kpc of the halo. We can also see this when we plot the maximum, mean and minimum temperature of the gas located in the central 10 kpc of the halo for both Runs 5 and 6 (Fig. \ref{fig:R56_Temp_Analysis}). Here we see the maximum, mean and minimum temperatures are generally higher in Run 6 than Run 5. Furthermore, the steep rise in the minimum temperature of the gas in Run 6 beyond 600 Myr is not present in Run 5 (while the maximum gas temperature converges at this time), indicating the effect of the chimney seen in Fig \ref{fig:R5_xz_evoln} has been to prevent the coldest gas from being heated by an order of magnitude.   

\begin{figure}
	\includegraphics[width=\columnwidth]{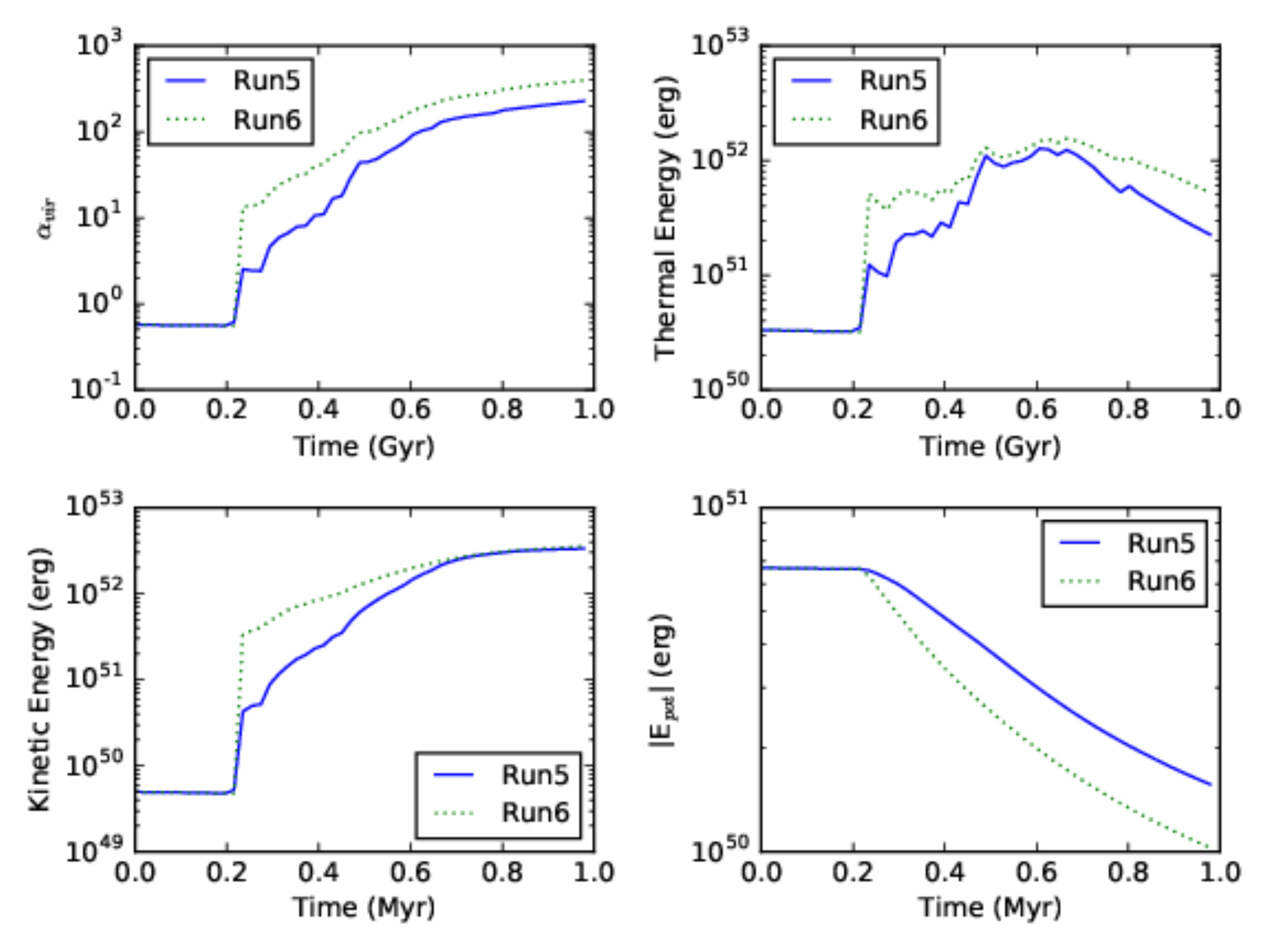}
    \caption{The time evolution of the virial parameter (top left plot, where $\alpha_{vir}$ = E$_{kin}$ + E$_{therm}$ / |E$_{pot}|$) and the total thermal (top right), kinetic (bottom left) and potential (bottom right) energy of the gas in Run 5 (solid, blue lines) and 6 (dotted, green lines).}
    \label{fig:R56_Eplots}
\end{figure}

\begin{figure}
	\includegraphics[width=\columnwidth]{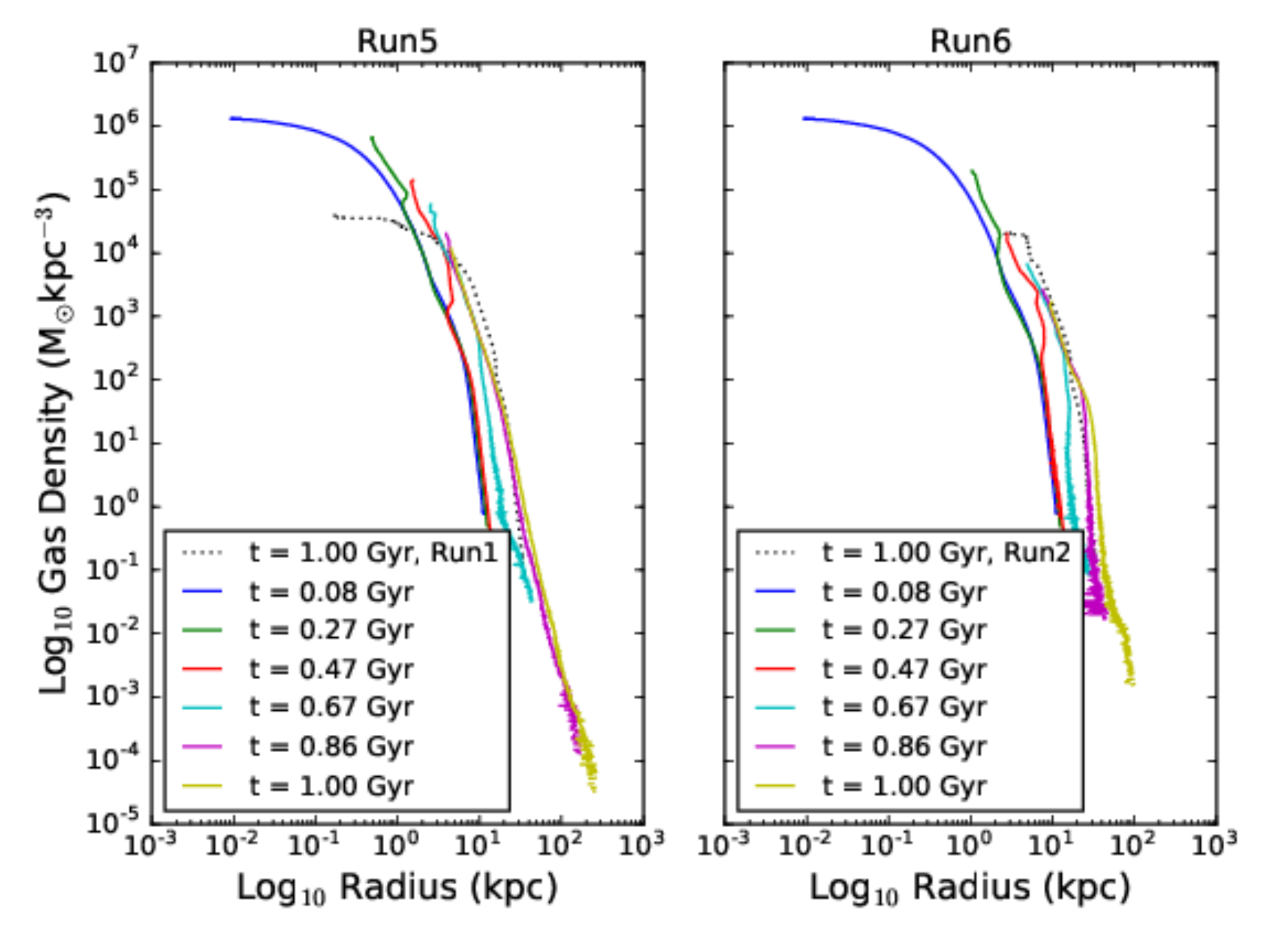}
    \caption{The density profile of the gas in Runs 5 (left) and 6 (right) taken at various times into each simulation. The final density profiles (taken at 1 Gyr) for Run 1 and 2 (dotted lines) have been over-plotted on Run 5 and 6 respectively.}
    \label{fig:R56_Gdens}
\end{figure}

 Fig. \ref{fig:R56_Eplots} shows the total thermal and kinetic energy of the gas in Run 5 is lower than in Run 6, until beyond 600 Myr, where the total kinetic energies of the two runs converge. Moreover, due to the fact the thermal energy of the gas in Run 6 is higher throughout the simulation, the virial parameter is also consistently higher. Looking instead at the gas density profile (Fig. \ref{fig:R56_Gdens}), we see the gas density in the inner 10 kpc of the halo is uniformly higher in Run 5 than Run 6, however beyond this radius the gas in Run 5 extends to lower densities than the gas in Run 6 - indicating the low density, warm/hot outflowing gas seen in Fig. \ref{fig:R56_yz}. Moreover, contrary to the simulations with the larger halo mass, there is no evidence for inflowing gas from the density profile of the smaller halo runs.

In summary, the inclusion of gradual feedback (on top of SNe feedback) in a smaller galaxy of halo mass 1.5$\times$10$^{7}$ M$_\odot$ has increased the time it has taken to unbind the majority of the gas. It has also facilitated the production of chimneys, which have vented hot gas from the centre of the halo and lowered the temperature of the gas in the central 10 kpc. 

\subsection{Evaluating the relative impact of HMXBs and stellar winds (Runs NoHMXB and NoWinds)}\label{sec:NoWinds}
\begin{figure}
	\includegraphics[width=\columnwidth]{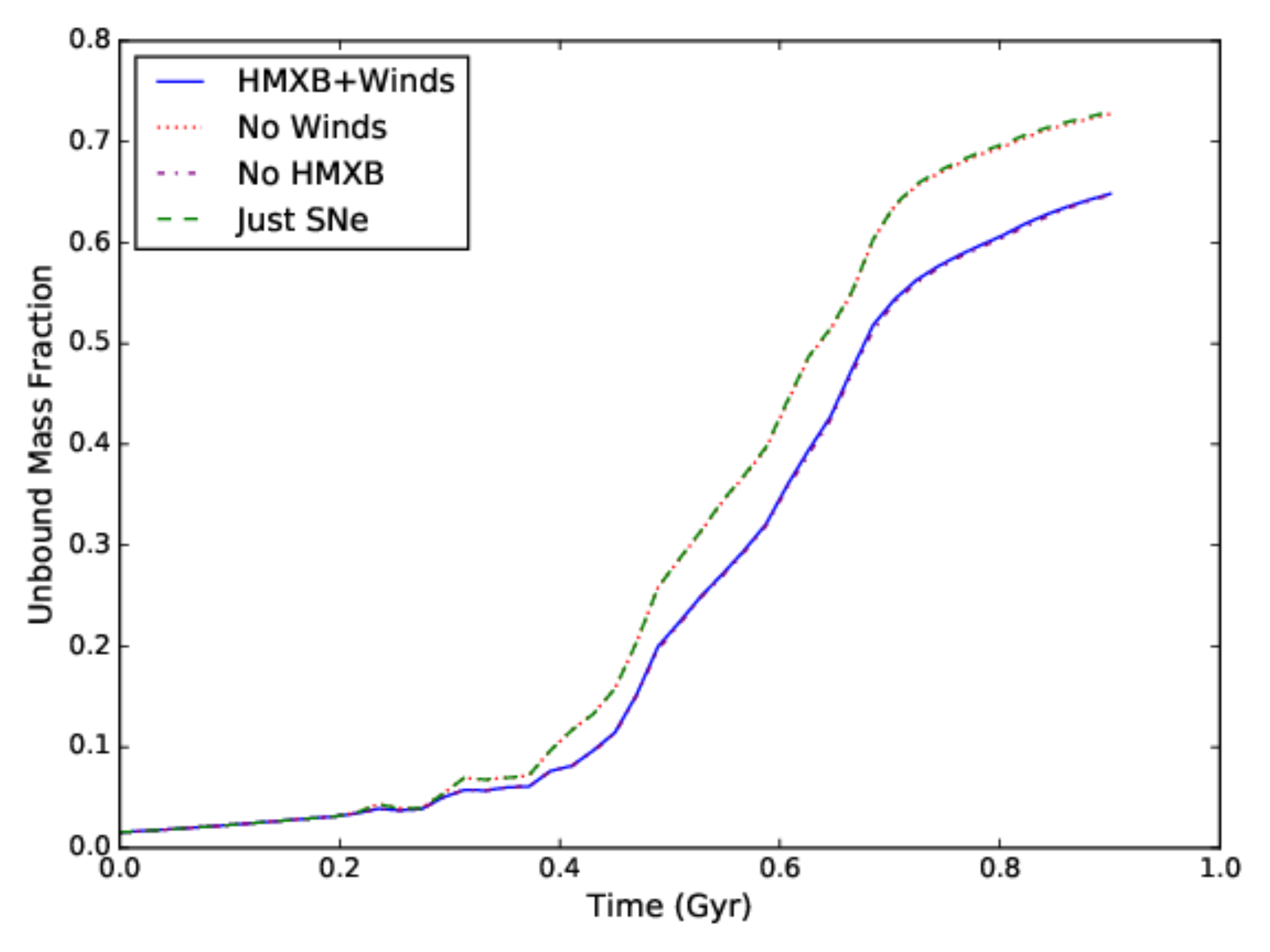}
    \caption{The time evolution of the fraction of the initial gas mass that has been unbound in Run 1 (which includes SNe, HMXBs and stellar winds, solid line), 2 (just SNe, dashed line), NoWinds (HMXBs and SNe, dotted line) and NoHMXB (winds and SNe, dot-dashed line). }
    \label{fig:Ubd_WindsVhmxb}
\end{figure}

Fig. \ref{fig:Ubd_WindsVhmxb} plots the unbound gas mass as a fraction of the initial mass in Runs 1, 2,  NoWinds and NoHMXB. All runs had the same initial conditions (halo mass - 1.1$\times$ 10$^8$ M$_\odot$, [Fe/H] = - 6), however used different combinations of stellar feedback (see Table \ref{tab:Runs}). We can see Runs 1 and NoHMXB are in good agreement, indicating the decrease in the amount of gas unbound via the addition of gradual feedback in Run 1 (compared with Run 2) is due to stellar winds, rather than HMXB feedback. Furthermore, by excluding stellar winds the unbound mass fraction has converged with Run 2, which just includes SNe feedback. Again, this points to the fact the ubiquity of stellar winds (seen in Fig. \ref{fig:Rel_Energy}), despite their relatively low power compared with HMXBs, means they have a far more significant impact on the fate of the gas in the galaxy. 
\begin{figure}
	\includegraphics[width=\columnwidth]{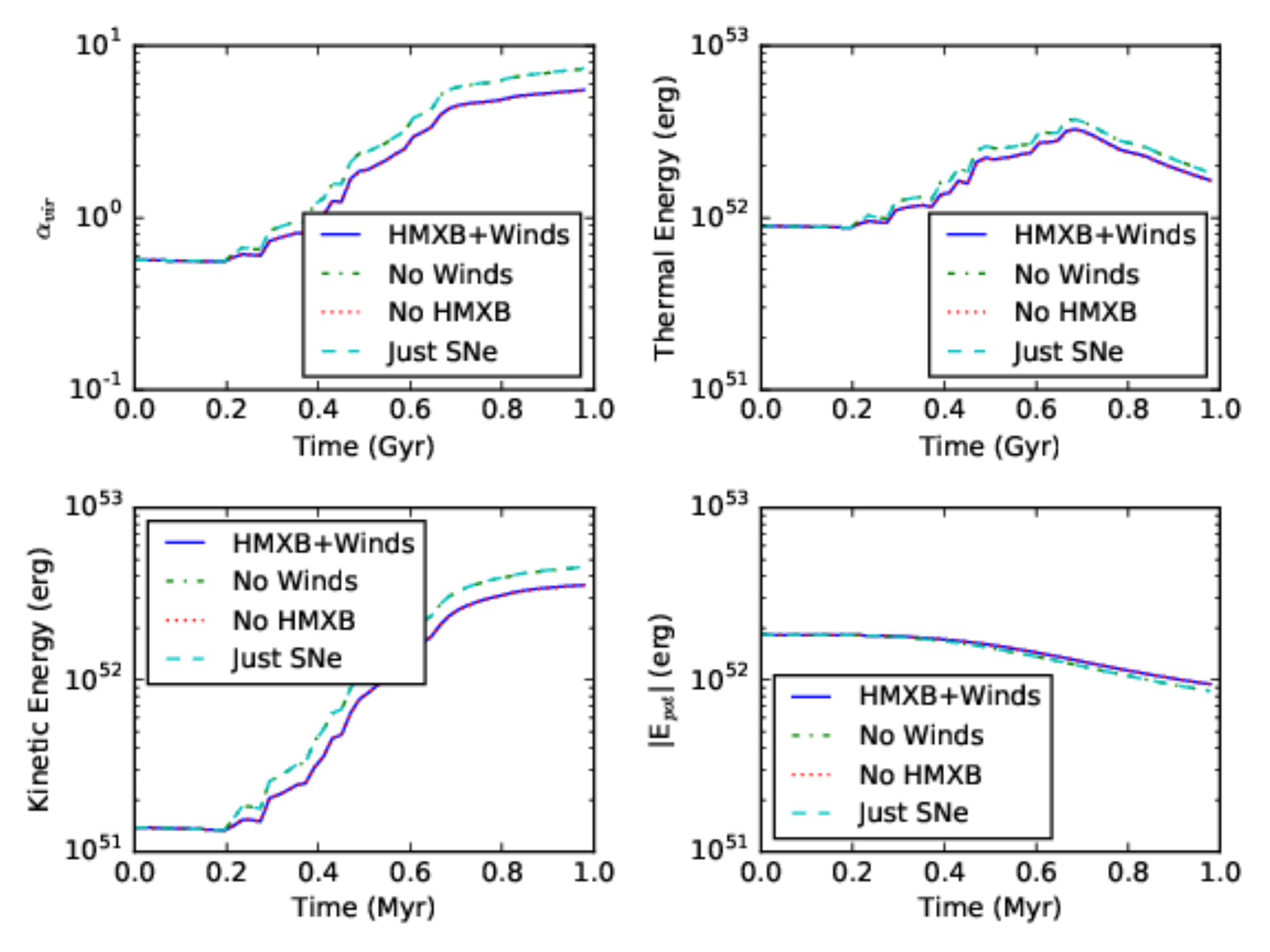}
    \caption{Plots to compare $\alpha_{vir}$ ($E_{therm} + E_{kin} / |E_{pot}|$ - top left plot), along with the total thermal (top right), kinetic (bottom left) and potential (bottom right) energies of the gas particles in Runs 1 (solid line), NoWinds (dot-dashed line line), NoHMXB (dotted line) and 2 (dashed line).}
    \label{fig:Eplots_WVHmxb}
\end{figure}

The same trends can be seen in the virial parameter, along with the total thermal, kinetic and potential energies of the gas in Runs 1, 2, NoWinds and NoHMXB (Fig. \ref{fig:Eplots_WVHmxb}). Once again, Runs 2 and NoWinds are converged, indicating HMXBs have had little impact on the energetics of the gas when added on top of SNe feedback. Moreover, Run 1 and NoHMXB also converge, indicating including stellar winds is the defining factor in determining the total kinetic, thermal energy of the galactic ISM.

\newnote{The fact it is the stellar wind feedback that governs the energetics of the galaxies in our simulations, indicates the importance of the timing of individual feedback events. Stellar winds act prior to both SNe and HMXB feedback and this pre-processing of the local ISM appears to have the greatest impact on the ability of SNe to drive gas out of the galaxy. This is despite HMXBs acting for longer and with higher power. Stellar wind events are also more numerous, giving them higher spatial coverage of the galaxy. However, how this result would change were star formation to be followed self-consistently, or if HMXBs were to be modelled non-isotropically, is an interesting question beyond the scope of this paper.}

Overall, it is the inclusion of stellar winds that has driven the observed trends in the fate of the ISM that are the key results of this paper. \note{This is due to both their ubiquity and the fact they act to pre-process the ISM prior to the onset of SNe feedback.}

\section{Conclusions and Discussion}
In this paper we investigated the inclusion of `gradual' feedback types on top of SNe feedback in high redshift isolated dwarf galaxies with halo masses between 1.1$\times$10$^8$ M$_\odot$ and 1.5$\times$10$^{7}$ M$_\odot$. The main results of this paper are as follows:
\begin{enumerate}
\item The fraction of the initial gas mass that has been unbound after 1 Gyr starburst is uniformly lower in the galaxies that include gradual feedback mechanisms on top of SNe feedback. This result holds across the galaxy masses we tested, along with varying metallicity, halo concentration and the duration of the starburst. 
\item By varying the standard deviation ($\sigma_{star}$) of the `wake-up' times of the stars (and therefore the duration of the starburst), we found the total kinetic energy of the gas in the simulations with varying $\sigma_{star}$ converged beyond 800 Myr. Moreover, the total internal energy of the gas was marginally lower in the 0.5 $\sigma_{star}$ runs, indicating the gas was able to cool more efficiently in the shorter, more violent starburst. The fact the energetics of both the $\sigma_{star}$ runs and the 0.5 $\sigma_{star}$ runs are so similar at the end of the starburst, indicates radiative cooling is \newnote{as ineffective at lower metallicity as it is at high metallicity, most likely due to the low density of the gas present towards the end of the starburst}.  
\item \newnote{Similarly,} altering the metallicity of the gas in the galaxy from [Fe/H] = -6 to [Fe/H] = - 1.2 had very little impact on the state of the gas \newnote{at the end of the starburst}, \newnote{again, most likely due to the low densities of the gas present in each simulation.}
\item Decreasing the duration of the starburst (while keeping the total number of feedback events the same) increased the gas density below 10 kpc at the end of the runs which just included SNe feedback, along with inside the inner 1 kpc of runs which also included gradual feedback.
\item As expected, increasing the concentration parameter lowered the amount of gas unbound by the stellar feedback by increasing the potential at the centre of the halo. 
\item Altering the size of the galaxy had the largest impact on our results. In these runs the majority of the gas was unbound in both simulations. However, the addition of gradual feedback acted to delay the unbinding of the gas. Furthermore, the addition of stellar winds facilitated the production of chimneys; which are holes in the high density shell surrounding the volume-filling, pressurised gas that vent the hot gas from the centre of the halo. These chimneys globally reduce the temperature of the gas in the central 10 kpc of the galaxy.
\item Finally, we evaluated the relative impact of HMXB feedback and stellar winds on our results, finding the ubiquity of stellar winds means they are more or less solely responsible for changes in the unbound mass and gas energetics seen in the 1.1$\times$10$^{8}$ M$_{\odot}$ galaxy. How this changes with galaxy mass and metallicity would be interesting follow-up work however is beyond the scope of this paper. It would also be interesting to investigate how this result varies if the HMXB feedback is no longer implemented isotropically.
\end{enumerate}
\subsection{Our results in context}
Our results point to the importance of considering time-resolved feedback events, including both instantaneous, spatially resolved SN, along with the continuous low-powered energy injection of winds. For example, the chimney seen in Run 5 was formed via colliding wind-fed bubbles, a process which is only possible if individual massive stars are resolved spatially. This result links to work by \citet{Su2017}, who found modelling discrete SNe events was essential to capturing the multiphase ISM along with forming galactic winds with physical mass loading factors. Moreover, since the mean temperature of the gas in our simulations was typically below 10$^4$ K, our results also highlight the need to include low temperature cooling prescriptions in these low mass galaxies. 

\newnote{Our results highlight the need include further feedback mechanisms and a model for the complex, multi-phase ISM on top of SNe feedback when considering the star formation histories and evolution of dwarf galaxies. This is in line with recent work by \citet{Smith2019}, who input a novel SNe feedback prescription in cosmological zoom-in simulations of dwarf galaxies, finding the stellar mass is overestimated at all redshifts due to additional feedback mechanisms and ISM physics (such as turbulent pressure support) being neglected.}

There is an existing large volume of work on SNe-generated superbubbles, similar to the 10 kpc-scale bubble of hot pressurised gas seen in Fig. \ref{fig:R56_xz_05Gyr}. This includes both analytical work \citep[e.g.][]{Weaver1977, Tomisaka1981}, along with multiple observations of superbubbles such as \citet{Pidopryhora2007, Ochsendorf2015}. Typically the radii of these superbubbles are an order of magnitude lower than the superbubble seen in Run 6, however once we include gradual feedback this radius is reduced.

The superbubbles seen in Run 5 and 6 (see Fig. \ref{fig:R56_xz_05Gyr}) have similar properties to the superbubbles investigated in \citet{Kim2017}. Like the superbubbles presented in \citet{Kim2017}, the gas in the centre of the bubble has been heated to between 10$^{6-7}$ km/s and accelerated to $\sim$ 10$^{2}$ km/s. The expansion velocity of between 10-100 km/s is enough to unbind the gas from the shallow potential wells of the dwarf galaxy. In our work the bubble in Run 6 is also surrounded by a cooler high density shell, which was also present in \citet{Kim2017}. In \citet{Kim2017}, this shell was formed of warm, swept-up gas that has cooled. The range of (cool) temperatures seen in this shell in our simulations would also fit with this hypothesis. In Run 5, the shell has a more complicated morphology, while the shell is less spatially extended. This makes it easier for the hot, feedback-heated gas to clear low density channels or `chimneys' in the shell. These chimneys have the capacity to launch hot, high velocity gas to a radius of $\sim$ 10$^2$ kpc (as seen in Fig. \ref{fig:R56_yz}). 

`Chimneys' have been previously linked to stellar winds in MCs. In particular, \citet{Rogers2013} found winds acting prior to SNe feedback preferentially escape the inner molecular cloud via paths of least resistance, creating low density channels through which the proceeding SNe energy can escape with weak coupling to the higher density molecular clumps at the centre of the cloud. We can see this process in action in our simulations, in particular Fig. \ref{fig:R5_xz_evoln}. \newnote{As well as this, there have been numerous observations of 10$^7$ K plasma in the interstices between star-forming regions, fed by the winds from OB stars \citep[e.g.][]{Townsley2018, Townsley2014}.}

\note{Our result that the energetics of the gas in the galaxy varies little with metallicity is due to the fact our simulations do not self-consistently form stars; instead they are implemented as a pre-determined population. \newnote{If we were} to form stars naturally out of the gas this would create differences between simulations, for example in stellar masses, formation times and locations. However, since the stars associated with feedback are the same between runs of varying metallicity, the main difference is that the cooling rate of the gas is altered due to different metal species being present. Moreover, in the case of these simulations, the density of the gas is low enough that the relative cooling rate between gas of different metallicities is similar.}

Furthermore, our result that the addition of HMXB feedback on top of SNe feedback does not significantly alter the state of the gas in dwarf galaxies, beyond the effects of SNe, \newnote{is broadly consistent with work by \citet{Artale2015}, who found including HMXB on top of SNe feedback did not significantly alter the mass of gas retained within the optical radius of a galaxy for galaxies with mass less than 10$^{10}$ M$_{\odot}$ (see Fig. 5 of their paper)}. However, our halo masses of 10$^{7}$ M$_\odot$ and 10$^8$ M$_\odot$ are below their mass range and hence are not directly comparable. Moreover, it is also possible our results are dependent on metallicity, as was seen in P08.

\note{In Appendix \ref{appendix:conv_test} we briefly explore the impact of varying the locations of stars (or in this case where their hot feedback-generated bubbles collide), finding this can have a significant impact on the overall unbinding of the gas in the galaxy. This stochasticity may help to explain the wide variety of star formation histories seen in dwarf galaxies \citep[e.g.][]{Tolstoy2009, McConnachie2012} and is worthy of further exploration in future papers.}

Fundamentally, the fact the dwarf galaxies retain more gas if gradual feedback is included, means they can fuel further star formation. This can help explain the star formation histories seen in dwarf satellite galaxies such as Fornax, Leo I, And VI and Carina, which show multiple starbursts across $\sim$ 12 Gyr \citep{Weisz2014}.

\section*{Acknowledgements}
During this work, LGS was supported by a Science and Technology facilities council (STFC) PhD studentship. CP  is  supported  by  Australia Research Council (ARC) Future  Fellowship FT130100041. CJN is supported by the Science and Technology Facilities Council (STFC) (grant number ST/M005917/1). This work used the DiRAC Complexity system, operated by the University of Leicester IT Services, which forms part of the STFC DiRAC HPC Facility. This equipment is funded by BIS National E-Infrastructure capital grant ST/K000373/1 and STFC DiRAC Operations grant ST/K0003259/1. DiRAC is part of the National E-Infrastructure. Figures \ref{fig:TD_xz_plane_R12}, \ref{fig:R9v10xz}, \ref{fig:R3v4_yz}, \ref{fig:R56_yz}, \ref{fig:R6_DensVrad}, \ref{fig:R6_vr}, \ref{fig:R5_DvR}, \ref{fig:R56_xz_05Gyr}, \ref{fig:R5_xz_evoln} and \ref{fig:conv_proj_dens} were produced using SPLASH \citep{Price2007}. 

\bibliographystyle{mnras}
\bibliography{Galactic_Chimney}

\appendix
\section{Changing the dark matter softening length}\label{appendix:soft}
In order to ensure our choice in softening length for the dark matter did not determine the results in this paper, we ran simulations varying the softening length between 1pc to 100 pc. 

\begin{figure}
	\includegraphics[width=\columnwidth]{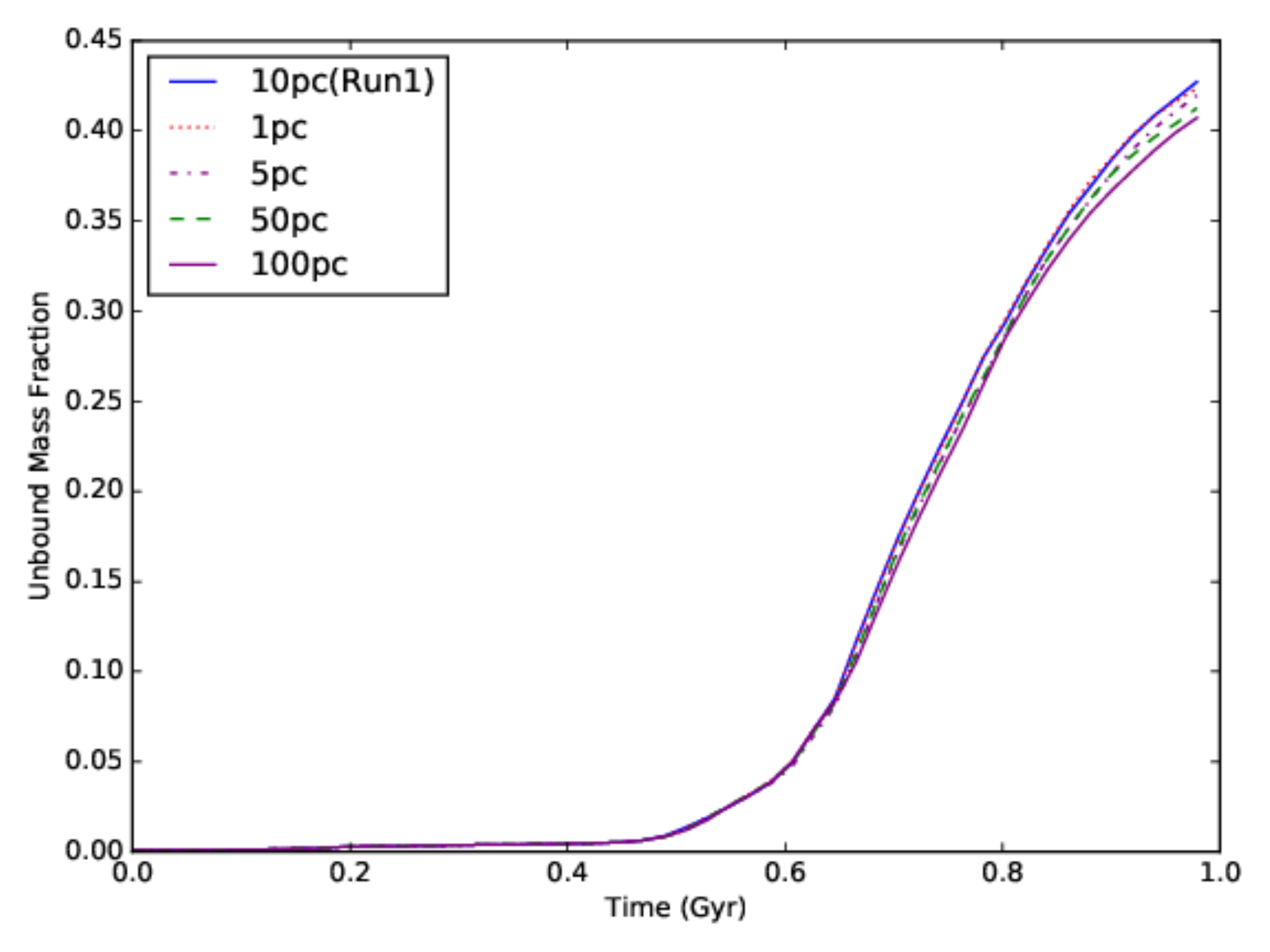}
    \caption{The time evolution of the fraction of the initial gas mass that has been unbound in simulations that have the same initial conditions (as Run 1), however dark matter softening lengths varying from 1 pc to 100 pc. }
    \label{fig:Ubd_Hsml}
\end{figure}

\begin{figure}
	\includegraphics[width=\columnwidth]{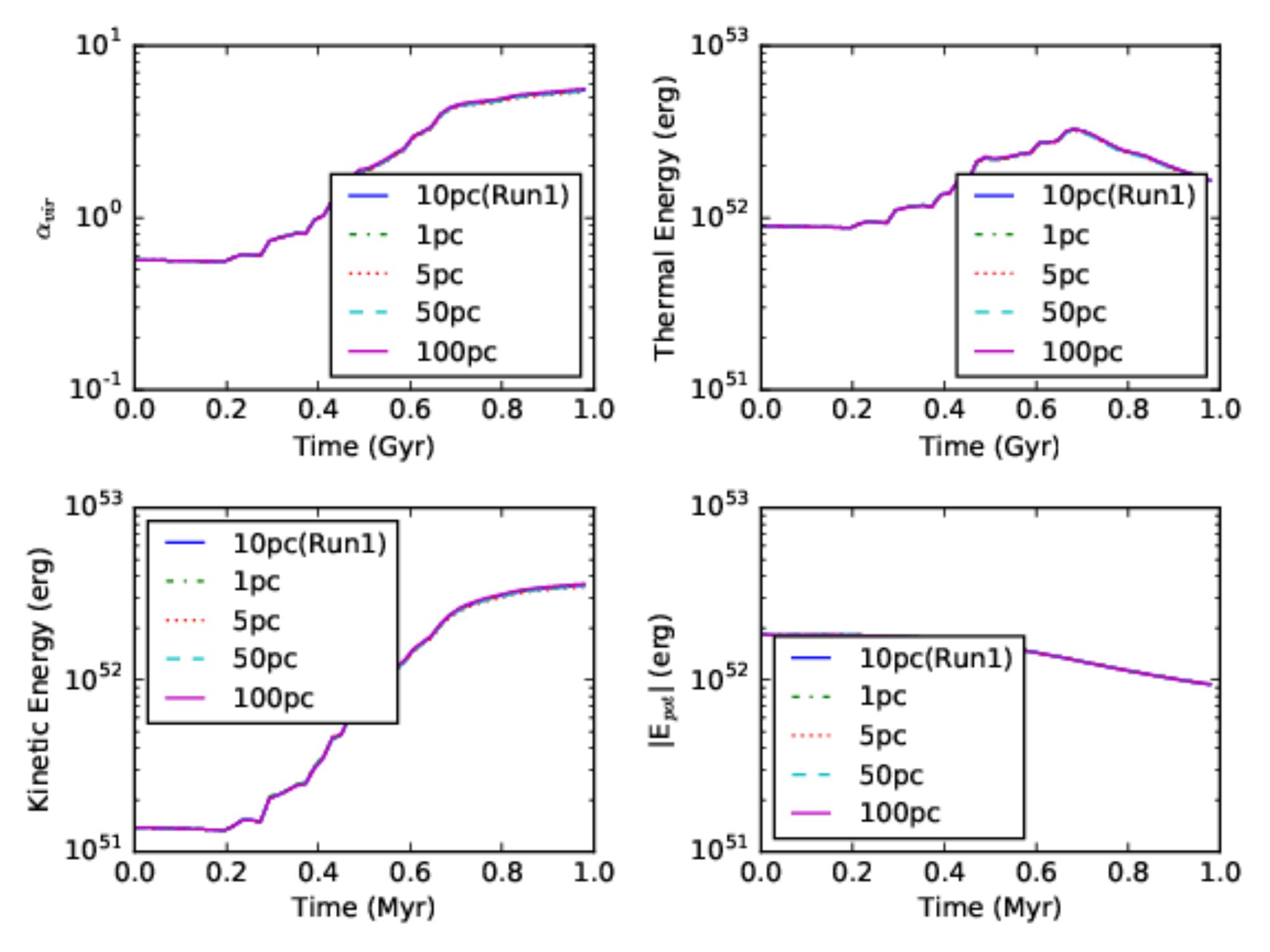}
    \caption{The time evolution of the virial parameter (evaluated as $\alpha_{vir}$ = E$_{kin}$ + E$_{therm}$ / |E$_{pot}$|), along with the total thermal, kinetic and potential energy of the gas in simulations with the same initial conditions however using different softening lengths for the dark matter.}
    \label{fig:Eplots_Hsml}
\end{figure}

We plot the time evolution of the unbound mass fractions of each run in Fig. \ref{fig:Ubd_Hsml}. Here we see the total unbound mass fraction varies by 0.025, however the run at the highest resolution (with a softening length of 1 pc) is converged with our results run (which uses a softening length of 10 pc). Moreover, when we plot the total kinetic, thermal and potential energies of the gas in each simulation, along with the virial parameter (Fig. \ref{fig:Eplots_Hsml}), we see all runs are converged - indicating changing the softening length of the dark matter has had little effect on the final state of the gas in the simulation. 

\begin{figure}
	\includegraphics[width=\columnwidth]{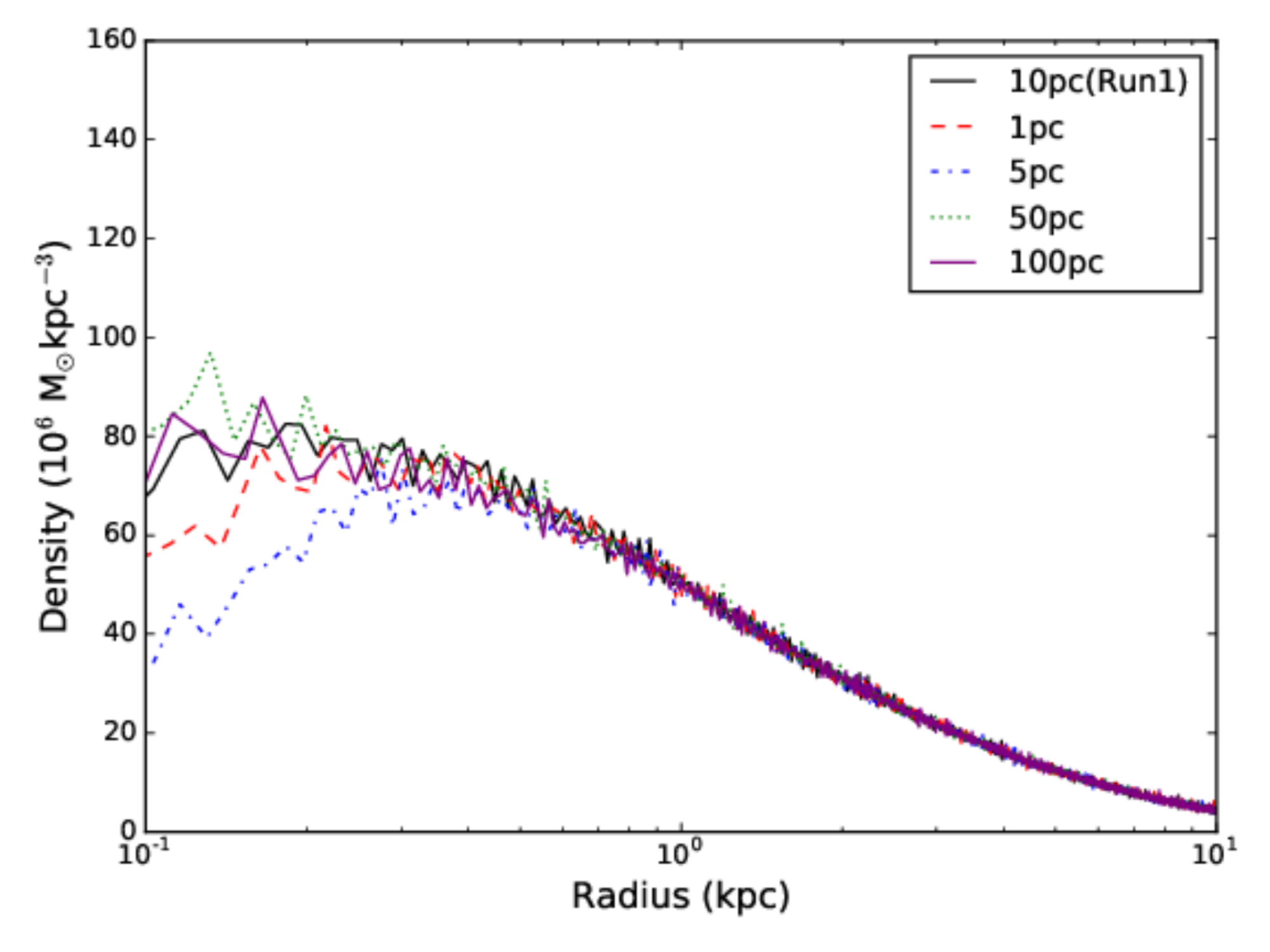}
    \caption{The dark matter density profiles taken at 1 Gyr for simulations with the same initial conditions, however different softening lengths for the dark matter.}
    \label{fig:DMDens_Hsml}
\end{figure}

Furthermore, in Fig \ref{fig:DMDens_Hsml} we plot the density profile of the dark matter density profile in each run. We see the profiles are converged above $\sim$ 0.5 kpc, however at the smallest scales (below 200 pc) the runs diverge, with no clear trend with softening length. This indicates a degree of noise on these scales. 

\section{Simulations with no stellar feedback included}\label{appendix:Nofb}
In order to check the changes associated with the gas and the dark matter we see in our results are entirely the result of stellar feedback, we ran three simulations with the initial conditions of Runs 1 (labelled 1.1$\times$10$^{8}$ M$_{\odot}$, referring to the mass of the galaxy), 3 (labelled 1.1$\times$ 10$^{8}$ M$_{\odot}$ - double c) and 5 (labelled 1.5$\times$10$^{7}$ M$_{\odot}$), however with no feedback included.

\begin{figure}
	\includegraphics[width=\columnwidth]{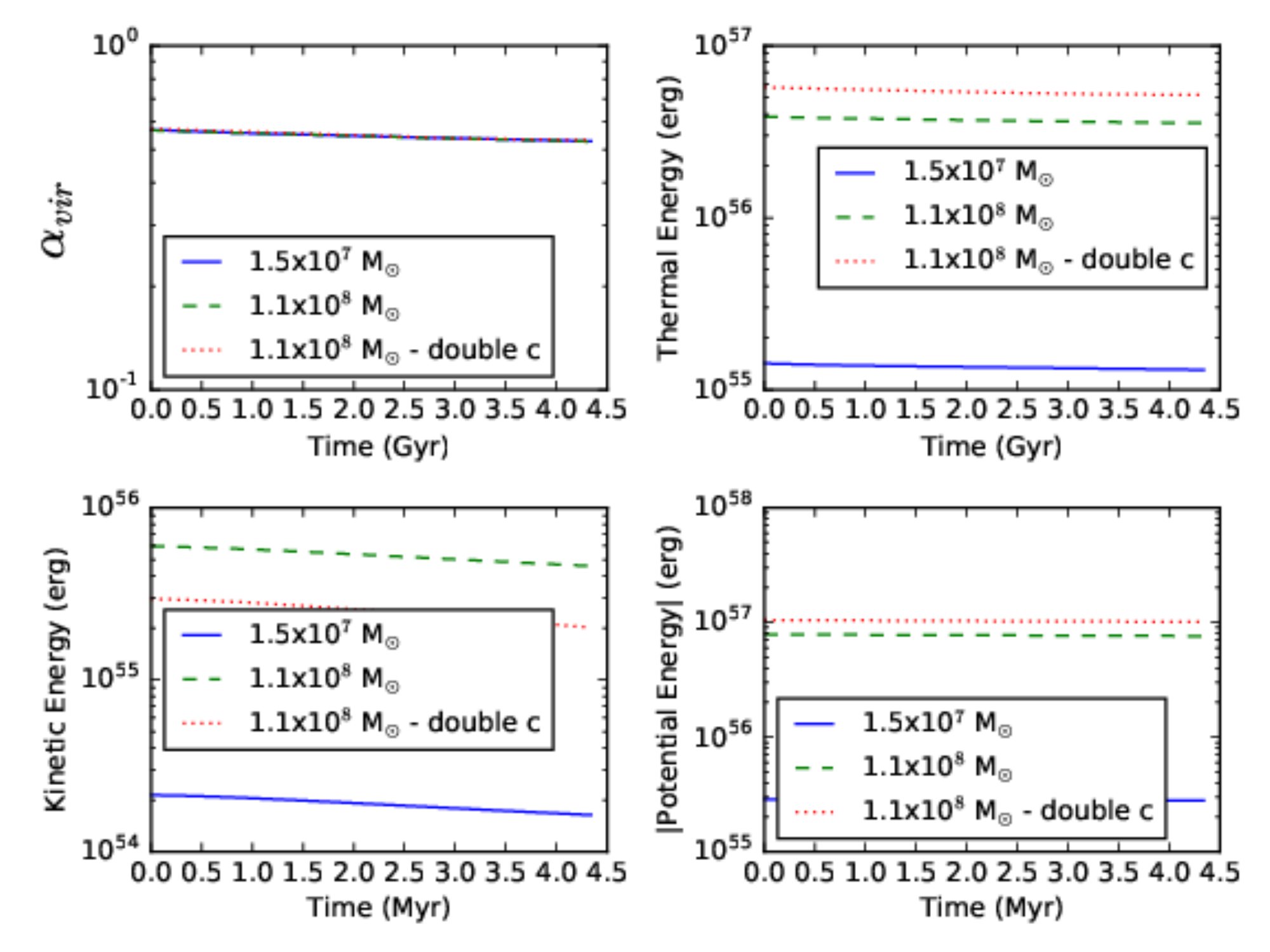}
    \caption{The time evolution of the virial parameter (evaluated as $\alpha_{vir}$ = E$_{kin}$ + E$_{therm}$ / |E$_{pot}$|), along with the total thermal, kinetic and potential energy of the gas in simulations which do not include feedback.}
    \label{fig:Eplots_nofb}
\end{figure}

\begin{figure}
	\includegraphics[width=\columnwidth]{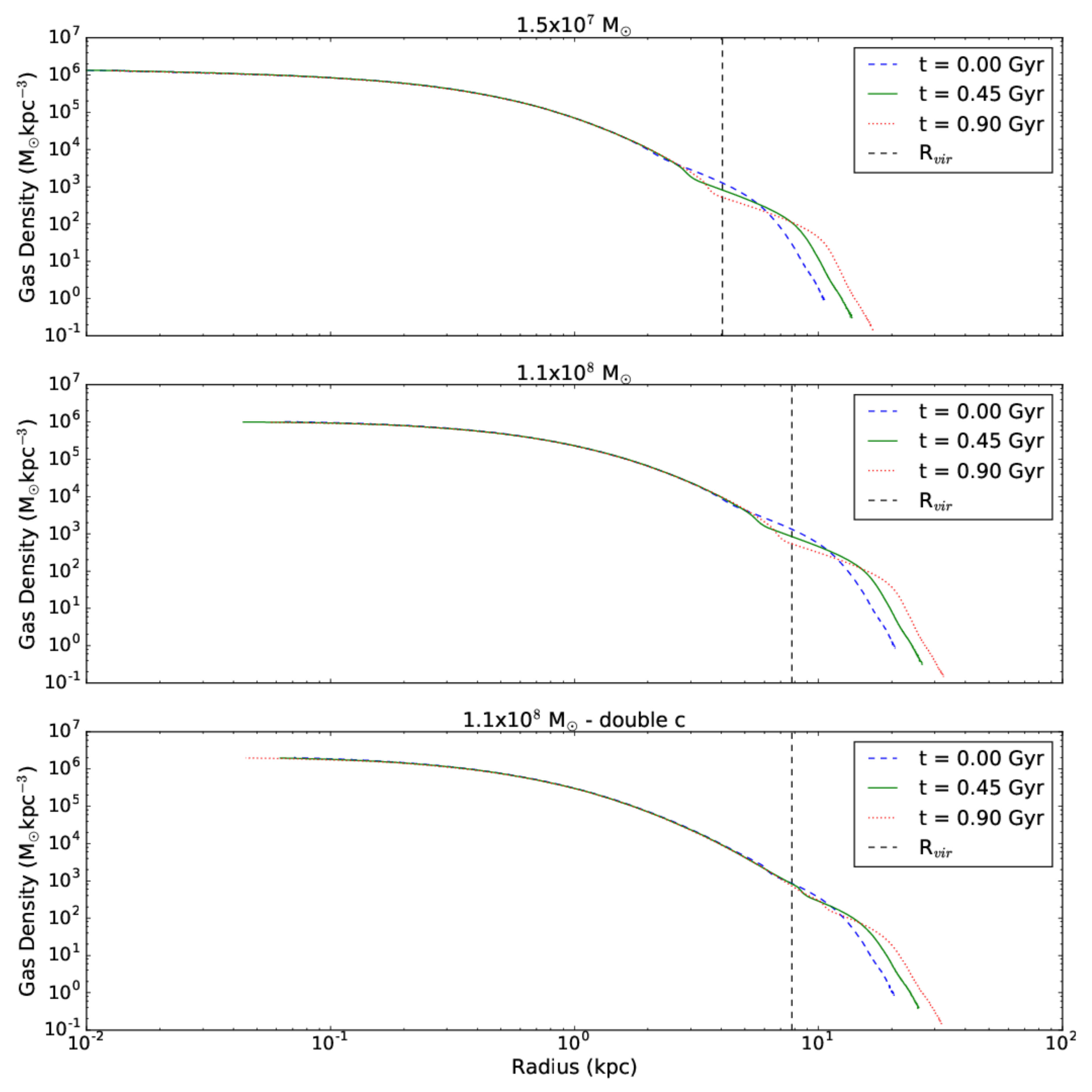}
    \caption{The gas density profile at varying times in simulations without feedback included.}
    \label{fig:GasDens_Nofb}
\end{figure}

\begin{figure}
	\includegraphics[width=\columnwidth]{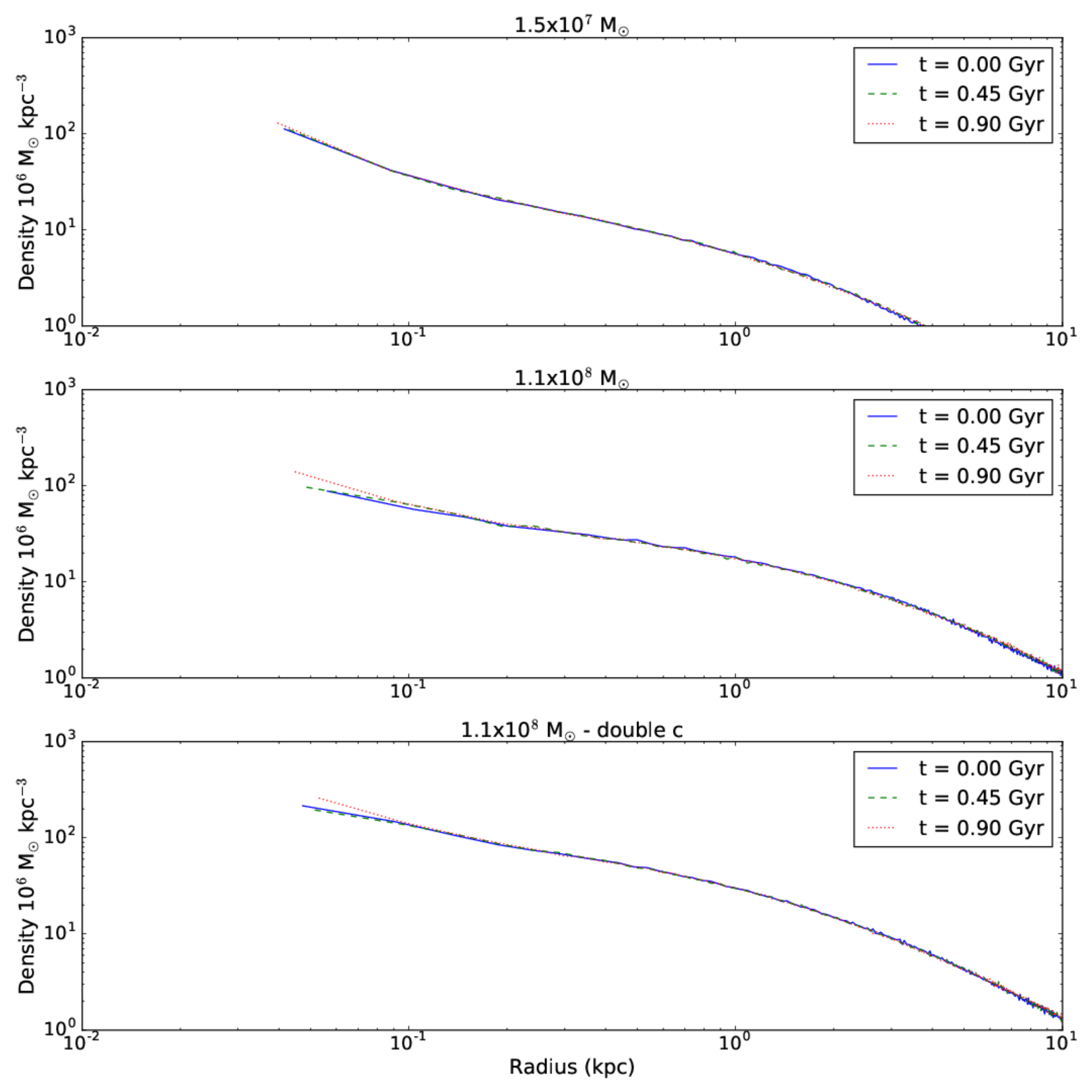}
    \caption{The dark matter density profile at varying times in simulations without feedback included.}
    \label{fig:DMDens_Nofb}
\end{figure}

In Fig. \ref{fig:Eplots_nofb} we plot the time evolution of the energetics of the gas, finding this is unchanging in all three simulations, indicating the gas is in a steady state. Furthermore, the virial parameter is consistently at $\sim$ 0.5, indicating all three systems are virialised.

We also check the density profile of the gas (Fig. \ref{fig:GasDens_Nofb}) and dark matter (Fig. \ref{fig:DMDens_Nofb}) in all three simulations. We find both are in good agreement across all times towards the centre of the halo, however the gas density profile is noisy towards the virial radius (marked on the plot as a vertical dashed line). This noise is more apparent in the smallest halo, however insignificant in the halo with the largest concentration parameter.

\section{Numerical Convergence Testing}\label{appendix:conv_test}
\begin{figure}
	\includegraphics[width=\columnwidth]{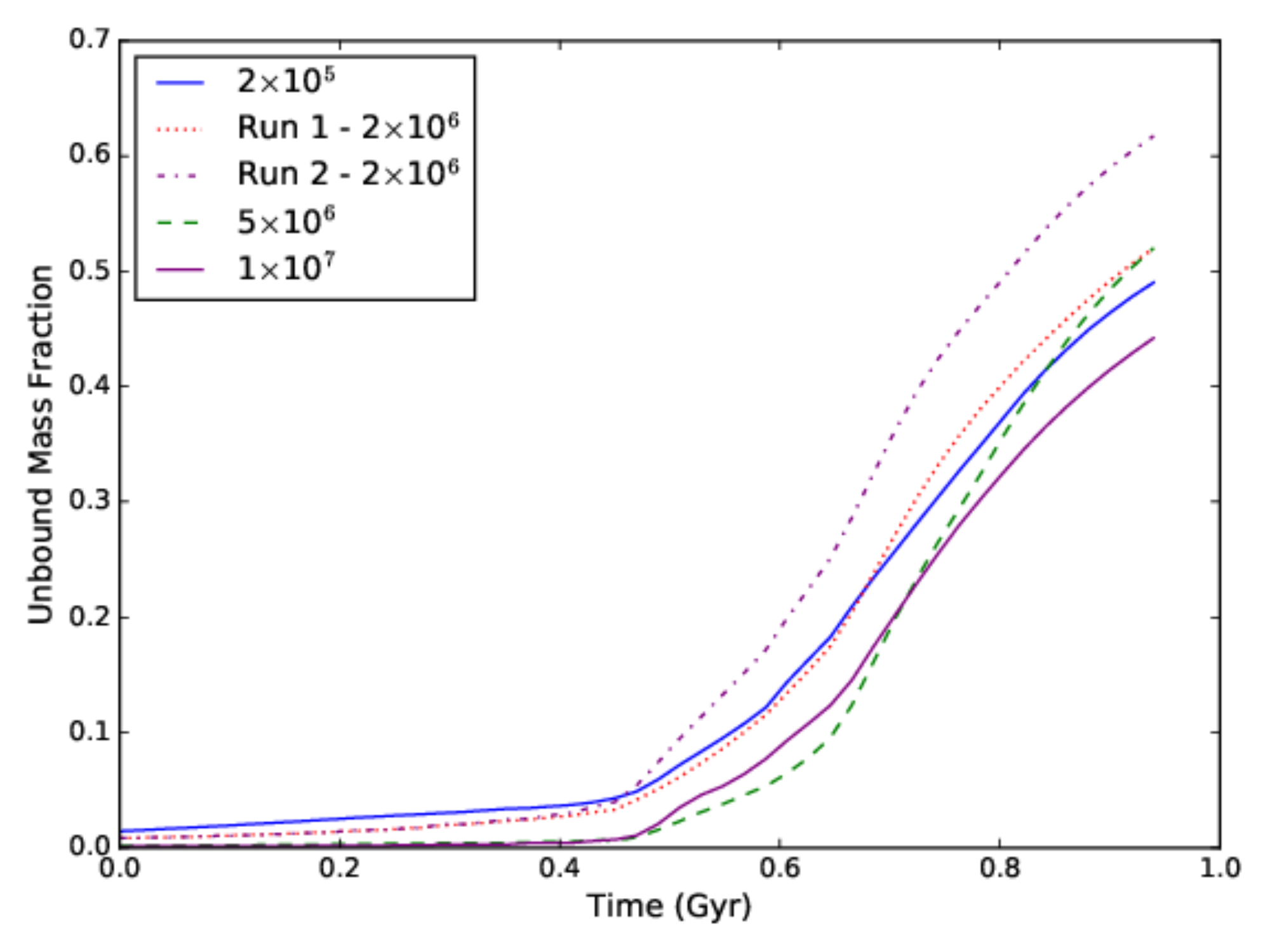}
    \caption{The time evolution of the fraction of the initial gas mass that has been unbound in runs of varying mass resolution (ranging from a particle number of 2$\times$10$^5$ to 1$\times$10$^7$). Run 2 (just including SNe feedback) has been added for comparison, however all other simulations have identical feedback prescriptions (HMXBs, SNe and winds), just altered resolution.}
    \label{fig:Conv_Test_Ubd_Frac}
\end{figure}

In order to test how numerical resolution effects the results of this paper, we re-ran Run 1 (which included SNe, HMXBs and stellar winds), with an SPH particle mass resolution of $\sim$ 90 M$_\odot$, 9 M$_{\odot}$, 3.5 M$_{\odot}$ and 2 M$_\odot$ (containing approximately 2$\times$10$^{5}$, 2$\times$10$^{6}$, 5$\times$10$^{6}$ and 1$\times$10$^{7}$ particles respectively). All other variables were kept constant between runs, including the stellar feedback prescriptions. In Fig. \ref{fig:Conv_Test_Ubd_Frac} we plot the time evolution of the unbound mass fraction for each resolution, alongside the results for Run 2 for comparison. We can immediately see the evolution of the unbound mass fraction differs between runs of varying resolution, along with the final unbound mass fraction at the end of each simulation. However, there does not appear to be a clear correlation with resolution, indicating by changing the resolution of the simulations we have introduced a degree of stochasticity to our results, which we will discuss below. Moreover, the dominant effect is still the addition of gradual feedback (comparing Runs 1 and 2), since the addition of gradual feedback has had the largest impact on the final unbound mass fraction. This indicates the primary result of this paper is robust; by adding gradual feedback mechanisms the mass of gas unbound in a 1 Gyr starburst can be reduced in dwarf spheroidal galaxies of typical z$=$0 mass 1$\times$10$^{9-10}$ M$_{\odot}$. 

\begin{figure}
	\includegraphics[width=\columnwidth]{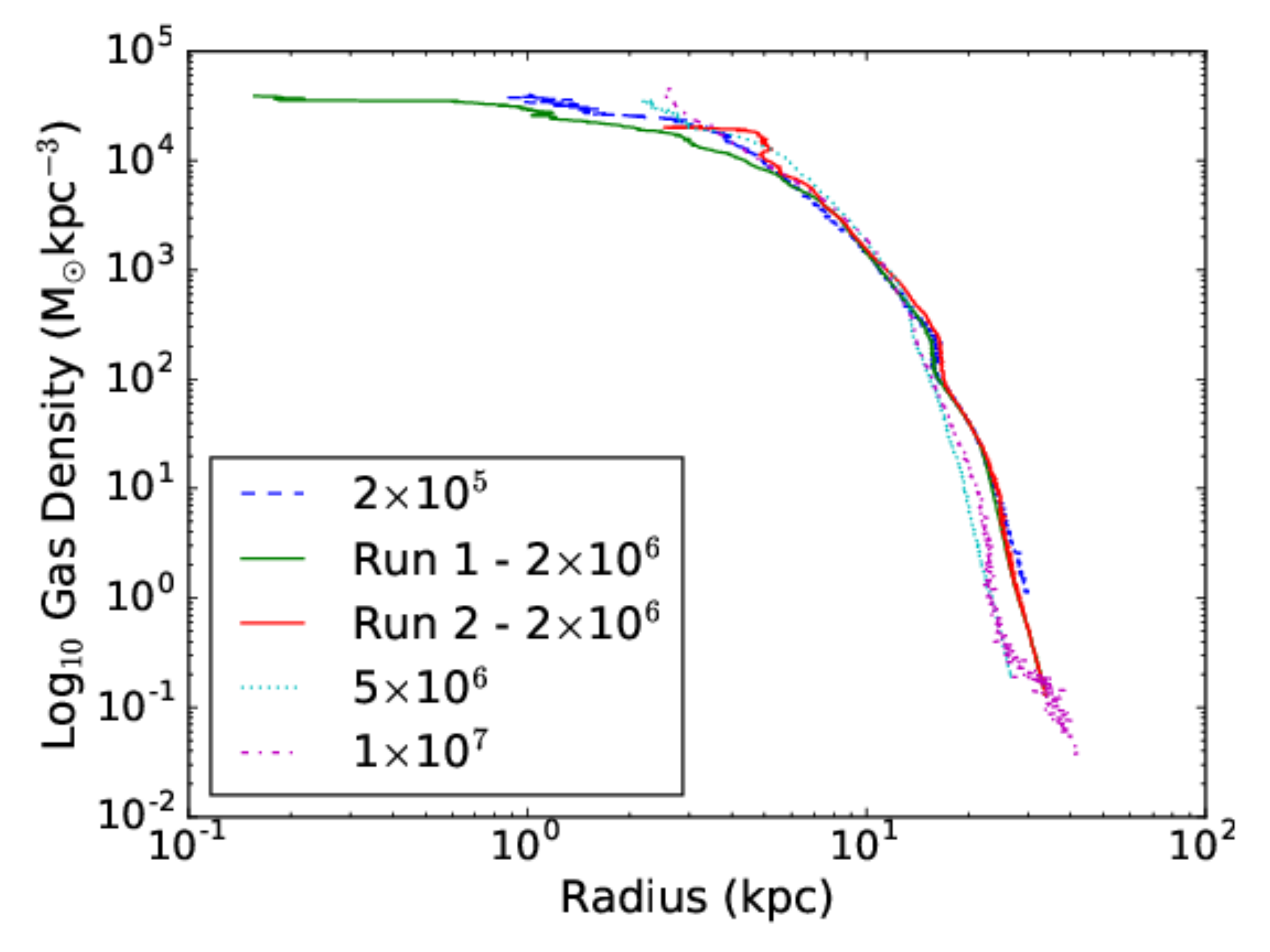}
    \caption{The gas density profiles taken at 1 Gyr into runs of varying mass resolution (ranging from a particle number of 2$\times$10$^5$ to 1$\times$10$^7$). Run 2 (just including SNe feedback) has been added for comparison, however all other simulations have identical feedback prescriptions (HMXBs, SNe and winds), just altered resolution.}
    \label{fig:GasDensity_Conv_Test}
\end{figure}

The differences seen in Fig. \ref{fig:Conv_Test_Ubd_Frac} are also present in Fig. \ref{fig:GasDensity_Conv_Test}, which plots the gas density profile at 1 Gyr for runs of varying resolution. Again, there is no clear trend with resolution, however the run without gradual feedback (Run 2) has a lower peak density than all other runs. From the particle data the peak density for Run 2 is 2.1$\times$10$^{4}$ M$_{\odot}$kpc$^{-3}$, while for the other runs plotted in Fig. \ref{fig:GasDensity_Conv_Test} it varies between 3.7$\times$10$^{4}$ - 4.7$\times$10$^{4}$ M$_{\odot}$kpc$^{-3}$. This implies the results of this paper are unaffected by numerical resolution; the factor that has had the largest impact on the gas density profile is whether or not gradual feedback has been included. It is also noticeable the two highest resolution runs have a density profile which diverges from the lower resolution runs beyond 10 kpc. This part of the profile is resolved by approximately 100 gas particles per kpc$^{3}$ for the result run and hence just 10 gas particles per kpc$^{3}$ for the lowest resolution run, hence it is likely to be subject to numerical noise. However, the primary concern of this paper is the inner parts of the galaxy, below the virial radius, where the majority of star formation takes place. 

The degree of stochasticity seen in figures \ref{fig:Conv_Test_Ubd_Frac} and \ref{fig:GasDensity_Conv_Test} is most likely a result of us keeping the feedback prescription the same, independent of resolution. This means the mass over which feedback energy is distributed differs between different resolution runs, along with the radius of influence. Focusing on the SNe feedback, this effectively means we are resolving different stages of the Sedov-Taylor expansion of a SN blast wave. If the Sedov-Taylor timescale inferred from the energy injection radius (given a set energy injection of 10$^{51}$ erg) is comparable to the time between individual feedback events, this would introduce a large degree of stochasticity in where the hot, feedback generated bubbles of gas collide and interact. 

\begin{figure}
	\includegraphics[width=\columnwidth]{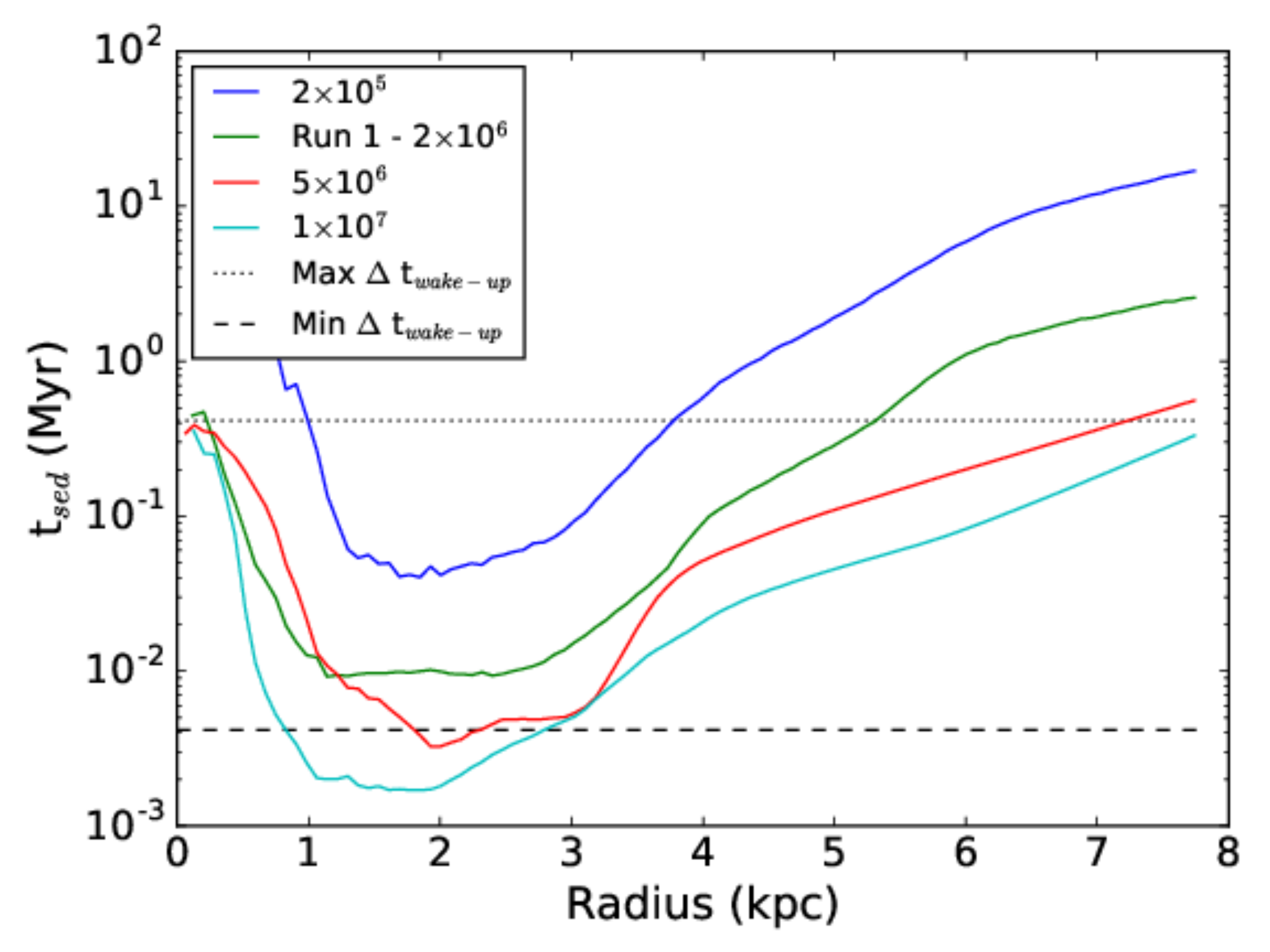}
    \caption{Plot to show the mean value of t$_{sed}$ in radial bins at 500 Myr into each simulation, where t$_{sed}$ represents the evolution time of a Sedov-Taylor blast wave of energy 10$^{51}$ erg, assuming it has reached a radius defined by the smoothing length of each gas particle. This has been plotted for runs with the same total gas mass, however different numbers of gas particles (ranging from 2$\times$10$^{5}$ to 1$\times$10$^{7}$). Also plotted are horizontal lines indicating the minimum (dashed line) and maximum (dotted line) time delay between successive star particle 'wake-up' times in all simulations.}
    \label{fig:tsedov}
\end{figure}

In order to investigate this hypothesis, we plot the mean t$_{sed}$ (the evolution timescale of a Sedov blast wave, assuming it had reached the smoothing length of the particle and had an associated energy of 10$^{51}$ erg) for gas particles in radial bins, 500 Myr into each simulation (Fig. \ref{fig:tsedov}). We only plot below the virial radius of the galaxy since in our simulations star particles cannot undergo feedback if they are located beyond this radius. We also plot the minimum and maximum time delay time between successive 'wake-up' times of individual star particles (see section \ref{sec:feedback_prescription}). From Fig. \ref{fig:tsedov} we can see only the highest resolution runs show t$_{sed}$ times below the minimum time difference between star particle 'wake-up' times, while the majority of gas particles (and hence star particles) sit above this line by over an order of magnitude. This means altering the radius of influence of a star particle between different resolutions (and hence t$_{sed}$) is likely to have a large impact on the interaction between spatially separated feedback events. This in turn will effect where/ when low density voids/chimneys develop and hence introduce a degree of stochasticity to the unbinding of the gas. 

\begin{figure*}
	\includegraphics[trim={0 4cm 0 0}, clip, width=\textwidth]{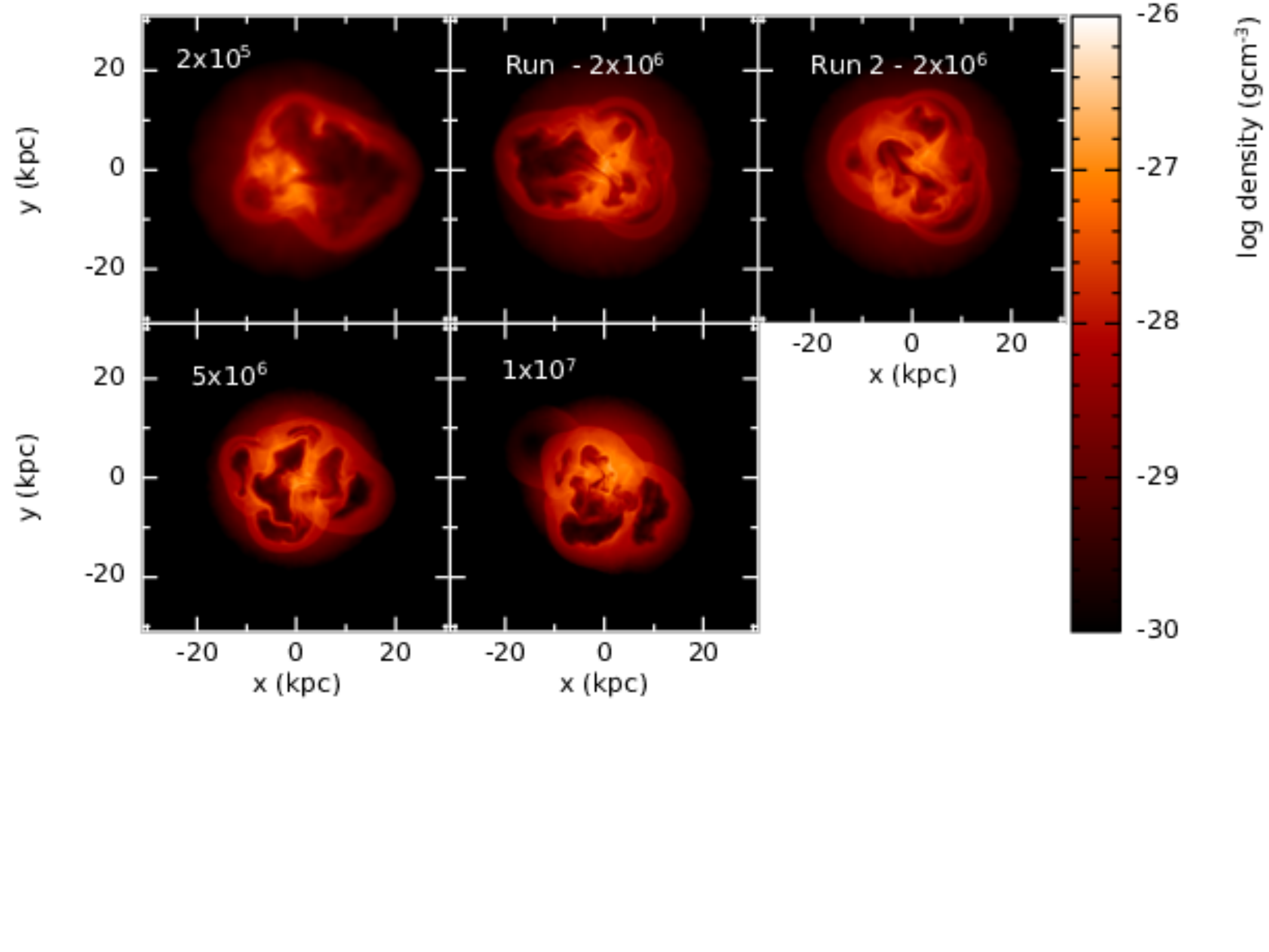}
    \caption{\newnote{Density slices in the x-y plane taken at z = 0, 1 Gyr into Run 1 and Run 2,} along with repeat simulations of Run 1 at varying mass resolution; using the same total gas mass, however different numbers of gas particles ranging between 2$\times$10$^{5}$ - 1$\times$10$^{7}$ particles.}
    \label{fig:conv_proj_dens}
\end{figure*}

This stochasticity in the location of feedback-generated bubble interactions can be seen in Figure \ref{fig:conv_proj_dens}, which shows a large variation in the gas morphology seen in the \newnote{density slices taken in the x-y plane at z = 0} from the runs of varying resolution, 1 Gyr into each simulation. This variation in morphology is larger for runs of varying resolution, compared to the differences seen when the feedback prescription is varied (comparing Runs 1 and 2). Given the difference between Runs 1 and 2 is whether or not gradual feedback mechanisms are present, this suggests SNe are primarily responsible for the morphology of the gas and hence the locations of low density voids/ chimneys.

It is worth noting the lowest resolution run may also be entering the regime where the Sedov-Taylor expansion can no longer be resolved, hence it will begin to suffer from numerical over-cooling. For Sedov-Taylor testing of the SN prescription we use in this paper, see Appendix A in GS18.


\bsp	
\label{lastpage}
\end{document}